\documentclass[12pt]{iopart}
\usepackage{graphicx}
\usepackage{siunitx}
\usepackage[toc,page]{appendix}
\usepackage{hyperref}
\newcommand{\angstrom}{\mbox{\normalfont\AA}}
\usepackage{amssymb}

\usepackage{pdflscape}

\usepackage{rotating}

\usepackage[margin=1in,letterpaper]{geometry} 
\usepackage{caption}
\usepackage{subcaption}
\usepackage[rightcaption]{sidecap}

\usepackage[utf8]{inputenc}
\usepackage{newunicodechar}
\newunicodechar{≤}{\ensuremath{\leq}}
\DeclareUnicodeCharacter{2264}{$\leq$}
\DeclareUnicodeCharacter{2032}{'}
\DeclareUnicodeCharacter{2033}{''}
\DeclareUnicodeCharacter{2212}{-}

\begin{document}

\title[Sciortino et al. 2021]{Experimental Inference of Neutral and Impurity Transport in Alcator C-Mod Using High-Resolution X-Ray and Ultra-Violet Spectra}

\author{F. Sciortino\textsuperscript{1}\footnote{Email: sciortino@psfc.mit.edu}, N.T. Howard\textsuperscript{1}, R. Reksoatmodjo\textsuperscript{2}, A.R. Foster\textsuperscript{4}, J.W. Hughes\textsuperscript{1}, E.S. Marmar\textsuperscript{1}, M.A. Miller\textsuperscript{1}, S. Mordijck\textsuperscript{2}, T. Odstr\v{c}il\textsuperscript{3}, T. P{\"u}tterich\textsuperscript{5}, M.L. Reinke\textsuperscript{6}, J.E. Rice\textsuperscript{1}, P. Rodriguez-Fernandez\textsuperscript{1}}

\address{$^1$ Massachusetts Institute of Technology, Cambridge, MA, 02139, USA}
\address{$^2$ Department of Physics, College of William \& Mary, Williamsburg, VA, USA}
\address{$^3$ General Atomics, P.O. Box 85608, San Diego, CA 92186-5608, USA}
\address{$^4$ Center for Astrophysics $\vert$ Harvard \& Smithsonian, 60 Garden St, Cambridge MA 02138, USA}
\address{$^5$ Max-Planck-Institut für Plasmaphysik, Boltzmannstraße 2, D-85748 Garching, Germany}
\address{$^6$ Commonwealth Fusion Systems, Cambridge, MA 02139, USA}

\vspace{10pt}

\date{\today}

\begin{abstract}
We present experimental inferences of cross-field impurity transport coefficients for Alcator C-Mod plasmas without edge-localized modes, leveraging a novel forward model for the entire Ca K$_\alpha$ spectrum, including satellite lines within the spectral range, to compare to high-resolution X-ray Imaging Crystal Spectroscopy (XICS). These measurements are complemented by Extreme Ultra-Violet (EUV) spectroscopy that constrains transport closer to the edge. Using new atomic data sets for both XICS and EUV analysis has enabled consideration of line ratios across both spectral ranges and has increased the accuracy of inferred transport coefficients. Inclusion of charge exchange between edge thermal neutrals and impurities is shown to be extremely important in C-Mod pedestals. Atomic D neutral densities from experimental D Ly$_\alpha$ measurements at the midplane are compared to SOLPS-ITER simulations, finding good agreement. Bayesian inferences of impurity transport coefficients are presented for L-, EDA H-, and I-mode discharges, making use of the Aurora package for forward modeling and combining our spectroscopic constraints. Experimentally inferred diffusion profiles are found to match turbulent transport models at midradius within uncertainties, using both quasilinear gyro-fluid TGLF SAT-1 and nonlinear ion-scale gyrokinetic CGYRO simulations. Significant discrepancies in convection are observed in some cases, suggesting difficulties in predictions of flat or hollow impurity profiles.
\end{abstract}

%
%
%
%
%



\section{Introduction} \label{sec:intro}
It has long been appreciated that impurities set complex operational limits on tokamaks~\cite{Putterich2019DeterminationFactors} and they play a central role in the attainment of high fusion performance in the core as well as in divertor heat flux mitigation~\cite{Reinke2017HeatTokamaks}. Neutral behavior has also been the subject of significant research~\cite{Mordijck2020OverviewTransport}, but substantial uncertainties persist in making projections to future devices~\cite{Dunne2017ImpactELMs}. Developing accurate predictions requires advanced models that remain valid across a wide range of plasma conditions. It is also crucial that these are effectively integrated to enable studies of core-edge solutions.

In this work, the coupling of edge neutrals and impurity transport is investigated, comparing experimental observations and state-of-the-art transport models. To sidestep difficulties incurred in quantifying experimental cross-field particle \emph{fluxes} ($\Gamma$), the common parametrization using diffusion ($D$) and convection ($v$) is adopted, based on the definition $\Gamma \equiv - D \ \nabla n + v \ n$, where $n$ is the density of a chosen plasma species. This allows one to compare theory and experiment without over-reliance on exact characterizations of particle sources and other experimental details~\cite{Sciortino2020InferenceSelection}. The inference of $D$ and $v$ radial profiles is a challenging \emph{inverse problem}, where one must iterate over a \emph{forward model} that predicts spatio-temporal profiles of impurity charge state densities until matching some experimental observations. Key elements to this process are (a) the quality and detail of available experimental data, (b) the fidelity of the adopted forward model (simulation code), and (c) the algorithm according to which one iterates over free parameters. Estimates of $D$ and $v$ radial profiles have been obtained on many fusion devices, including ASDEX-Upgrade~\cite{Dux2003ImpurityUpgrade, Dux2003InfluenceUpgrade, Putterich2011ELMUpgrade,Bruhn2020Erratum:10.1088/1361-6587/aac870}, DIII-D~\cite{Grierson2015ImpurityDIII-D,Odstrcil2020DependenceTokamak}, HL-2A~\cite{Cui2018StudyDepositions}, JET~\cite{Dux2004ImpurityJET, Giroud2007MethodJET}, MAST~\cite{Henderson2015ChargeMAST}, NSTX~\cite{Delgado-Aparicio2009ImpurityPlasmas}, TCV~\cite{Scavino2003EffectsTCV}, Tore Supra~\cite{Parisot2008ExperimentalPlasma, Villegas2014ExperimentalSupra}, W7-X~\cite{Geiger2019ObservationIron}, and Alcator C-Mod~\cite{Rice1997XPlasmas,Rice2000ImpurityPlasmas, Pedersen2000RadialC-Mod, Rice2007ImpurityPlasmas,Howard2012QuantitativePlasma, Chilenski2018EfficientExperiments}.

This paper presents advances for the inference of experimental impurity transport coefficients focusing on spectroscopic constraints and forward model fidelity. We significantly expand on previous work presented in Ref.~\cite{Sciortino2020InferenceSelection}, where a fully-Bayesian framework was applied for the first time to an Alcator C-Mod high-performance discharge. Here, new statistical techniques that have recently been added to this Bayesian workflow are presented. Most importantly, this paper describes a novel forward model for the entire K$_\alpha$ spectrum of Ca ($Z=20$) measured via high-resolution X-ray Imaging Crystal Spectroscopy (XICS) on Alcator C-Mod. This synthetic diagnostic offers constraints on core impurity transport after Laser Blow-Off (LBO) injections, which provide a clear \emph{non-perturbative} source of \emph{trace}, \emph{non-recycling} ions~\cite{Howard2011CharacterizationSystem, Marmar1975SystemPlasmas}. The XICS measurements are complemented by Extreme Ultra-Violet (EUV) spectroscopy, which constrains the transport of multiple charge states closer to the edge. To the authors' knowledge, this is the first time that the impact of charge exchange (CX) between neutrals and impurities is included in an iterative framework to infer radial profiles of particle transport coefficients. This is done based on atomic D neutral density predictions from SOLPS-ITER~\cite{Wiesen2015ThePackage,Bonnin2016PresentationModelling} simulations using the EIRENE Monte Carlo neutral model~\cite{Reiter2005TheCodes}, which are compared to experimental Ly$_\alpha$ measurements at the midplane for three discharges.

Section~\ref{sec:spectroscopy} presents our spectroscopic analysis and its use to validate transport models. Section~\ref{sec:xics} describes the forward model of Ca K$_\alpha$ spectra, and Section~\ref{sec:euv} of EUV spectra. In Section~\ref{sec:neutrals}, experimental D Ly$_\alpha$ measurements are compared to SOLPS-ITER predictions for D neutral densities near the Last Closed Flux Surface (LCFS). Section~\ref{sec:forward} discusses the use of Aurora 1.5D simulations~\cite{Sciortino2021ModelingAurora_2} in inferences of impurity transport coefficients. Inferences in low- (L-), improved- (I-), and high- (H-) mode discharges are presented in Section~\ref{sec:inferences}. In Section~\ref{sec:validation}, experimentally-inferred transport coefficients are presented and compared with neoclassical and (quasilinear and nonlinear) turbulent models. Finally, in Section~\ref{sec:conclusion} the results of this study are summarized before discussing their impact.

\section{Spectroscopic Analysis} \label{sec:spectroscopy}

This paper focuses on data from Alcator C-Mod~\cite{Hutchinson1994FirstAlcator-C-MOD}, a compact ($a=0.22$ m, $R = 0.68$ m), high field, diverted tokamak that operated until the end of 2016. C-Mod could run with up to 5.5~\si{MW} of Ion Cyclotron Resonance Heating (ICRH), typically in an H-minority scheme, and did not utilize neutral beam injection, hence making edge neutrals the only source of D fueling. C-Mod parameters have ranged within $B_T=[2.0,8.1]$~\si{T}, $I_p=[0.4, 2.0]$~\si{MA}, $n_e(r=0)=[0.2, 6.0]\times10^{20}$~\si{m^{-3}}, and $T_e(r=0)$ up to $9$~\si{keV}~\cite{Greenwald201420Tokamak}. The experimental setup of the work described here is the same as presented in Ref.~\cite{Sciortino2020InferenceSelection}. All the data of interest is from quasi-steady phases of plasma discharges with no Edge-Localized Modes (ELMs). 



\subsection{Analysis of High-Resolution K$_\alpha$ spectra} \label{sec:xics}

The analysis of x-ray spectra arising from $n=2\rightarrow1$ transitions of He-like ions, referred to as K$_\alpha$ radiation, plays a central role in the inferences of impurity transport discussed in Section~\ref{sec:inferences}. He-like ions, having a complete electron shell, are particularly stable over a wide range of temperatures and represent the dominant charge state across most of the plasma radius in all the tokamak discharges considered in this work. The $n=2\rightarrow1$ transitions have been the subject of intense study, both in fusion devices~\cite{Kallne1985HighPlasmas, Rice1995X-rayTokamak, Bitter2003NewIons, Marchuk2006ComparisonTEXTOR,  Rice2014X-rayPlasmas, Rice2015X-rayPlasmas} and astrophysics~\cite{Gabriel1969InterpretationIntensities, Mewe19722InterpolationNe-Sequences,  MeweR.andSchrijver1978Helium-likeIntensities,  Porquet2000X-rayNuclei, Porquet2001LinePlasmas, Porquet2010He-likeNuclei}. The brightest features of K$_\alpha$ spectra are the resonance line w $\left(1\mathrm{s} 2\mathrm{p} \ {}^{1} \mathrm{P}_{1}-1\mathrm{s}^{2} \ {}^{1} \mathrm{S}_{0}\right)$, the
forbidden line z $\left(1\mathrm{s} 2\mathrm{p} \ {}^{3}\mathrm{S}_{1}-1\mathrm{s}^{2} \ {}^1\mathrm{S}_{0}\right)$, and the intercombination lines x $\left(1\mathrm{s} 2\mathrm{p} \ {}^{3}\mathrm{P}_{2}-1\mathrm{s}^2 \ {}^{1}\mathrm{S}_{0}\right)$ and y $\left(1\mathrm{s} 2\mathrm{p} \ {}^{3} \mathrm{P}_{1}-1\mathrm{s}^{2} \ {}^{1}\mathrm{S}_{0}\right)$, identified using Gabriel notation~\cite{Gabriel1969InterpretationIntensities}. Among these, there is a large number of \emph{satellite lines} that correspond to transitions from doubly-excited states of Li-like ions where one electron effectively acts as a ``spectator'' that slightly perturbs the nuclear potential perceived by the transitioning electron~\cite{Vainshtein1978WavelengthsIons, Bely-Dubau1982DielectronicSpectra}. 

\begin{figure}[!htb]
    \centering
    \begin{subfigure}[b]{0.9\textwidth}
       \includegraphics[width=1\linewidth]{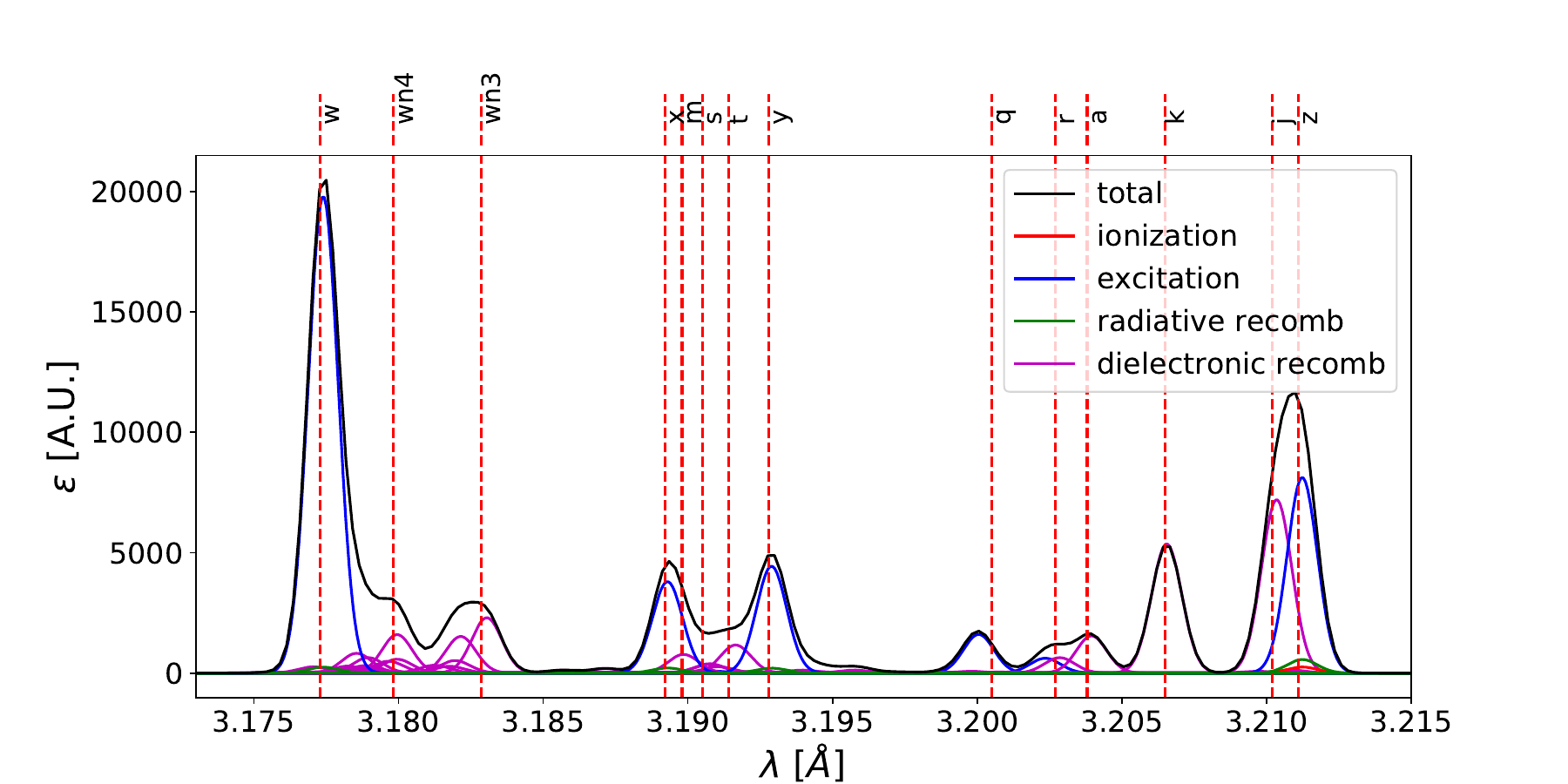}
       \caption{}
       \label{fig:Helike_Ca_spectrum_ne1e20m3_TeTi1keV} 
    \end{subfigure}
    
    \begin{subfigure}[b]{0.9\textwidth}
       \includegraphics[width=1\linewidth]{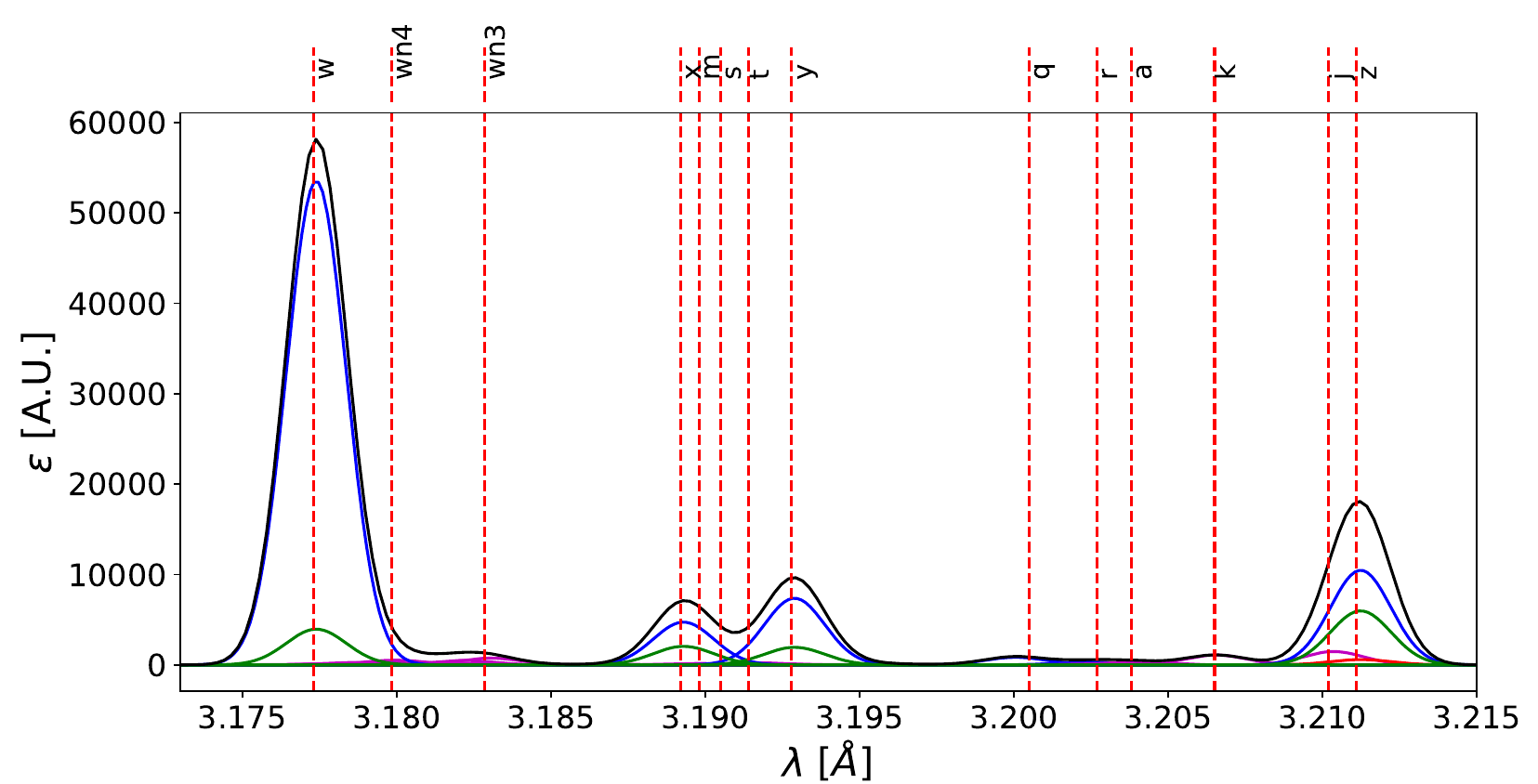}
       \caption{}
       \label{fig:Helike_Ca_spectrum_ne1e20m3_TeTi3500eV}
    \end{subfigure}
    
    \caption[Local K$_\alpha$ Ca synthetic spectra at $n_e=10^{20}$~\si{m^{-3}}, with (a) $T_e=T_i=1$~\si{keV} and (b) $T_e=T_i=3.5$~\si{keV}.]{Local K$_\alpha$ Ca synthetic spectra at $n_e=10^{20}$~\si{m^{-3}}, with (a) $T_e=T_i=1$~\si{keV} and (b) $T_e=T_i=3.5$~\si{keV}. Colors are used to distinguish emissivity from upper level populations driven by ionization (red), excitation (blue), radiative recombination (green), and dielectronic recombination (magenta). Black continuous lines show the total over all processes, weighed by charge state fractional abundances at ionization equilibrium. Vertical dashed lines identify major spectral features using Gabriel notation~\cite{Gabriel1969InterpretationIntensities}.}
\end{figure}

In this work, impurity transport is constrained with high-resolution measurements of Ca K$_\alpha$ spectra from the XICS diagnostic~\cite{Ince-Cushman2008Spatiallyinvited,Reinke2012X-rayResearch} on Alcator C-Mod. For the first time, detailed forward modeling of the entire measured spectral range is demonstrated, making use of a new compilation of atomic rates within the atomDB database~\cite{Foster2020PyAtomDB:Uncertainties} that includes radiative recombination, dielectronic recombination, electron impact excitation, inner shell excitation, and inner shell ionization. These rates were mostly derived from the existent literature~\cite{Vainshtein1978WavelengthsIons, Liang2011R-matrixDamping, Badnell2006RadiativePlasmas, Bautista2007DielectronicSequence, Palmeri2008RadiativeSequences, NIST_ASD, Li2015Radiative42} and complemented with calculations using the Flexible Atomic Code~\cite{Gu2008TheCode}. These atomic data can be used to create synthetic spectra such as those in Figs.~\ref{fig:Helike_Ca_spectrum_ne1e20m3_TeTi1keV} and~\ref{fig:Helike_Ca_spectrum_ne1e20m3_TeTi3500eV}, each computed at a combination of local electron density and temperature ($n_e=10^{20}$~\si{m^{-3}} and $T_e=T_i=1$~\si{keV} for Fig.~\ref{fig:Helike_Ca_spectrum_ne1e20m3_TeTi1keV}; $n_e=10^{20}$~\si{m^{-3}} and $T_e=T_i=3.5$~\si{keV} for Fig.~\ref{fig:Helike_Ca_spectrum_ne1e20m3_TeTi3500eV}), assuming pure Doppler broadening and charge state fractional abundances in conditions of ionization equilibrium. A finite instrumental broadening of 200~\si{eV} characterizing the diagnostic~\cite{Reinke2012X-rayResearch} was included. Of course, when including the effect of impurity transport (see Section~\ref{sec:inferences}), fractional abundances are set by the balance of atomic and transport effects, rather than the ionization equilibrium alone. 

Vertical red dashed lines in Figs.~\ref{fig:Helike_Ca_spectrum_ne1e20m3_TeTi1keV} and~\ref{fig:Helike_Ca_spectrum_ne1e20m3_TeTi3500eV} indicate the Gabriel nomenclature for some of the most important lines in the spectrum. Colors of continuous curves distinguish emissivity contributions from different physical processes. Collisional excitation (blue) is seen to account for most of the upper level population of the w, x, and y lines, while the z line can be similarly dependent on excitation and radiative recombination (green) at high temperatures. The j satellite is nearly degenerate with the z line, meaning that the region between 3.208~\si{\angstrom} and 3.213~\si{\angstrom} is dependent on Li-like, He-like, and H-like Ca stages. Fig.~\ref{fig:Helike_Ca_spectrum_ne1e20m3_TeTi1keV} shows the strong influence of hundreds of satellite lines in the colder part of the confined region; most of these satellites have amplitude smaller than 1\% of the w line emissivity, but some of them significantly contribute to the main features of these spectra. Radiative recombination components are seen to only be significant at multi-keV temperatures, as a result of the low fraction of H-like Ca at lower temperatures. Note that this observation does not need to hold in real experimental plasmas where particle transport can bring H-like Ca out of ionization equilibrium and into the edge region, thus enhancing the radiative recombination components of the w, x, y, and z lines~\cite{Rice1997XPlasmas}. It is also worth remarking that the spectra shown in Figs.~\ref{fig:Helike_Ca_spectrum_ne1e20m3_TeTi1keV} and~\ref{fig:Helike_Ca_spectrum_ne1e20m3_TeTi3500eV} have only a weak dependence on electron density, except for the x and y lines~\cite{Porquet2010He-likeNuclei}. Investigations on the TEXTOR tokamak~\cite{Schlummer2014ChargeTEXTOR} have also shown that CX contributions to K$_\alpha$ spectra can be important whenever background D densities are significant. This effect is negligible when modeling the C-Mod measurements because the XICS lines of sight have tangency radii only inside of $\rho_p\approx0.7$, where neutral densities are low.

\begin{figure}[!htb]
	\centering
	\includegraphics[width=0.8\textwidth]{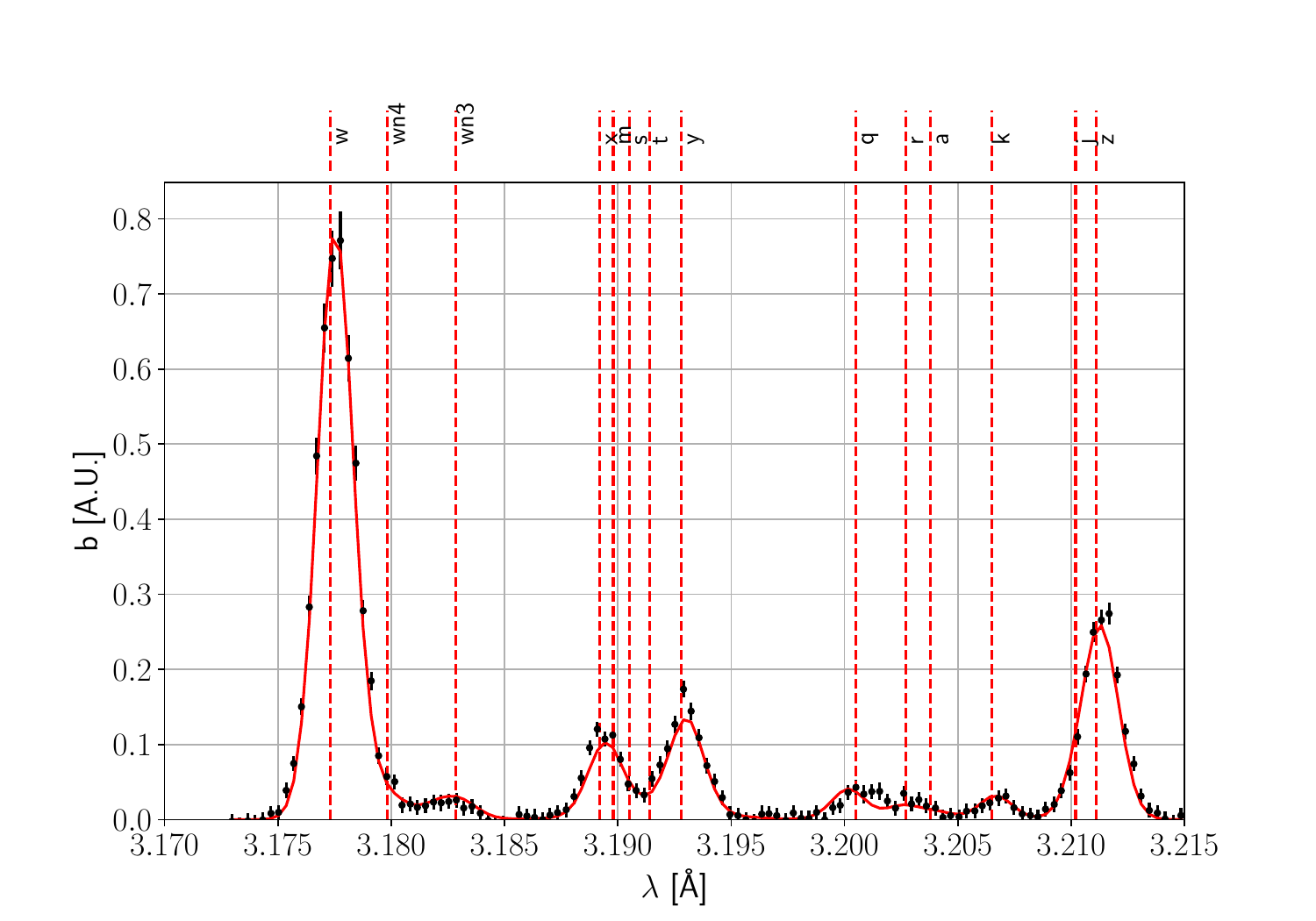} 
    	\caption[Comparison of an experimental Ca K$_\alpha$ spectrum and forward modeling from the I-mode case discussed in Section~\ref{sec:inferences}.]{Comparison of an experimental Ca K$_\alpha$ spectrum from a C-Mod I-mode discharge (here, spatial channel \#0, slightly below the midplane, near the time of peak brightness) and forward modeling. Here, values of free parameters in our forward model correspond to the Maximum A Posteriori estimate of the Bayesian inference presented in Section~\ref{sec:inferences}.}
	\label{fig:cmod_xics_match}
\end{figure}

Fig.~\ref{fig:cmod_xics_match} shows an experimental (line-integrated) XICS Ca spectrum (here, spatial channel \#0, slightly below the midplane, near the time of peak brightness) for the I-mode discharge discussed in Section~\ref{sec:inferences}. Black data points with error bars show the experimental data, while the continuous red line shows the result of forward modeling using free parameter values inferred for this discharge, as discussed in Section~\ref{sec:inferences}. Matches of analogous quality have been obtained in all the experimental inferences discussed below, across 32 spatial chords and over the time of evolution of LBO injections. The spectrum is well matched across the entire wavelength range, with small discrepancies observed near the intercombination (x and y) lines, where the m, s, and t satellites also exist. The region near the q and r lines is affected by slight imperfections in background subtraction since He-like Ar (present in the plasma for background ion temperature and toroidal rotation measurements) also emits in this region~\cite{Rice1987RadialTokamak, Rice1999ThePlasmas}. 

The excellent match in Fig.~\ref{fig:cmod_xics_match} is made possible by comprehensive collisional-radiative modeling as well as Aurora simulation capabilities, enabling reliable inferences of core impurity transport. This demonstration of forward modeling of an entire high-resolution x-ray spectrum within an inference of radial particle transport is a first-of-its-kind in fusion research. The next section describes how Extreme Ultra-Violet (EUV) spectroscopy is used to complement transport constraints from XICS.


\subsection{Line Ratios from EUV Spectroscopy} \label{sec:euv}
The XICS lines of sight do not permit transport effects to be clearly discerned in the outer part of the plasma. However, the XEUS spectrometer~\cite{Reinke2010VacuumTokamak}, with a single line of sight at a small angle to the midplane, does measure emission from Be- and Li-like Ca in this region within its 1-7~\si{nm} wavelength range. Previous work on Alcator C-Mod~\cite{Rice2007ImpurityPlasmas,Howard2014ImpurityPlasmas, Chilenski2017ExperimentalSimulations} and DIII-D~\cite{Odstrcil2020DependenceTokamak} focused on reproducing the time history of individual EUV lines. In the present work, the relative amplitude of lines within chosen wavelength bins is used to constrain transport from charge states that emit in different radial regions in the plasma. This approach is enabled by the recent inclusion of high-quality R-matrix calculations for a number of isoelectronic sequences within the Atomic Data and Analysis Structure (ADAS) database~\cite{Summers2006IonizationElements} database.~\footnote{In particular, the ADF04 files \texttt{lilike\_lgy10\#ca17.dat} and \texttt{belike\_lfm14\#ca16.dat} are primarily responsible for our ability to accurately match the lines reported in Table~\ref{table:EUV_lines}.} Emission from B-like, C-like, F-like, Ne-like, Na-like, and Mg-like charge states has also been included in the analysis, although it proves less relevant for the final results.

\begin{table}[t!]
    \centering
    \begin{tabular}{llll}
    \hline 
    Label & $\lambda$ [\si{\angstrom}] & Main Contributions \\
    \hline 
    XEUS-0 & 186.0-188.0 &  2 Li-like lines  \\
    XEUS-1 & 195.0-197.0 &  2 Be-like lines, 3 Li-like lines \\
    XEUS-2 & 197.2-199.0 &  1 Be-like line, 5 Li-like lines \\
    XEUS-3 & 202.5-205.2 &  6 Be-like lines, 2 Li-like lines \\
    XEUS-4 & 211.0-213.0 & 3 Be-like lines  \\
    XEUS-5 & 220.5-222.5 & 1 Be-like line  \\
    \hline
    \end{tabular}
    \caption[EUV line identification used for the analysis of Ca LBO injections.]{EUV line identification used for the analysis of Ca LBO injections.}
    \label{table:EUV_lines}
\end{table}

\begin{figure}[b!]
	\centering
	\includegraphics[width=0.8\textwidth]{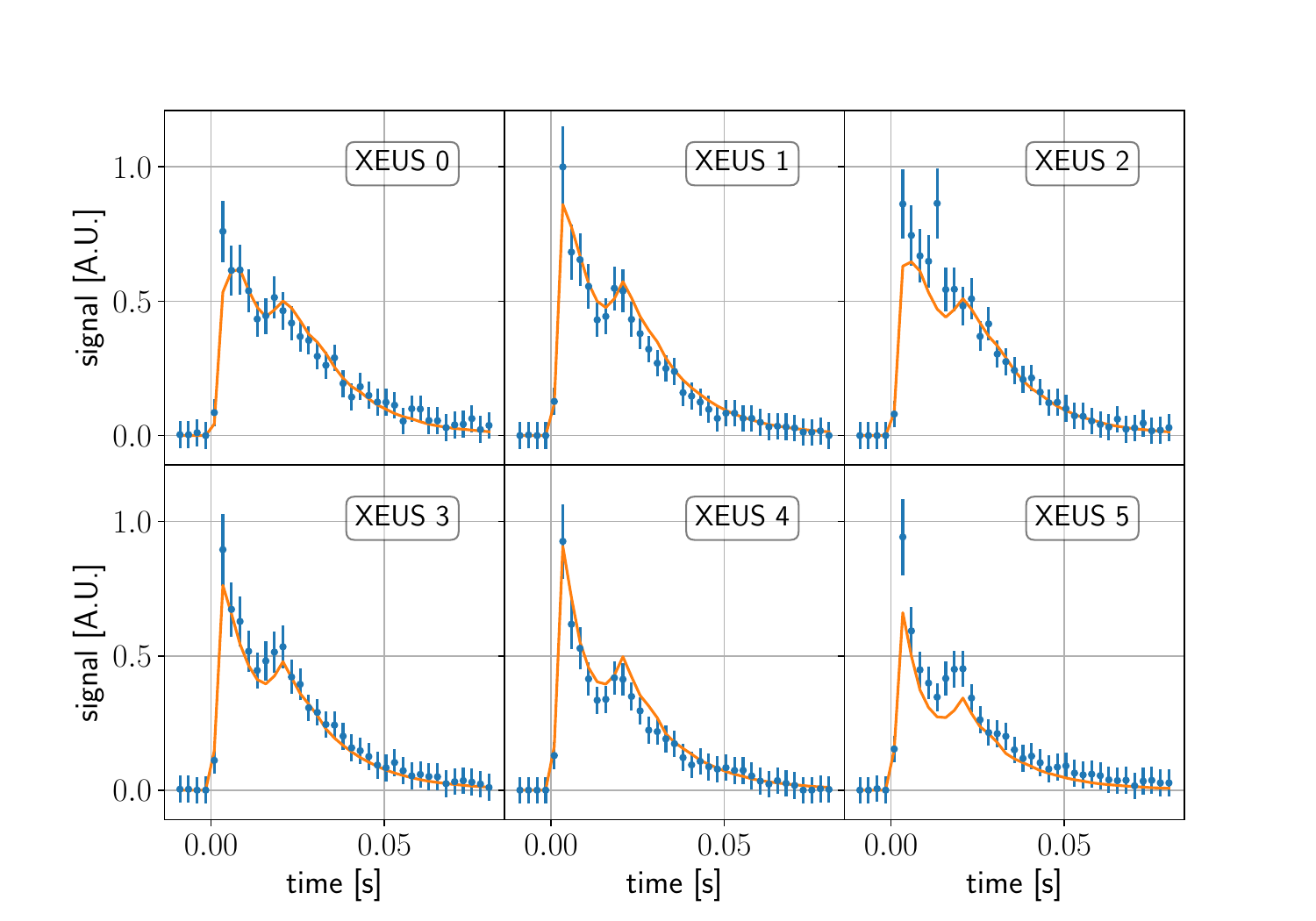}
	\caption{Comparison of EUV measurements for the 6 line groups in Table~\ref{table:EUV_lines} and forward modeling, using inferred free parameters for the L-mode discharge discussed in Section~\ref{sec:inferences}. }
	\label{fig:EUV_matches}
\end{figure}

Fig.~\ref{fig:EUV_matches} shows the result of our fitting procedure on the L-mode discharge that will be later described in Section~\ref{sec:inferences}. Contributions from Be-like states are seen to rise sharply after the LBO injection and decay more rapidly than those from Li-like states. The bump seen approximately 20~\si{ms} after the LBO injection is due to slower clusters of particles deriving from an imperfect ablation~\cite{Marmar1975SystemPlasmas}, as evidenced by an edge filterscope measuring Ca I emission near the LCFS~\cite{Howard2011CharacterizationSystem}. It is worth highlighting that since these lines are very close in wavelength, and therefore near each other on the XEUS detector, wavelength calibration issues are likely negligible. Use of Ca observations elsewhere in the spectra will be the subject of future work. 

In order to combine multiple diagnostics in inferences of impurity transport, one must account for the vastly larger number of data points from XICS spectra. In all the cases analyzed here, the ratio of numbers of data points from EUV and XICS is $\approx 0.004$. Weighing down the XICS contribution to the total likelihood by this factor roughly equilibrates the importance of the two diagnostics, but is by no means an accurate procedure to account for information content from the two diagnostics in different discharges. In practice, an XICS weight of $10^{-2}$ appears to be sufficient to obtain radial profiles of transport coefficients that best match XICS in the core and EUV further towards the edge ($\rho_p \gtrsim 0.75$), where each diagnostic offers greatest experimental constraints.

\subsection{Measurement \& Modeling of Edge Neutrals} \label{sec:neutrals}

Whereas the importance of CX is well appreciated in Scrape-Off Layer (SOL) and divertor plasmas, neutrals are often taken to be negligible for core and pedestal analysis. Here, this assumption is assessed via experimental measurements of D Ly$_\alpha$ ($n=2-1$) at the outboard midplane of Alcator C-Mod, as well as via modeling using SOLPS-ITER.

The D Ly$_\alpha$ data discussed here are from three discharges with no ELMs: one L-mode, one Enhanced D-Alpha (EDA) H-mode~\cite{Greenwald2000StudiesC-Mod}, and one I-mode~\cite{Whyte2010I-mode:C-Mod}. These plasmas have very similar engineering and physics parameters to those of discharges that will be later discussed in the context of impurity transport. The three cases have different plasma currents ($I_p=0.8$~\si{MA} for the L-mode, $I_p=0.55$~\si{MA} for the EDA H-mode, and $I_p=1.0$~\si{MA} for the I-mode), hence their measurements are not easily comparable among themselves, but they are each individually comparable to SOLPS-ITER. Electron density, $n_e$, and temperature, $T_e$, in the pedestal and SOL were measured via Thomson Scattering~\cite{Hughes2001High-resolutionTokamak,Hughes2013PedestalC-Mod} (TS), complemented by Electron Cyclotron Emission~\cite{Basse2007DiagnosticC-Mod} (ECE) $T_e$ data further into the core. To ensure reliable profile fitting across the LCFS, a \emph{modified-tanh} parametric function similar to the one introduced by Groebner and Carlstrom~\cite{Groebner1998CriticalDIII-D} is adopted. The impact of any misalignment between diagnostic measurements and magnetic equilibrium reconstructions from EFIT~\cite{Lao1985ReconstructionTokamaks} is limited by constraining kinetic profile fits using the 2-point model~\cite{stangeby_book} prediction for $T_e$ at the separatrix. This requires estimates of the total power going into the SOL, given by $P_{\text{sol}} = P_{\text{Ohm}} + P_{\text{aux}} - P_{\text{rad}} - dW/dt$, where $P_{\text{Ohm}}$ is the Ohmic power, $P_{\text{aux}}$ is the auxiliary power, $P_{\text{rad}}$ is the total radiated power within the confined plasma, and $W$ is the plasma stored energy. 
The simplest 2-point model assumes that $50\%$ of the SOL power is transported to the divertors by electrons via parallel heat conduction. All power is taken to exit the LCFS at the outer midplane and to decay exponentially over a width $\lambda_q$. This quantity is estimated via the experimental Brunner scaling, developed from a database of divertor measurements across C-Mod confinement regimes and given by $\lambda_q$ [\si{m}] $=(C_f/\bar{p})^{0.5}$, with the constant $C_f = 0.06$ [\si{N}] and $\bar{p}$ being the volume-averaged plasma pressure [Pa]~\cite{Brunner2018High-resolutionTokamak}. One can then compute the 2-point model prediction for $T_e$ at the LCFS with
\begin{equation} \label{eq:Te_LCFS}
    T_{e, L C F S}=\left(\frac{7}{2} \frac{q_{\|} L}{\kappa_{0, e}}\right)^{2 / 7} \qquad \text{with} \qquad q_{\|}=\frac{\frac{1}{2} P_{sol}}{2 \pi R \lambda_{q} \frac{B_{\theta}}{B}},
\end{equation}
where $\kappa_{0,e}$ is the Spitzer-H\"arm electron heat conduction coefficient and $L$ is an estimate of parallel connection length in the SOL. 

The Ly$_\alpha$ signals discussed here are from an array comprising 20 spatial chords going across the Low-Field Side (LFS) midplane. Radially-localized emissivity estimates have been obtained by assuming toroidal symmetry and using a regularized Abel inversion as in previous C-Mod work~\cite{Boivin2000EffectsPlasmas,Hughes2006AdvancesTokamak}. Atomic D neutral densities are calculated by dividing local emissivities by the Ly$_\alpha$ transition ($n=2-1$) energy and the appropriate rate coefficients, taken from ADAS~\cite{Summers2006IonizationElements} for both excitation and recombination.

These experimental measurements provide a valuable opportunity to compare the SOLPS-ITER Monte Carlo neutral model, EIRENE~\cite{Reiter2005TheCodes}, with high-quality pedestal midplane data. The SOLPS-ITER results discussed here have been obtained by iteratively modifying input heat and particle diffusivities until matching TS $n_e$ and $T_e$ profiles. The setup of these runs is similar to the one presented in Ref.~\cite{Reksoatmodjo2021TheExperiments}, except for the fact that here kinetic profiles have been matched further into the core. Up to 50,000 Monte Carlo neutrals were included in each EIRENE simulation for good statistics. For simplicity, these SOLPS-ITER runs included only 2 species, they did not include drifts and they did not aim to match measured divertor heat fluxes. These simplifications are expected to have little influence on the penetration of D neutrals through the pedestal at the midplane. In this sense, the present comparison to Ly$_\alpha$ data is only weakly affected by the most complex numerical challenges of SOLPS-ITER simulations.  

\begin{figure}[ht]
    \centering
    \subfloat{%
        \includegraphics[clip,width=0.73\columnwidth]{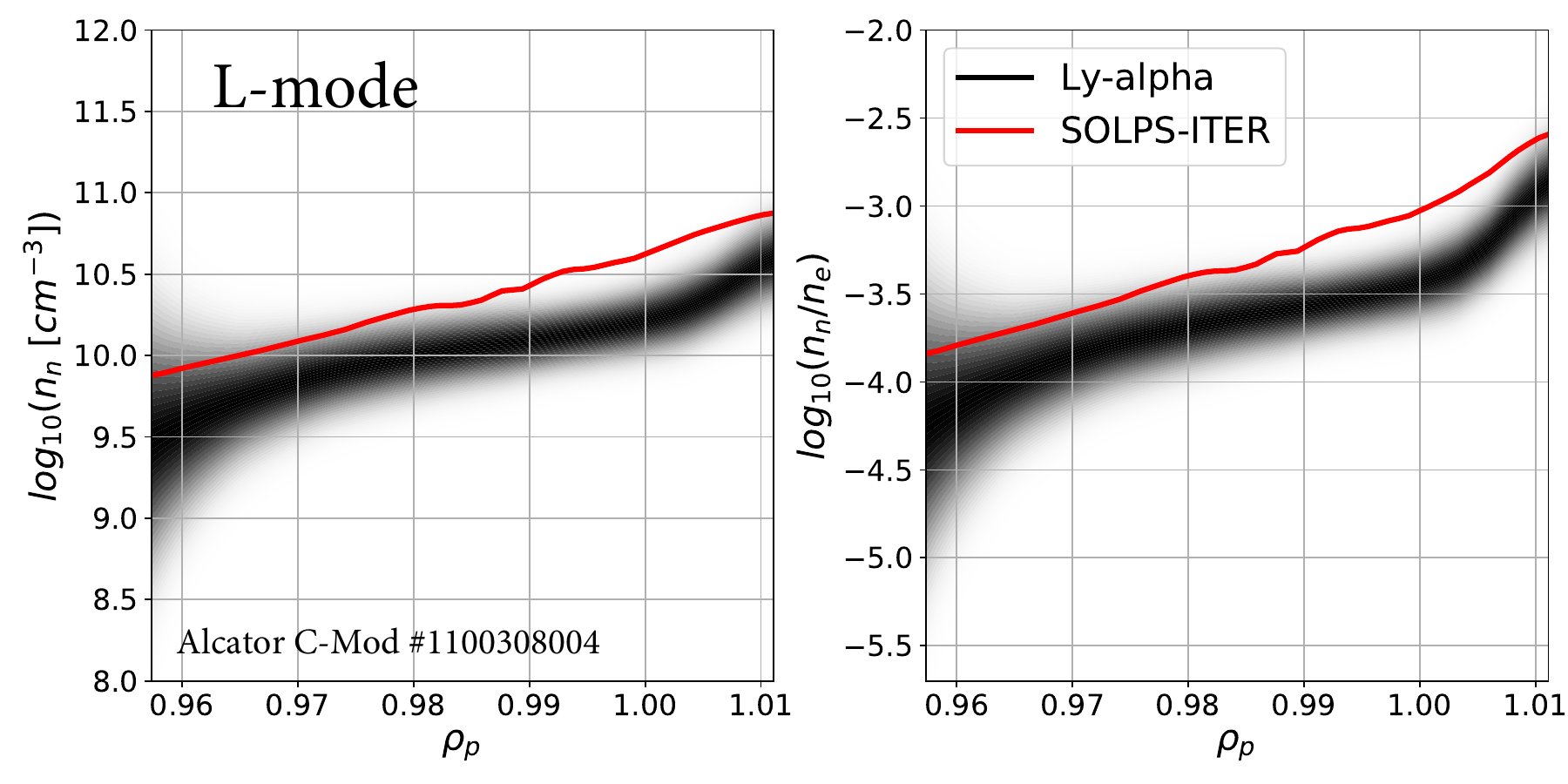}%
    } \\
    \subfloat{%
        \includegraphics[clip,width=0.73\columnwidth]{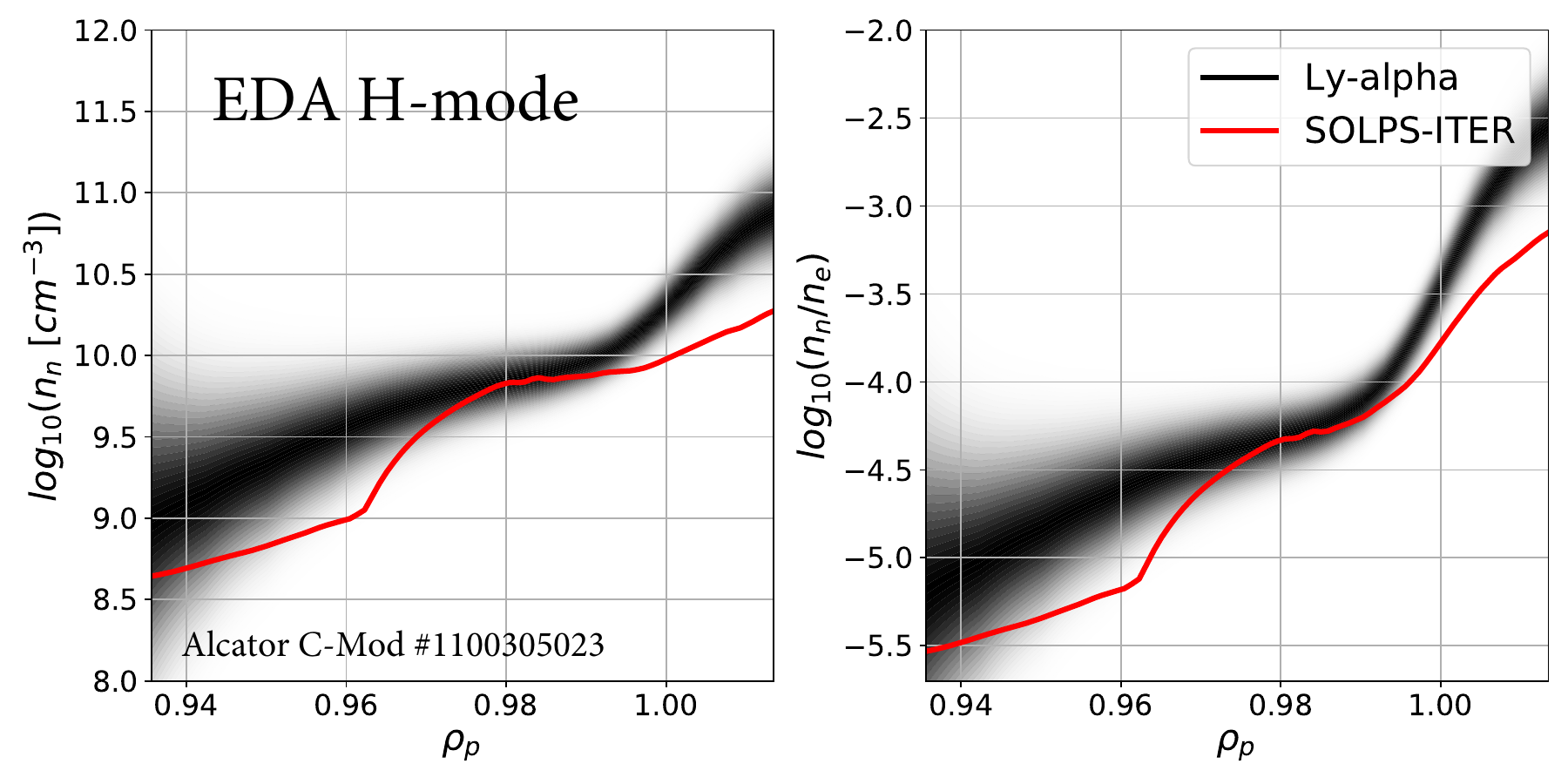}%
    } \\
    \subfloat{%
        \includegraphics[clip,width=0.73\columnwidth]{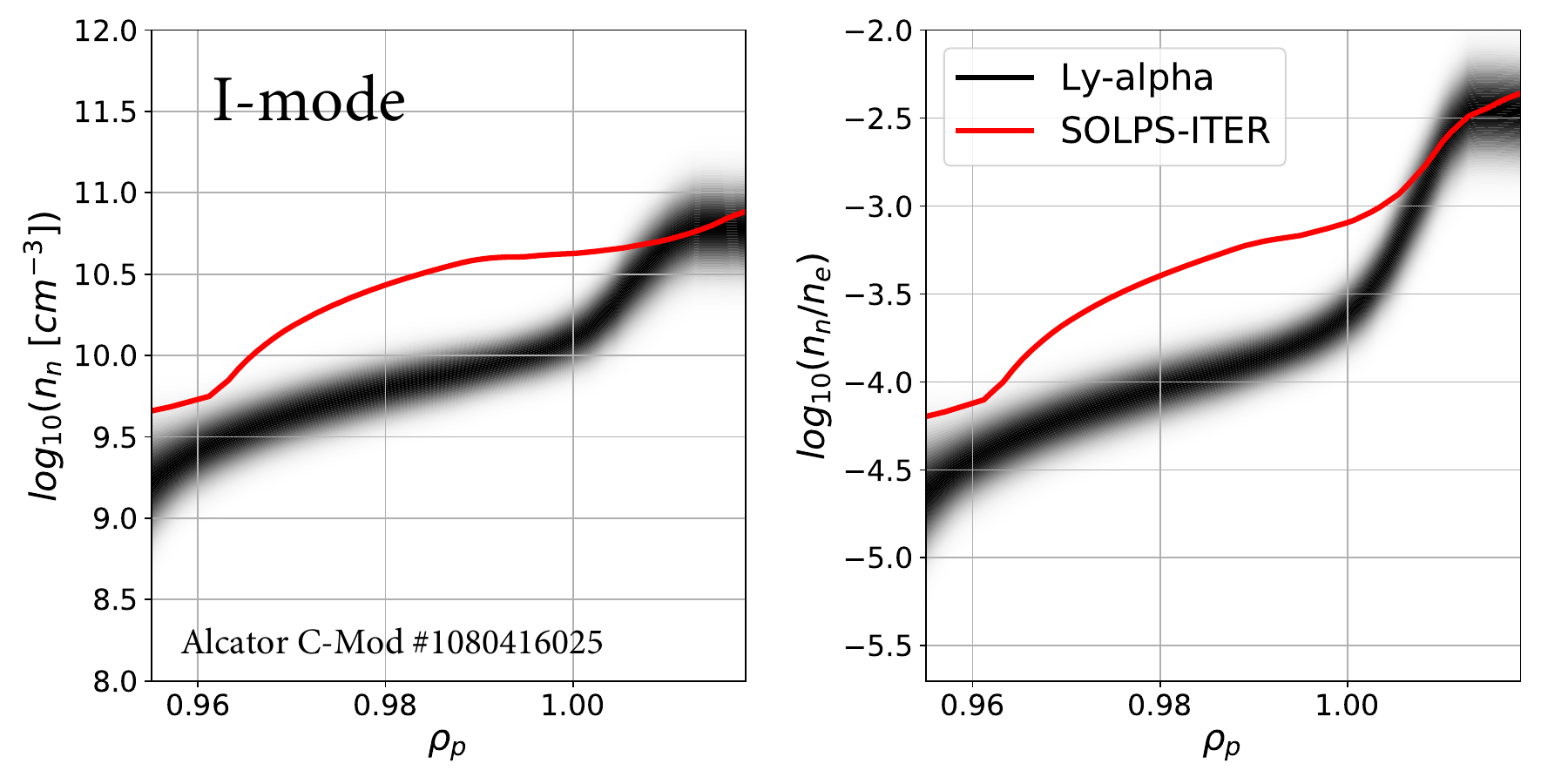}%
    }
    \caption{Comparison of SOLPS-ITER predictions for neutral midplane radial profiles, obtained using Monte Carlo neutrals from EIRENE, with Ly$_\alpha$ spectroscopy in the C-Mod (a) L-mode, (b) EDA H-mode, and (c) I-mode shots discussed in Section~\ref{sec:neutrals}. Note that radial ranges differ for the three cases to focus on regions that are well covered by Ly$_\alpha$ and Thomson scattering data.}
    \label{fig:lya_solps_comp}
\end{figure}

Fig.~\ref{fig:lya_solps_comp} shows the comparison of the inferred midplane neutral densities, $n_n$, with the SOLPS-ITER results, interpolated to the low-field side, for the three shots of interest. Red lines represent SOLPS-ITER results, while the black regions represent the experimental uncertainty range, with color intensity representing the Gaussian probability density function. Figures on the left show the base-10 logarithm of atomic D neutral density, those on the right use a normalization by the local $n_e$. In evaluating these comparisons, one should consider that uncertainties here were solely evaluated by propagating uncertainties in $n_e$, $T_e$, and Ly$_\alpha$ emissivity profiles. By comparing the ADAS rates to those from the Janev \& Smith database~\cite{Janev1993CrossIons}, $\sim 10\%$ discrepancies have been found in neutral densities, which may therefore be interpreted as an additional source of uncertainty. 
Most importantly, an uncertainty on the absolute calibration of the Ly$_\alpha$ system for the three shots, taken over a span of two years, is not available and is likely to affect neutral densities by up to a factor of 2. We highlight therefore that the uncertainties in Fig.~\ref{fig:lya_solps_comp} should be interpreted as purely \emph{statistical}, based on Ly$_\alpha$ random scatter and kinetic profile fitting uncertainties, and are not attempting to represent the entire range of reasonable experimental values. These complications do not preclude a valuable comparison with SOLPS-ITER, whose results are themselves affected by significant uncertainties, for example related to the choice of transport coefficients used to match experimental $n_e$ and $T_e$ data points. Numerical choices of SOLPS-ITER modeling will be detailed in an upcoming publication~\cite{RR_2021}.

Overall, SOLPS-ITER appears to be rather successful at matching the experimental data in Fig.~\ref{fig:lya_solps_comp}. Significant discrepancies are found at some locations, but this is not unexpected. For the reasons discussed above, profile shapes may be argued to be more important than absolute densities themselves. In the L-mode case, the profile gradients are very well recovered. The EDA H-mode case also shows a relatively good match. On the other hand, the I-mode case presents features that are not well matched between the Ly$_\alpha$ data and SOLPS-ITER, with a difference of up to a factor of 5 between the two. 

The results in Fig.~\ref{fig:lya_solps_comp} should be considered as ``spot checks'' for SOLPS-ITER's 2D model at the LFS midplane. This work is currently being expanded in a comprehensive validation exercise, adding to a growing series of research efforts focused on D neutral particles, ranging from the analysis of D$_\alpha$ spectroscopy~\cite{Hughes2006AdvancesTokamak, Scotti2021OuterNSTX-U} to high-n x-ray transitions populated by CX~\cite{Rice1986ObservationPlasma, Kato1991EffectsTokamak,Rice2021ThePlasmas}. In the rest of this paper, the SOLPS-ITER results shown in Fig.~\ref{fig:lya_solps_comp} are taken to give a sufficiently good representation of neutral density profiles, and justify some corrections in the inferences of impurity transport by introducing a free rescaling parameter. As described in Section~\ref{sec:inferences}, this allows one to adapt neutral densities to best match XICS and VUV measurements, using SOLPS-ITER results as well as a range of credible scaling factors, encapsulated by an appropriate Bayesian prior that is heuristically informed by the comparison in Fig.~\ref{fig:lya_solps_comp}.



\section{Forward Modeling of Impurity Transport} \label{sec:forward}
Previous sections described experimental spectroscopic data based on which we quantify neutral densities and impurity transport coefficients. In this section, we first describe our approach to simulating impurity transport and then demonstrate the importance of including CX of impurities and neutrals for an accurate assessment of charge state balance near the pedestal.

The forward model developed in this study is based on Aurora~\cite{Sciortino2021ModelingAurora_2}, a toolbox for particle transport, neutrals and radiation that grew out of the work presented in this paper, as well as research in Refs.~\cite{Sciortino2020InferenceSelection} and~\cite{Odstrcil2020DependenceTokamak}. Aurora's impurity transport simulations use a 1.5D geometry to model the time-dependent evolution of charge states of a chosen ion. In many ways, Aurora is similar to STRAHL~\cite{Dux2006STRAHLManual, Dux2004ImpurityPlasmas}, with which it has been thoroughly benchmarked. Crucially, Aurora's modern interface coupling Python and Fortran allows modeling within an iterative framework such as the one needed for the Bayesian inferences discussed in Section~\ref{sec:inferences}. Atomic rates are taken from ADAS~\cite{Summers2006IonizationElements}. Magnetic geometry reconstructions from EFIT~\cite{Lao1985ReconstructionTokamaks} are handled via the \emph{omfit\_classes} package\footnote{\url{https://omfit.io/classes.html}}~\cite{Meneghini2015IntegratedOMFIT}. Aurora also allows application of impurity \emph{superstages}, which can significantly reduce the number of charge states included in forward modeling. In this work, we apply superstaging by \emph{bundling} the low charge states of Ca (Ca$^{1+}$ to Ca$^{9+}$), incurring negligible errors and almost a $2\times$ speed up, as demonstrated in Ref.~\cite{Sciortino2021ModelingAurora_2}. Aurora is also used to process PECs from ADAS and create synthetic diagnostics.

\begin{figure*}
    \centering
    \begin{subfigure}[b]{0.475\textwidth}
        \centering
        \includegraphics[width=\textwidth]{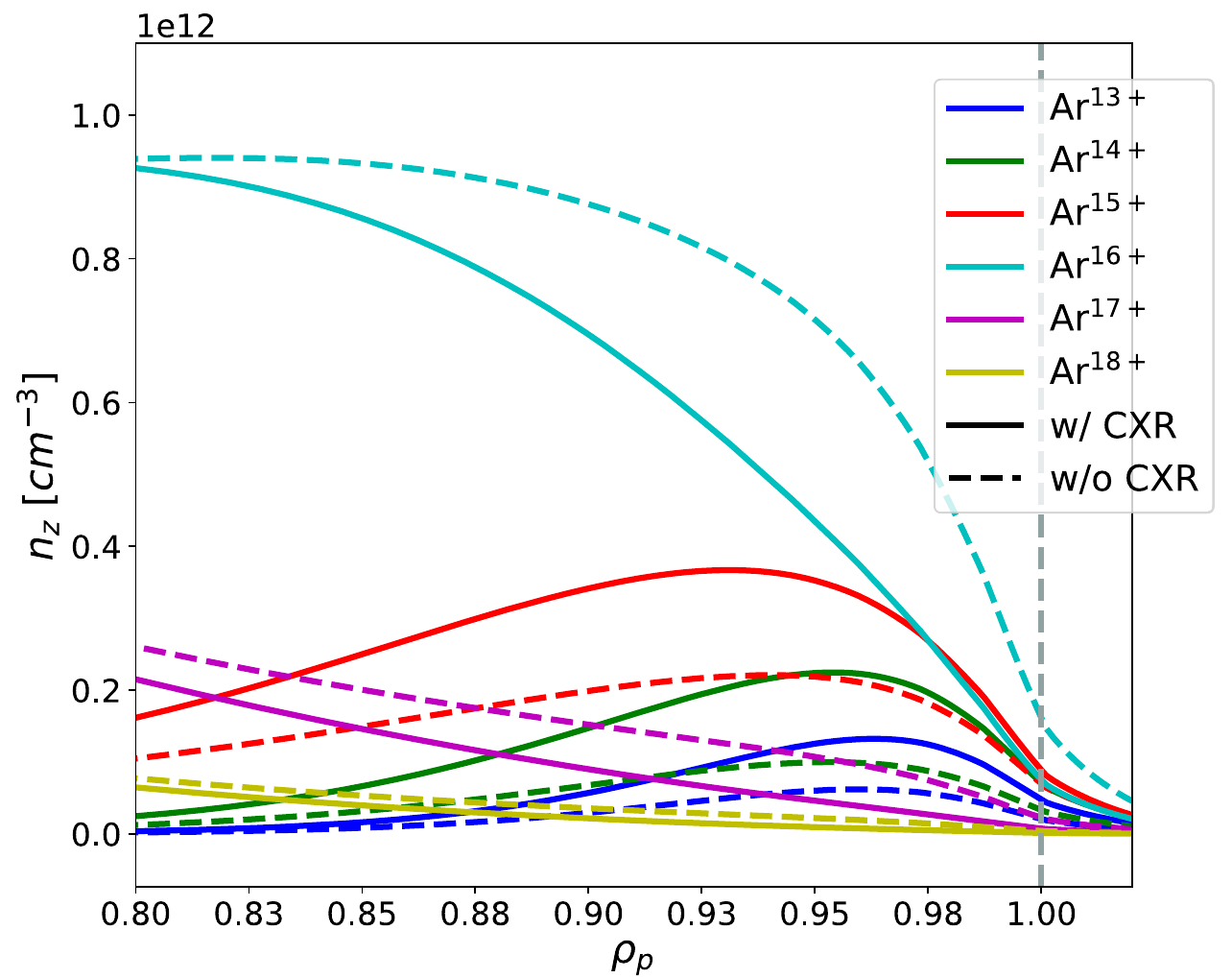}
        \caption{}
        \label{fig:nz_cxr_imode}
    \end{subfigure}
    \hfill
    \begin{subfigure}[b]{0.475\textwidth}  
        \centering 
        \includegraphics[width=\textwidth]{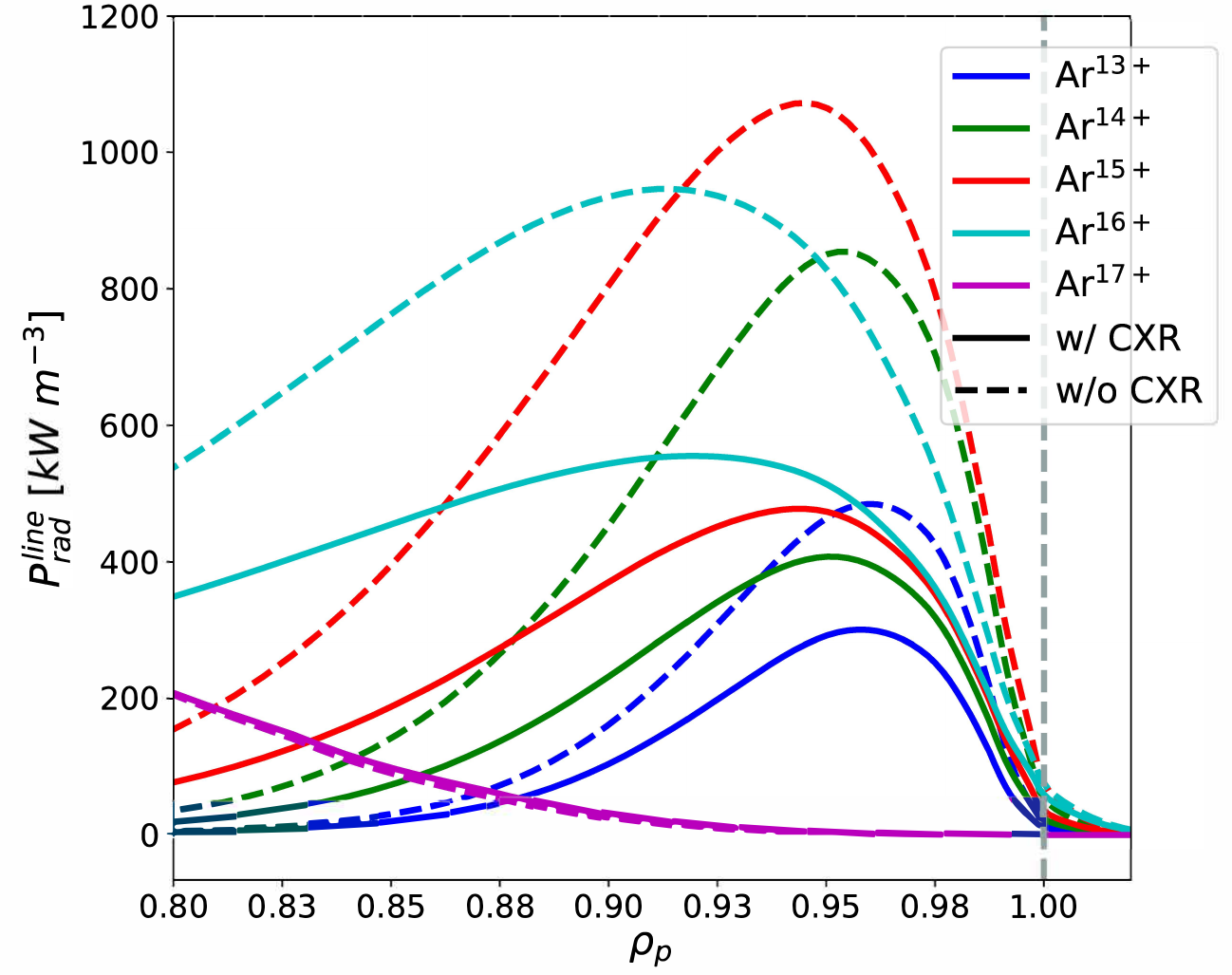}
        \caption{}
    \label{fig:Pline_cxr_imode}
    \end{subfigure}
    \caption[Illustration of the impact of edge neutrals on Alcator C-Mod for identical Aurora simulations in an I-mode discharge (\#1080416025) with a constant Ar source and realistic $D$ and $v$ profiles.]{Illustration of the impact of edge neutrals on Alcator C-Mod for identical Aurora simulations in an I-mode discharge (\#1080416025) with a constant Ar source of $10^{20}$~\si{particles/s} and realistic $D$ and $v$ profiles. Panel (a) shows charge states densities; (b) displays corresponding line radiated power. Continuous curves indicate the result obtained including charge exchange with background neutrals; dashed curves give the result without charge exchange. Neutral densities are from SOLPS-ITER simulations.}
    \label{fig:cxr_imode}
\end{figure*}

Here, the effect of including charge exchange in Aurora simulations is demonstrated by considering the I-mode scenario described in Section~\ref{sec:neutrals} as an example. For this purpose, we make use of flux-surface averaged (FSA) neutral profiles from SOLPS-ITER, rather than experimental Ly$_\alpha$ data, because the latter are localized at the LFS. As discussed by Dux \emph{et al.}~\cite{Dux2020InfluenceUpgrade}, since the average ionization rate over charge states is much smaller than their inverse transit time on flux surfaces, impurities average out the effect of CX everywhere except extremely close to the LCFS. Hence, even though neutral densities are by no means poloidally symmetric, it is appropriate to consider FSA neutral densities for the inference of impurity transport. 
Fig.~\ref{fig:cxr_imode} shows the result of Aurora simulations with and without CX, at a time long after an Ar injection, when charge state profiles are decaying in amplitude while maintaining their shapes. The radial profiles of $D$ and $v$ used for these I-mode simulations are similar to those inferred from experiment in Section~\ref{sec:inferences}, but their details are not important for the main message of Fig.~\ref{fig:cxr_imode}. The left hand side of this figure shows the charge state densities of the highest charge states of Ar, while the right hand side shows the corresponding line radiation. Continuous (dashed) curves represent profiles obtained with (without) CX.

Clearly, it is hard to overstate the importance of considering charge exchange of neutrals and impurities for particle transport studies in the pedestal region. We highlight that the plasma condition shown here is from Alcator C-Mod, a tokamak where electron densities are routinely much higher than on other devices, thus reducing neutral penetration from the edge. The example of Fig.~\ref{fig:cxr_imode} is from an I-mode discharge, where there exist a temperature pedestal but no density pedestal. As shown in Fig.~\ref{fig:lya_solps_comp}, neutral penetration is generally lower in H-mode due to a steep density pedestal, but is similar in the L-mode case since D ionization rates are only a weak function of $T_e$ inside the confined plasma. In other current devices, neutral penetration is generally expected to be significantly higher than on C-Mod, hence making the impact of charge exchange even more significant, often also much further inside of the pedestal. This significantly differentiates analysis of today's experiments from predictions of future devices, where neutral penetration is expected to be lower~\cite{Mordijck2020OverviewTransport}. In Ref.~\cite{Sciortino2021ModelingAurora_2} SOLPS-ITER results were shown to be in agreement with this expectation, reflecting the likely negligible role that charge exchange will have in setting pedestal charge state balance in a reactor.

\section{Bayesian Inference of Impurity Transport} \label{sec:inferences}

In this section, advances in forward modeling of XICS and EUV data are brought together with predictions of background edge D neutrals for CX and improvements to the Bayesian inference methods presented in Ref.~\cite{Sciortino2020InferenceSelection}. In order to infer impurity transport coefficients, a chosen set of inputs to Aurora (\emph{free parameters}) are iteratively modified, each time evaluating synthetic diagnostics and quantifying a metric of ``goodness of fit'' to experimental data (XICS and EUV signals). As in Refs.~\cite{Sciortino2020InferenceSelection, Sciortino2021ParticleLines, Chilenski2019OnProfiles}, the orchestration of new samples is done via \emph{nested sampling}~\cite{Skilling2006NestedComputation} (NS) and in particular via the \texttt{MultiNest} algorithm~\cite{Feroz2008MultimodalAnalyses, Feroz2013ImportanceAlgorithm, Buchner2014X-rayCatalogue}. A major objective of NS is the evaluation of the Bayesian \emph{evidence}, also called \emph{marginal likelihood}, which is the denominator in Bayes Theorem:
\begin{equation} \label{eq:Bayes}
p(\theta | \mathcal{D}, \mathcal{M}_i) = \frac{ p(\mathcal{D}|\theta, \mathcal{M}_i) p(\theta | \mathcal{M}_i)}{\mathcal{Z}(\mathcal{D}|\mathcal{M}_i)}.
\end{equation}
The evaluation of the evidence, $\mathcal{Z}(\mathcal{D}|\mathcal{M}_i)$, allows the selection of \emph{models}, $\mathcal{M}_i$, intended as both parametrizations and physical descriptions of reality, providing rigorous means to avoid under- or over-fitting to experimental data. In the context of this work, $\mathcal{Z}$ is used to select the appropriate number of radial knots used in the splines describing transport coefficients. The term $p(\theta | \mathcal{D}, \mathcal{M}_i)$ in Eq.~\ref{eq:Bayes} is the \emph{posterior} distribution, which is the probability distribution for the free parameters of interest, $\theta$. The numerator on the right hand side is the product of the \emph{likelihood} distribution, $p(\mathcal{D}|\theta, \mathcal{M}_i)$, and the \emph{prior} distribution, $p(\theta | \mathcal{M}_i)$. Both of these terms are subject to modeling choices, for the former via the definition of a forward model, for the latter via the selection of appropriate distributions that represent an agnostic and yet useful description of knowledge about the free parameters before an inference. Readers are referred to Ref.~\cite{Sciortino2020InferenceSelection} for a detailed discussion of NS and the \texttt{MultiNest} algorithm, as well as an overview of our iterative framework. As in that work, here we make use of a High-Performance Computing (HPC) framework, typically running Aurora simulations (samples) on approximately $300$ CPUs within an MPI parallelization. Major differences from previous work are in the advanced synthetic diagnostics for both XICS (Section~\ref{sec:xics}) and EUV (Section~\ref{sec:euv}), inclusion of background atomic D neutral densities to model CX (Section~\ref{sec:neutrals}), and the more comprehensive Aurora forward model (Section~\ref{sec:forward}).

Two technical advances, from the perspective of statistical methods, are also worth highlighting, the first related to the \emph{physically-correlated} $D$, $v$, and $v/D$ samples, the second to \emph{forced-identifiability} free spline knots. The former technique, described in Appendix~\ref{app:triple_priors}, offers a simple way to obtain samples of $D$ and $v$ such that their ratio, $v/D$, is also constrained by a reasonable prior. This avoids, for example, simultaneous sampling of values such as $D=0.1$~\si{m^2/s} and $v=10$~\si{m/s}, which are both individually physical, but have a ratio $v/D=10/0.1=100$~\si{m^{-1}} that is unlikely to be realistic in the core plasma. The adopted method to prevent the exploration of unphysical parameter space is effectively to sample $D$ and $v$ in the complex plane, such that both the amplitude of $D$ and $v$ as well as the \emph{phase} between them can be constrained with only 2 free parameters per location. 

The second technique mentioned above is related to the desire to allow spline knots in $D$ and $v$ profiles to freely explore the entire radial profile, in order to find the locations where they can best allow an appropriate representation of experimental observations while avoiding over-parametrization. Previous work by Chilenski~\cite{Chilenski2017ExperimentalSimulations} observed that allowing spline knots to be completely free runs into the problem of knots swapping places, leading to \emph{unidentifiable} posterior distributions. The solution adopted here is to sample free knots from a unit hyper-triangle which statistically enforces the order of sampled knots. This effectively corresponds to non-separable Bayesian priors for different knots, making use of the \emph{symmetry} in sampling different knots. Details are provided in Appendix~\ref{app:knots}.

We demonstrate here inferences on Alcator C-Mod discharges that are very similar, in both engineering and measured physics parameters, to those analyzed in Section~\ref{sec:neutrals} from the standpoint of edge neutrals for each confinement regime. As discussed in Section~\ref{sec:forward}, inclusion of CX between neutrals and impurities in forward modeling is important to appropriately represent atomic sources and sinks for each charge state near the pedestal. As noted in Section~\ref{sec:neutrals}, the inferences presented here include a free parameter that allows for rescaling of the background D neutral densities (and therefore Ca CX rates), while maintaining the profile shapes given by SOLPS-ITER. Variation of this parameter within a log-normal prior with mean of 1 and $\sigma=0.25$ can lead to corrections of neutral densities by more than a factor of 2 to improve the likelihood. In practice, since this parameter is only weakly constrained by the XICS and VUV data, its inferred value is near 1 for all the cases discussed here. We note that these discharges have different plasma currents ($I_p=0.8$, $0.55$, and $1.0$~\si{MA} for the L-, EDA H-, and I-mode, respectively), hence one cannot easily compare transport coefficients between discharges. Nonetheless, it is worthwhile comparing each experimental result with predictions from theoretical transport models.

\begin{figure}[!t]
	\centering
	\includegraphics[width=0.8\textwidth]{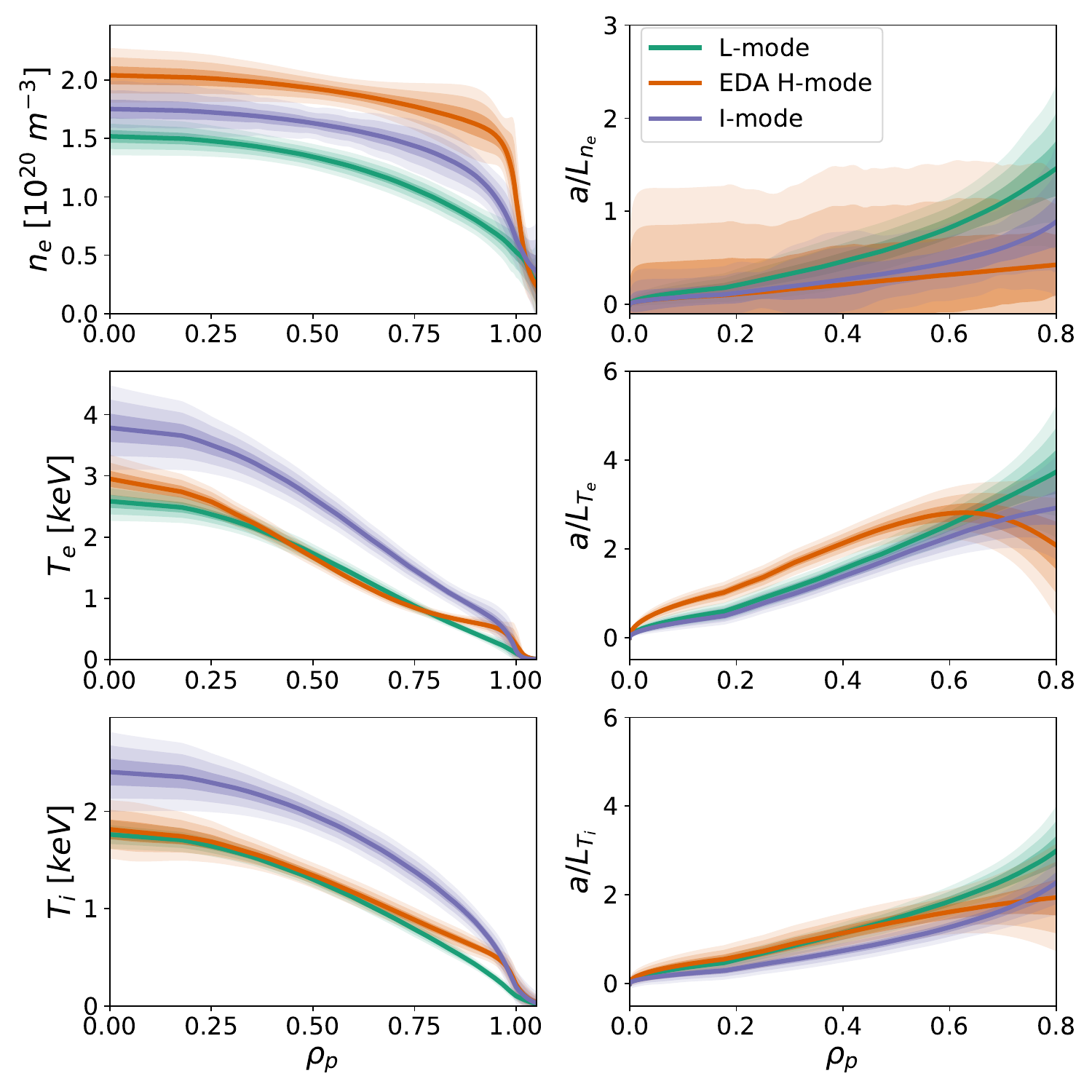}
	\caption{Kinetic profiles and normalized gradient scale lengths for the 3 Alcator C-Mod discharges where impurity transport was inferred. }
	\label{fig:cmod_3_kin_prof_fits}
\end{figure}

Fig.~\ref{fig:cmod_3_kin_prof_fits} shows kinetic profile fits of $n_e$, $T_e$, and $T_i$ for the three discharges of interest, obtained from Thomson scattering~\cite{Hughes2003ThomsonC-Mod} and Electron Cyclotron Emission~\cite{Basse2007DiagnosticC-Mod} measurements. Normalized gradient scale lengths are shown on the right hand side. Unlike in the analysis of Ly$_\alpha$ edge signals, these fits are from Gaussian Process Regression (GPR) using Markov Chain Monte Carlo (MCMC) sampling. This non-parametric procedure is particularly valuable to quantify uncertainties in gradients, since these are analytically derived from the fit~\cite{Chilenski2015ImprovedRegression}. The profile fits of Fig.~\ref{fig:cmod_3_kin_prof_fits} are therefore used for the theoretical transport modeling discussed below. On the other hand, experimental inferences of impurity transport coefficients require time-dependent kinetic profiles in order to appropriately model variations of atomic rates following an LBO injection, particularly near the magnetic axis during sawteeth~\cite{Sciortino2020InferenceSelection}. For this purpose, a parametric but robust fitting routine based on Radial Basis Functions (RBFs)~\cite{Odstrcil2020DependenceTokamak} is adopted.

\begin{figure}[!htb]
	\centering
	\includegraphics[width=0.75\textwidth]{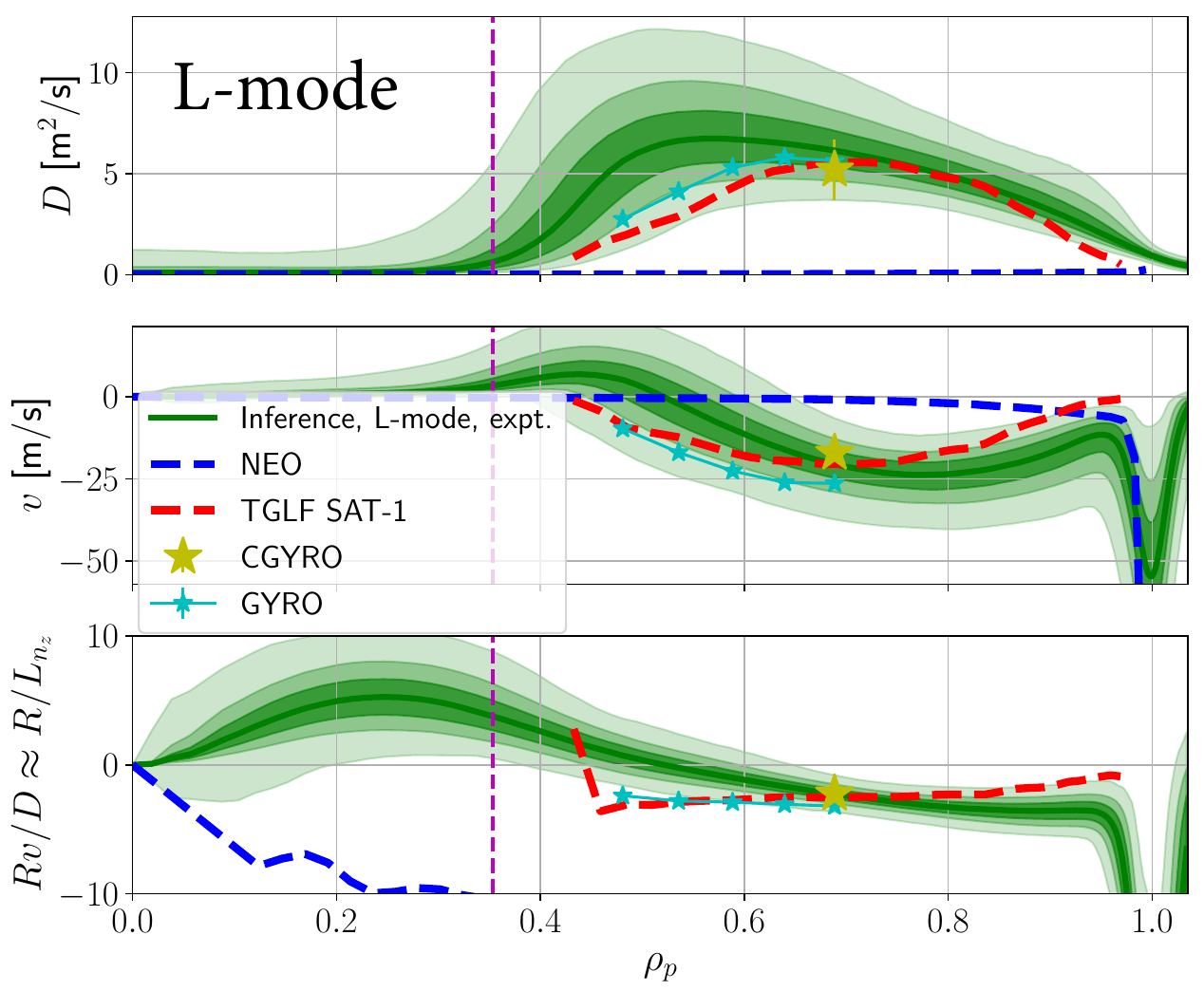} 
	\caption{Comparison of experimental Ca transport coefficients inferred in the C-Mod L-mode discharge 1101014006 with $D$, $v$, and $R v/D$ results from the neoclassical NEO, quasi-linear gyrofluid TGLF SAT-1, and nonlinear CGYRO models.} 
	\label{fig:inference_lmode}
\end{figure}

\begin{figure}[!htb]
	\centering
	\includegraphics[width=0.75\textwidth]{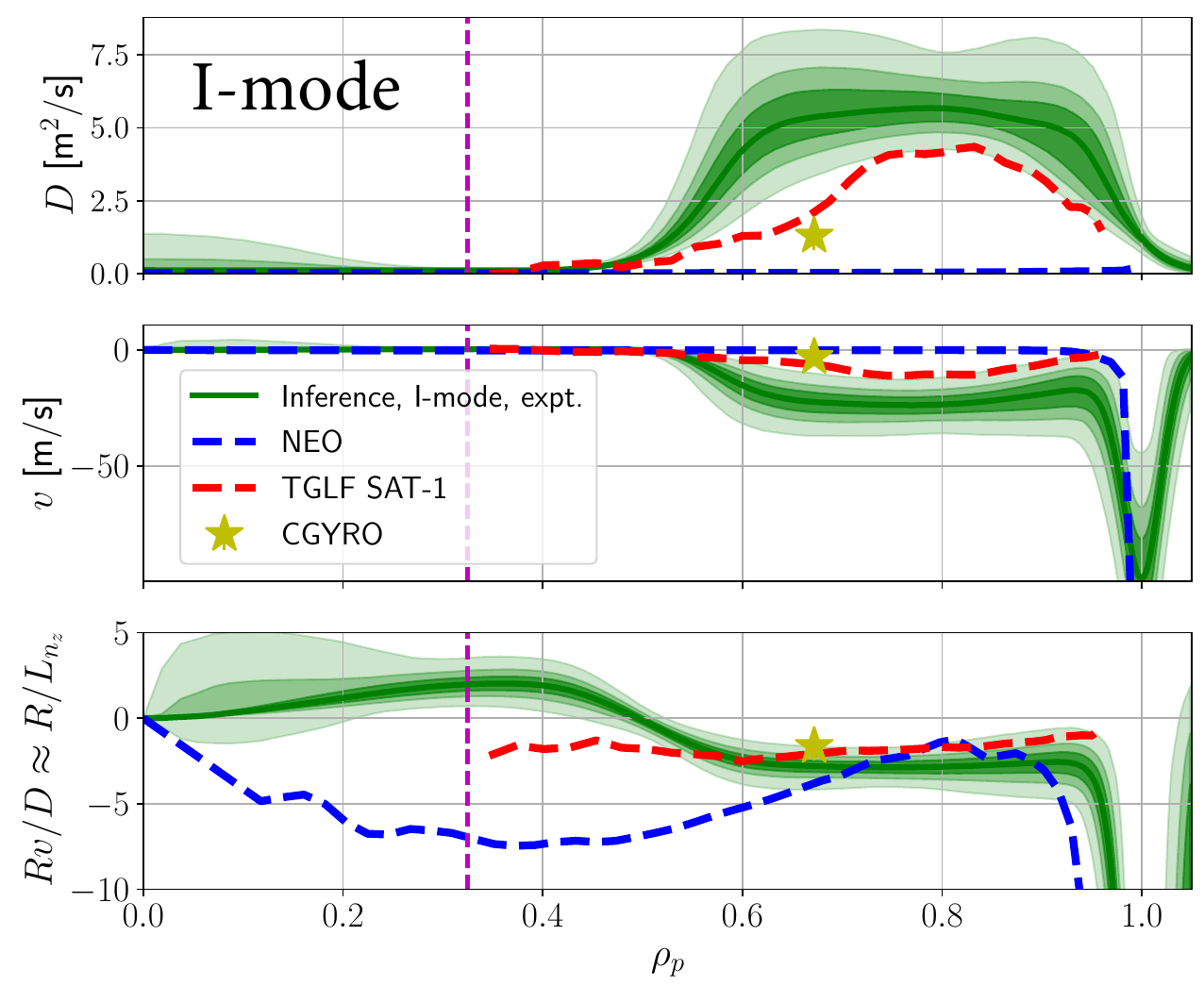} 
	\caption{Analogously to Fig.~\ref{fig:inference_lmode}, comparison of inferred Ca experimental transport coefficients in the C-Mod I-mode discharge 1101014030 with neoclassical (NEO), quasilinear gyrofluid (TGLF SAT-1), and nonlinear gyrokinetic (CGYRO) results.} 
	\label{fig:inference_imode}
\end{figure}

\begin{figure}[!htb]
	\centering
	\includegraphics[width=0.75\textwidth]{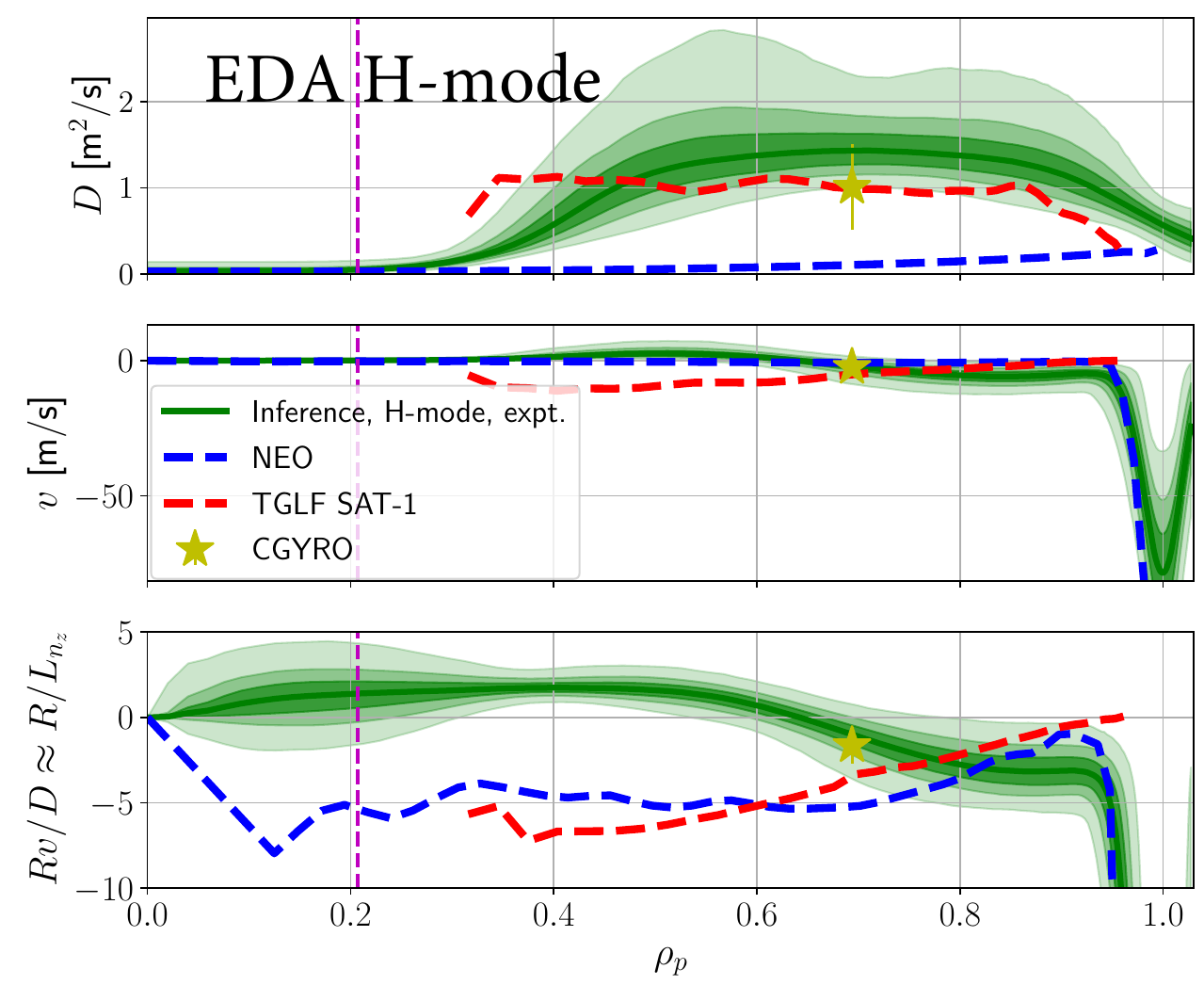} 
	\caption{Analogously to Figs.~\ref{fig:inference_imode} and~\ref{fig:inference_lmode}, comparison of inferred Ca experimental transport coefficients in the C-Mod EDA H-mode discharge 1101014019 with neoclassical (NEO), quasilinear gyrofluid (TGLF SAT-1), and nonlinear gyrokinetic (CGYRO) results.} 
	\label{fig:inference_hmode}
\end{figure}

Figs.~\ref{fig:inference_lmode}, \ref{fig:inference_imode}, and~\ref{fig:inference_hmode} show inferred radial profiles of $D$, $v$, and $R v/D$ for the L-mode, I-mode, and EDA H-mode discharges, respectively. In each case, the green shaded regions represent the 1-99, 10-90, and 25-75 percentiles of the posterior distribution for the inferred transport coefficients, with no assumptions made about the functional form (e.g. Gaussian) of uncertainties. In other words, these shaded regions show how likely it is for certain values of free parameters to correspond to the measured spectroscopic data. The vertical dashed magenta line shows the location of the sawtooth inversion radius, within which transport coefficients appear to transition to neoclassical-like levels. Experimentally-inferred results are compared to neoclassical predictions (including sonic rotation terms) from NEO~\cite{Belli2008KineticDynamics} (blue dashed lines), quasilinear gyrofluid (electromagnetic) TGLF SAT-1~\cite{Staebler2005Gyro-LandauParticles,Staebler2016TheTurbulence} (red dashed lines), and nonlinear gyrokinetic CGYRO~\cite{Candy2016APlasmas} (yellow stars, at a single radial location). CGYRO results, described in more detail in Section~\ref{sec:validation} and Appendix~\ref{app:cgyro}, are from ion-scale, heat-flux matched simulations. For the L-mode case, Fig.~\ref{fig:inference_lmode} also shows a set of GYRO~\cite{Candy2003AnomalousSimulation,Candy2003AnSolver} runs from previous work by Howard~\cite{Howard2012QuantitativePlasma} which used a completely separate data analysis workflow from the one discussed here. The total particle flux may be taken to be the sum of neoclassical and turbulent components, i.e. $\Gamma_{\text{tot}} = - (D_{\text{turb}}+D_{\text{neo}}) \nabla n + (v_{\text{turb}}+v_{\text{neo}}) n$. At midradius, where the comparison between experimental inferences and theoretical modeling is most robust, $Rv/D$ is entirely dominated by the turbulent contribution.

Good agreement is observed in $D$ at midradius for all cases. As discussed in Ref.~\cite{Sciortino2020InferenceSelection}, matches between turbulence modeling and experiment may be expected to have a 50\% discrepancy, since kinetic profile gradients (Fig.~\ref{fig:cmod_3_kin_prof_fits}) are themselves affected by uncertainties that can give such level of variability. Inside of the sawtooth inversion radius, experimental and theoretical convection predictions are seen to differ substantially, but this is also not particularly relevant since theory models were run on time-averaged kinetic profiles, although all these discharges are affected by sawteeth. Hence, while experimental inferences make use of a phenomenological sawtooth model in Aurora and attempt to model the detailed time dependence of the background plasma, NEO results can only be as good as the inputs that we provide to the code. Future work will run NEO on individual time slices, at different phases of the sawtooth cycle, to explore the variability of its predictions. 

\begin{SCfigure}[1.0][!htb]
    \centering
	\includegraphics[width=0.6\textwidth]{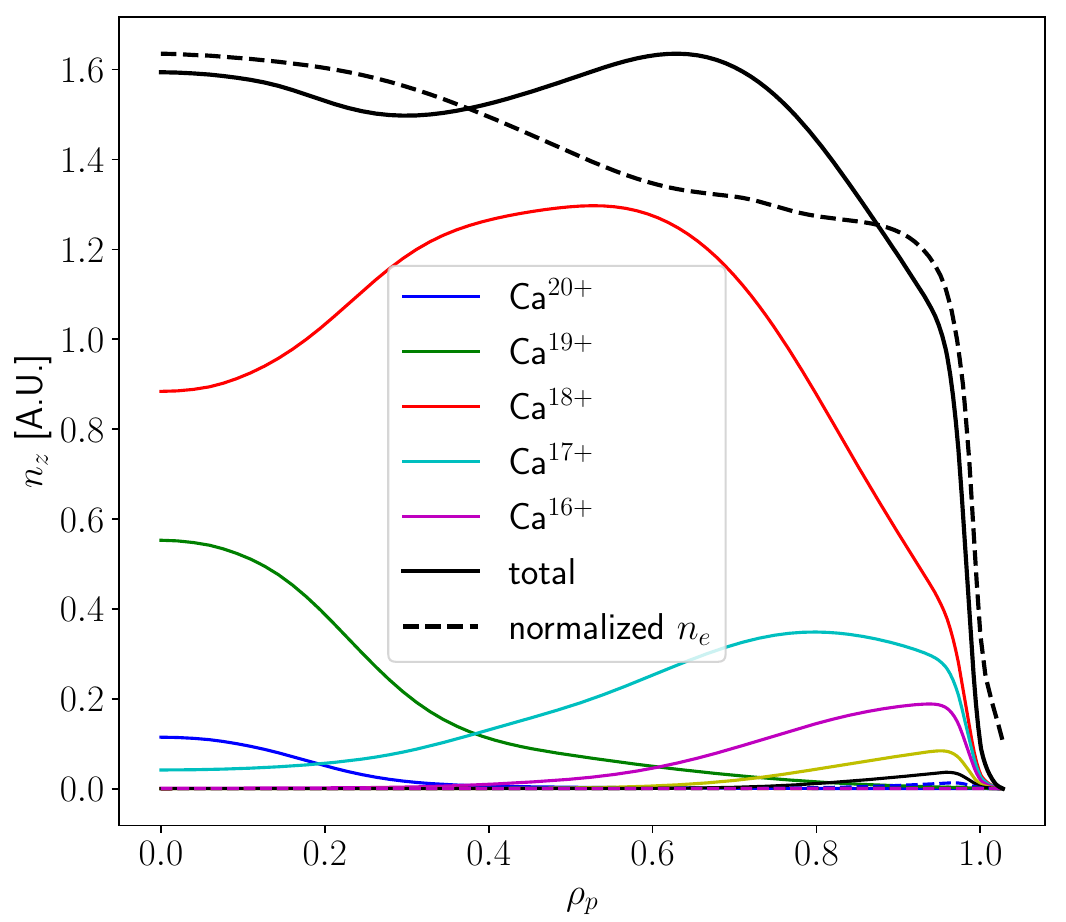}
	\caption[Steady-state radial density profiles of the five highest Ca charge states from the experimental EDA H-mode inference of transport coefficients.]{Steady-state radial density profiles of the five highest Ca charge states from the experimental EDA H-mode inference of transport coefficients of Fig.~\ref{fig:inference_hmode}. Dashed black lines show the rescaled electron density profile for comparison with the total impurity density profile.\\ \\ \\ \\}
	\label{fig:cmod_hmode_nz_profs}
\end{SCfigure}

The most significant discrepancy between theory and experiments in the reported inferences is the different sign of convection at midradius in the EDA H-mode case (Fig.~\ref{fig:inference_hmode}). This discharge has low plasma current ($I_p=0.55$~\si{MA}) and a small inversion radius, hence predictions at $\rho_p\approx 0.5$ are unlikely to be affected by inaccuracies in our sawtooth modeling. However, TGLF predicts a strongly peaked impurity density profile, with $Rv/D\approx -5$~\si{m^{-1}}, in clear contrast to our experimental observations. Fig.~\ref{fig:cmod_hmode_nz_profs} shows radial profiles from Aurora for some of the highest Ca charge states in this H-mode case, using the transport coefficients of Fig.~\ref{fig:inference_hmode}, long after the simulated LBO injection. Profiles inside of $\rho_p=0.2$ are affected by sawteeth and hence do not present the hollowness that may be expected based on Fig.~\ref{fig:inference_hmode}. The black continuous line shows the total impurity density profile (sum over all charge states). It is interesting to contrast this to the black dashed profile, which shows the electron density profile shape. An analogous pedestal structure is found, but core gradients differ substantially. The inferred profile of total impurity density is slightly hollow (almost flat), thus differing from what may be predicted from an impurity model that fixes profile shapes to be a constant fraction of the $n_e$ profile. 

We highlight that the L- and I-mode cases are found to have relatively similar transport coefficients, while the EDA H-mode has smaller $D$ and $v$, according to both theory and experiment. The similarity between L- and I-mode, which have close values of $I_p=0.8$ and $1.0$~\si{MA}, respectively, is in agreement with general observations of particle transport being similar between these two regimes, even though I-modes have a pedestal heat transport barrier~\cite{Whyte2010I-mode:C-Mod}. On the other hand, the EDA H-mode case has lower $I_p=0.55$~\si{MA}, which may suggest lower confinement, and therefore larger transport coefficients with respect to the L- and I-mode cases. The small $D$ and $v$ in Fig.~\ref{fig:inference_hmode} may seem to contradict this intuition. In fact, global impurity confinement times, $\tau_p$, appear to be dominated by the size of the density pedestal pinch in EDA H-mode, rather than by core $D$ and $v$. Previous work on Alcator C-Mod has shown a strong positive correlation between impurity confinement time, $\tau_p$, and $I_p$ in EDA H-mode~\cite{Rice2015CorePlasmas}, relating it to variations in pedestal density gradients~\cite{Hughes2007H-modeC-Mod}. The EDA H-mode in Fig.~\ref{fig:inference_hmode} has slow transport in the core, but a comparable pedestal pinch to the L- (Fig.~\ref{fig:inference_lmode}) and I-mode (Fig.~\ref{fig:inference_imode}) cases, resulting in larger $\tau_p$ overall. 


\subsection{Comparison to Neoclassical \& Turbulence Transport Models} \label{sec:validation}

In this section, we provide some remarks on the neoclassical and turbulent modeling shown in Figs.~\ref{fig:inference_lmode}, \ref{fig:inference_imode}, and~\ref{fig:inference_hmode}. All modeling choices for NEO, TGLF, and CGYRO simulations are the same as those described in detail in Ref.~\cite{Sciortino2020InferenceSelection}. Here we summarize some of these choices and briefly comment on the results of the validation effort presented earlier in this section.

All the theoretical transport model predictions shown in Figs.~\ref{fig:inference_lmode}, \ref{fig:inference_imode}, and~\ref{fig:inference_hmode} refer to Ca$^{18+}$, which is the dominant charge state across the entire plasma radius, as shown in Fig.~\ref{fig:cmod_hmode_nz_profs}. Variations in transport between Ca$^{18+}$ and adjacent charge states are taken to be negligible. In order to test whether any $Z$ dependence of \emph{pedestal} impurity transport may be constrained by our data, inclusion of a free parameter characterizing the edge convection dependence on $Z$ was tested. In agreement with work previously reported in Ref.~\cite{Sciortino2020InferenceSelection}, no significant evidence has been found for such scaling. This should not be interpreted as suggesting that there exists no $Z$ scaling of transport in the pedestal, but only that our data are unable to identify it. Direct measurements of multiple charge states spanning a wide range of $Z$ are likely necessary to determine whether neoclassical theory applies in the C-Mod pedestal, analogous to reports by P{\"u}tterich \emph{et al.} on AUG~\cite{Putterich2011ELMUpgrade}. The comparison of NEO results for Ca$^{18+}$ and the inferred $v$ pedestal profile is certainly suggestive of this hypothesis, which however awaits further confirmation.

TGLF results in Figs.~\ref{fig:inference_lmode}, \ref{fig:inference_imode}, and~\ref{fig:inference_hmode} were obtained including all electromagnetic components and using the SAT-1 saturation rule, since this appears to better match experimental kinetic profiles on C-Mod when attempting to match experimental heat fluxes from TRANSP~\cite{Breslau2018TRANSP} with the TGYRO~\cite{Candy2009TokamakSimulation} code. Fast ions (from ICRH) were excluded from simulations. In all the cases analyzed here, the Ion Temperature Gradient (ITG) mode was found to be strongly dominant, as indicated by strong sensitivity to variations of the ion temperature gradient scale length, $L_{T_i}$. In Ref.~\cite{Sciortino2020InferenceSelection}, TGLF scans of major turbulence drives were shown to indicate an uncertainty in theoretical predictions for $D$ and $v$ by approximately 50\%. In such scans, $D$ and $v$ variations were typically found to leave the $v/D$ ratio unchanged. Hence, comparison of $v/D$ from experimental inferences and turbulence predictions is expected to be more robust, although matching $D$ and $v$ separately within uncertainty is clearly a stronger validation constraint. Our results for the L-mode (Fig.~\ref{fig:inference_lmode}) and I-mode (Fig.~\ref{fig:inference_imode}) discharges show a favorable match between theory and experiment. On the other hand, the discrepancy seen in the H-mode case (Fig.~\ref{fig:inference_hmode}) clearly challenges our theoretical modeling at midradius.

The nonlinear, heat-flux matched CGYRO simulations are described in greater detail in Appendix~\ref{app:cgyro}. Here we highlight that, in all cases considered, impurity transport appears to be driven solely by ion-scale fluctuations, which are well resolved by our CGYRO runs. CGYRO results are generally found to be in agreement with TGLF SAT-1, although it consistently gives lower values of $D$ and $v$, with consistent $v/D$ values. Future work should obtain nonlinear CGYRO predictions at multiple radii, so as to identify any fortuitous agreement with experimental inferences at a single location. For example, it would have been useful to have additional predictions for $v/D$ at $\rho_p=0.4$ in Fig.~\ref{fig:inference_hmode}, to determine whether the discrepancy between inferred transport coefficients and TGLF is due to inadequacies of the experimental or theoretical modeling. 

\begin{SCfigure}[1.0][!hbt]
    \centering
	\includegraphics[width=0.6\textwidth]{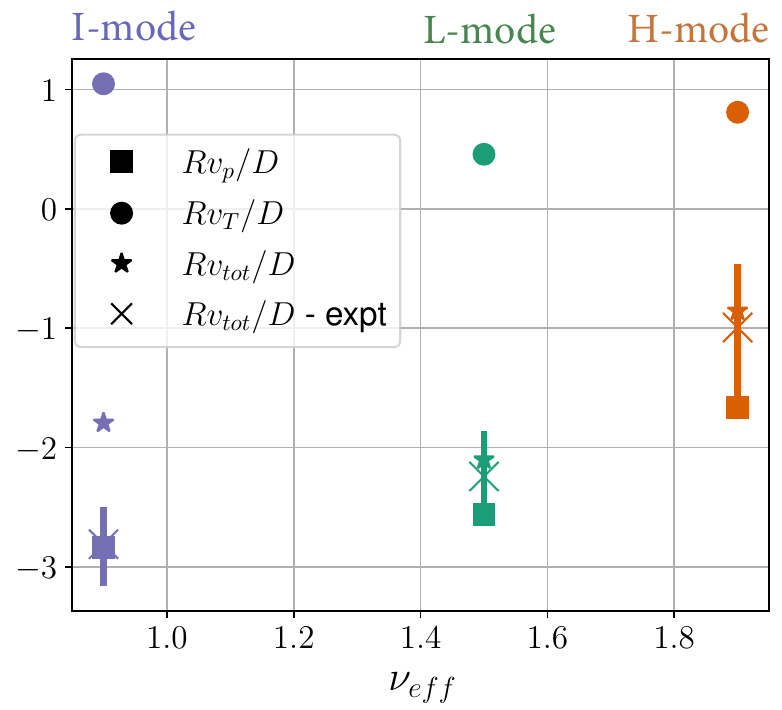}
	\caption[Estimates of $Rv/D$ components from nonlinear CGYRO simulations (pure convection, thermodiffusion and total) and from experimental inferences for the Alcator C-Mod L-mode, EDA H-mode, and I-mode discharges.]{Comparison of $Rv/D$ components from experimental inferences (crosses) and nonlinear CGYRO simulations, separating pure convection (squares), thermodiffusion (circles) and total $Rv/D$ (stars) for the Alcator C-Mod L-mode (green), EDA H-mode (orange), and I-mode (lilac) discharges. Vertical bars display the interquartile range for experimentally-inferred values. \\ \\ }
	\label{fig:cgyro_vod_shots}
\end{SCfigure}

Fig.~\ref{fig:cgyro_vod_shots} shows a comparison of $R v/D$ values from the experimental inferences of Figs.~\ref{fig:inference_lmode}, \ref{fig:inference_imode}, and~\ref{fig:inference_hmode}, together with theoretical components of $R v/D$ from the nonlinear CGYRO simulations for each shot. These are plotted as a function of effective collisionality, $\nu_{eff}$, defined as~\cite{Angioni2007ScalingObservations}
\begin{equation} \label{eq:nu_eff}
    \nu_{\mathrm{eff}}=\frac{0.1 Z_{\mathrm{eff}}\left\langle n_{e}\right\rangle R_{\mathrm{geo}}}{\left\langle T_{e}\right\rangle^{2}}.
\end{equation}
Here, $\left\langle n_{e}\right\rangle$ is the volume-averaged density (in $10^{19}$~\si{m^{-3}}),  $\left\langle T_{e}\right\rangle$ is the volume-averaged plasma temperature (in \si{keV}), $Z_{\text {eff }}$ is the effective charge, and $R_{\text {geo }}$ is the geometric plasma radius (in \si{m}). In Fig.~\ref{fig:cgyro_vod_shots}, as in Fig.~\ref{fig:cmod_3_kin_prof_fits}, L-mode values are green, I-mode ones are lilac, and H-mode ones are orange. Vertical bars display the interquartile range for experimentally-inferred values, shown by crosses. As in numerous previous works (e.g. Refs.~\cite{Casson2010GyrokineticPlasma,Angioni2012Off-diagonalAspects, Angioni2014TungstenModelling}), the effect of thermodiffusion, $D_T$, is separated from the remaining pinch term, $v_p$\footnote{We do not explicitly quantify the rotodiffusion term since this would require assigning different rotation to trace ions in our simulations, which is not possible in CGYRO.}, according to
\begin{equation} \label{eq:theory_fluxes}
\frac{R \Gamma}{n} =  D \frac{R}{L_{n}} + D_{T} \frac{R}{L_{T}} + R v_{p}
\end{equation}
where $L_n$ and $L_T$ are the density and temperature gradient scale lengths of the impurity species, and $R$ is the major radius. From this expression, a \emph{thermal convection} term in units of \si{m/s} can be computed as $v_{T} = D_{T}/L_{T}$. Fig.~\ref{fig:cgyro_vod_shots} shows that the I-mode, at the lowest collisionality, is predicted by CGYRO to have similar $R v_{\text{tot}}/D$ to the L-mode. On the other hand, the EDA H-mode has more than twice the I-mode value of $\nu_{eff}$ and is found to have much smaller peaking, according to both CGYRO and the experimental inference. The collisionality trend observed here matches the general observation that electron density peaking is lower in high-collisionality H-modes than in L- and I-modes~\cite{Angioni2007ScalingObservations, Greenwald2007DensityScalings}. We note that the EDA H-mode case appears to have an excellent match between theory and experiment, but as shown in Fig.~\ref{fig:inference_hmode} the nonlinear CGYRO simulation is actually at a radial location where there exist significant variation in $R v_{\text{tot}}/D$, thus suggesting that this agreement may be fortuitous. In the I-mode case, the experimental inference suggests a larger $R v_{\text{tot}}/D$ than CGYRO does. Finally, a very good match is observed for the L-mode case, with the CGYRO $R v_{\text{tot}}/D$ being well within the experimental interquartile range. 

\section{Summary \& Discussion} \label{sec:conclusion}

Experimental studies of neutrals and impurity transport in tokamaks are central to improving predictive capabilities and ensuring sustainable core-edge scenarios for future reactors. In this work, a detailed analysis of impurity transport has been presented, highlighting the large impact of charge exchange with edge neutrals on ionization balance and radiation in the pedestal. In Section~\ref{sec:xics}, a novel forward model has been introduced for the full Ca K$_\alpha$ spectrum measured by XICS, complemented in Section~\ref{sec:euv} by the analysis of EUV line ratios measured by XEUS. We highlight that the forward model for these diagnostics depends on multiple Ca charge states, effectively all those that have significant density in the confined plasma. This results in effective constraints on particle transport all the way to the separatrix. In order to accurately assess the impact of CX near the edge, in Section~\ref{sec:neutrals} predictions from the EIRENE Monte Carlo neutral model within SOLPS-ITER have been compared with experimental Ly$_\alpha$ data at the C-Mod midplane. Relatively good agreement between the two was found in the three cases examined here. Section~\ref{sec:forward} showed that making use of accurate predictions of background D atomic neutral densities is paramount for effective modeling of near-edge particle transport, especially in plasmas with low (or inexistent) density pedestals. Neutral densities from SOLPS-ITER have been used within the Aurora 1.5D forward model for impurity transport, which is at the core of our C-Mod inferences. While the C-Mod diagnostics used in this study do not have sufficient spatial and temporal resolution to examine pedestal transport in detail, the inference framework presented in this paper is clearly effective in the plasma core, having overcome many of the challenges identified in previous C-Mod work on the subject~\cite{Chilenski2019OnProfiles}.


Inferences of impurity transport for L-, I-, and EDA H-mode discharges have shown general agreement with theoretical models in diffusion profiles and partial agreement in convection profiles. The single most significant discrepancy observed in this work is in the EDA H-mode case, where the predicted peaked impurity density profile is in clear contrast to the hollow (or flat) profile of total impurity density found from experimental data. This observation suggests that there may be inaccuracies in predictions of impurity density peaking from plasma microturbulence, possibly leading to performance under-estimation in future reactors. The inferences presented in this work all employed Bayesian model selection, described in Ref.~\cite{Sciortino2020InferenceSelection}, which in each case indicated the optimal use of 4 spline coefficients for $D$ and 4 spline coefficients for $v/D$ (with $v$ being constrained by sampling in the complex plane, see Section~\ref{sec:inferences} and Appendix~\ref{app:triple_priors}), together with a Gaussian feature in $v$ centered near the LCFS. Although this parametrization was selected via a rigorous statistical procedure, it is likely that \emph{model inadequacy} still limits the accuracy of our experimental results. For example, radial features of experimental transport coefficients may be under-resolved by our diagnostics, or a $Z$ dependence of $D$ and $v$ may be unaccounted for. While significant progress has been made on the development of synthetic diagnostics for both XICS and VUV signals, future work should still expand spectroscopic constraints on impurity transport, aiming to obtain higher temporal and spatial resolution, especially in the pedestal region.

In Section~\ref{sec:validation}, a number of limitations of our theoretical modeling have been described, among these the use of time-averaged plasma profiles. While our approximations are expected not to affect comparisons of transport coefficients at midradius, this is certainly an area deserving further attention. Given the known stabilizing effects of fast ions on ITG turbulence~\cite{Wilkie2018FirstIons,Eriksson2019ImpactModeling}, it is possible that future work may further improve the agreement between theory and experiment shown in this work by including fast ions in theoretical modeling. However, we note that the fast H minority population has relatively low concentration ($\approx 5\%$) and it is almost entirely thermalized with the main D ions at $r/a=0.6$, where CGYRO was run in each case. 

The trend of impurity density peaking on collisionality shown in Fig.~\ref{fig:cgyro_vod_shots} is reminiscent of the well-known scaling for electron density peaking established by Angioni~\cite{Angioni2007ScalingObservations, Greenwald2007DensityScalings}, which has a logarithmic dependence on $\nu_{eff}$ and a linear one on plasma $\beta \sim p/B^2$. However, the simplicity of this scaling cannot recover the wealth of plasma physics that is observed across confinement regimes on C-Mod, where indeed we observe significant variations of electron density peaking that are not described by the scaling. Understanding such behavior via detailed experimental model validation is crucial, particularly when there exists a coupling of transport and atomic physics, as in impurity dynamics. Future work will extend the impurity transport analysis presented in this paper to other devices, attempting to provide clear evidence for any missing physics in predictive models.

\section*{Acknowledgements}
The authors would like to thank the Alcator C-Mod Team for operation of the device over many years. FS also gratefully acknowledges helpful conversations on Bayesian statistics with Y. Marzouk, and on ADAS data with M. O'Mullane. This work made use of the MIT-PSFC partition of the \emph{engaging} cluster at the MGHPCC facility, funded by DoE award No. DE-FG02-91-ER54109, as well as the National Energy Research Scientific Computing Center (NERSC)–a US DoE Office of Science User Facility operated under Contract No. DE-AC02-05CH11231. This material is based upon work supported by the Department of Energy under Award Numbers DE-SC0014264, DE-FC02-04ER54698, and DE-SC0007880.

\clearpage
\begin{appendices}


\section{Sampling of Physically-Correlated $D$, $v$ and $v/D$ Values} \label{app:triple_priors}

When sampling free spline values for $D$ and $v$ parameters, one must ensure that not only such parameters are physically reasonable by themselves, but their ratio, $v/D$, is physical too. This is important because $v/D$ is a physically meaningful quantity, indicative of impurity profile peaking in the absence of particle sources, and therefore we must exclude the possibility of our forward model exploring unreasonable parameter space. For example, if one were to sample $D$ and $v$ values of $0.1$ m$^2$/s and $10$ m/s, respectively, both individually reasonable parameters, the $v/D$ ratio would be $100$ $m^{-1}$, which is itself unlikely to be physical in the plasma core. Clearly, sampling of reasonable $D$ and $v$ values may be achieved by tuning hyper-parameters of each individual prior, but constraining the ratio of the two is not possible when using fully independent priors. The same issue occurs regardless of whether one chooses to sample the \{$D$, $v$\} pair, the \{$D$, $v/D$\} one, or \{$v/D$, $v$\}, since the same two degrees of freedom are effectively used in sampling. Correlations between priors must therefore be introduced.

\begin{figure}[htp]
  \centering
  \subfloat{\includegraphics[clip,width=0.45\columnwidth]{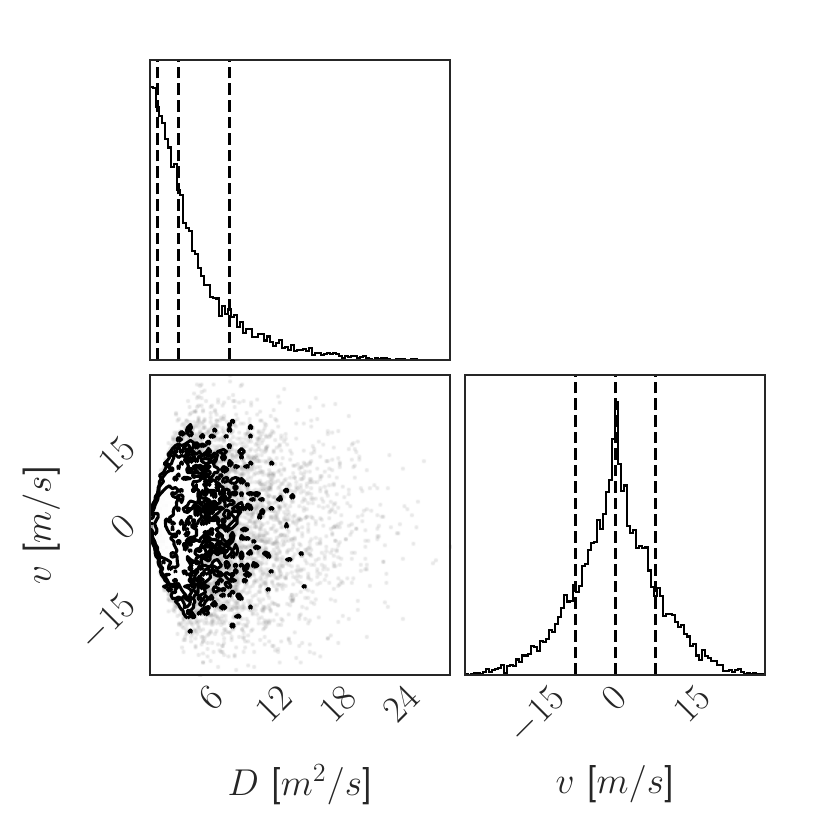}\label{fig:1}}
  \subfloat{\includegraphics[clip,width=0.45\columnwidth]{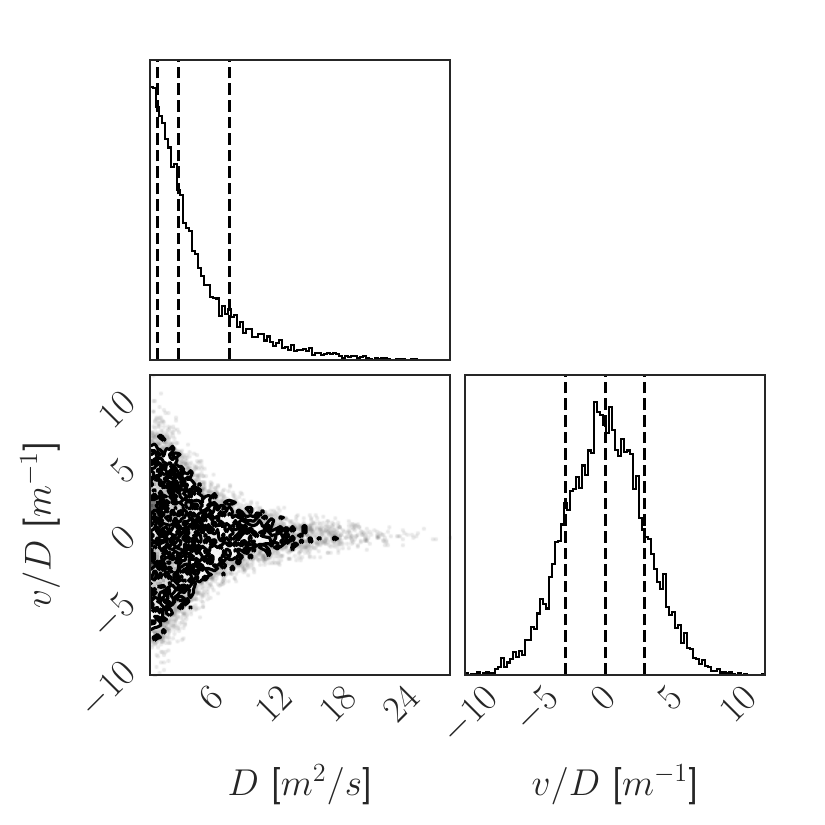}\label{fig:2}}\hspace{1em}
  \subfloat{\includegraphics[clip,width=0.45\columnwidth]{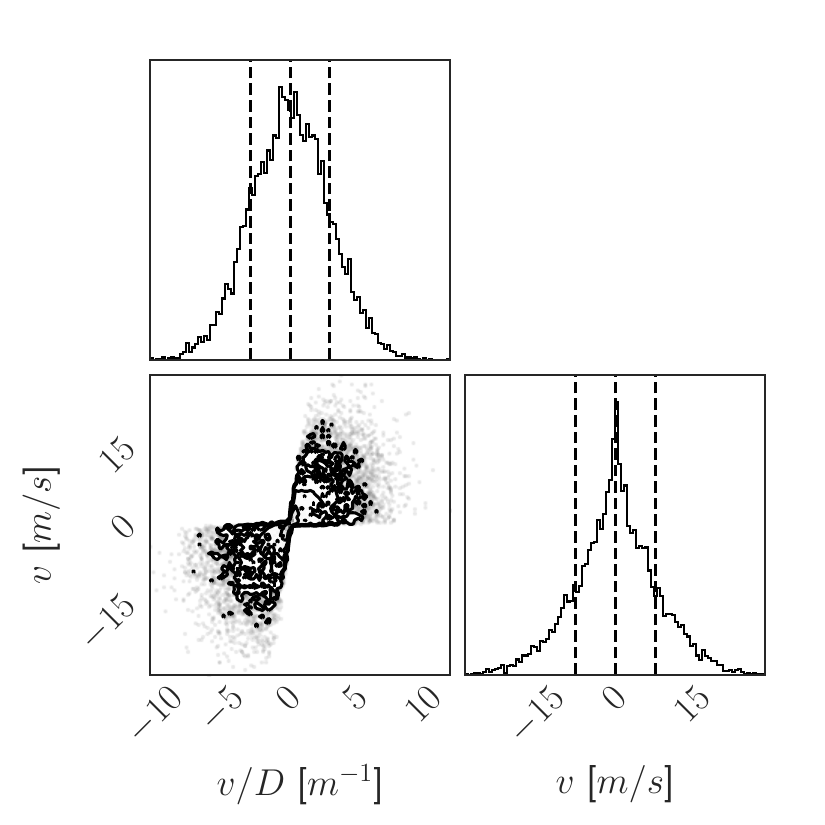}\label{fig:3}}
  \caption{Physical sampling of $D$, $v$ and $v/D$ for the case of $\chi=1$, $\mathcal{P}_{\chi v/D} = \mathcal{N}(0,5)$ and $\mathcal{P}_\psi = \mathcal{N}(0,10)$ based on the scheme described in the text.}
  \label{fig:correlated_DV_priors}
\end{figure}

Here we present a simple, yet effective method to overcome this difficulty. Rather than sampling any of the parameter pairs listed above, two abstract coordinates are sampled in a polar plane. We interpret one of these to correspond to the tangent of an angle, $\tan(\Theta)$, and the second to a radial coordinate, $\psi$. The $\tan(\Theta)$ variable is taken to be distributed according to an arbitrary prior distribution for the product $\chi \cdot v/D$, which we denote by $\mathcal{P}_{\chi v/D}$. Here, $\chi$ is a parameter that allows for scale separation of $D$ and $v$ priors. We sample a value of $\psi$ according to a chosen distribution $\mathcal{P}_{\psi}$. The $\psi$ and $\Theta$ samples can then be simply combined to give 
\begin{equation} \label{eq:triple_DV_priors_scheme}
    D = \psi \cos(\Theta) \qquad \qquad v = \psi \sin(\Theta) / \chi
\end{equation}
This scheme can be interpreted geometrically as follows: one degree of freedom is used to sample the ratio of $\chi v$ and $D$ (represented by the $\tan(\Theta)$ variable) and another degree of freedom to sample the value of $\psi = \sqrt{|\chi v|^2 + D^2}$. Eq.~\ref{eq:triple_DV_priors_scheme} then projects the $\psi$ radial distance onto the $v$ and $D$ dimensions. The parameter $\chi$ allows one to set different scales for $D$ and $v$ while still sampling their $\psi$ amplitude from the same distribution. A value of $\chi<1$ would indicate that sampling of the magnitude of $v$ must be allowed to explore large values. Fig.~\ref{fig:correlated_DV_priors} offers a visual representation of this sampling scheme for the case of $\chi=1$, $\mathcal{P}_{\chi v/D} = \mathcal{N}(0,5)$ and $\mathcal{P}_\psi = \mathcal{N}(0,10)$. These corner plots clearly show that the scheme effectively prevents sampling of unphysical values, trading prior freedom in $D$ and $v$ so as to maintain a reasonable $v/D$. One trade-off in using this is that $D$ and $v$ amplitudes must use the same prior distribution. This is partly a limitation for $D$, because a rigorous Jeffrey' prior for $D$ may be more appropriately set as a log-uniform distribution. However, this is a relatively arbitrary choice in our case, and the practical difference is expected to be always negligible in realistic inferences. Another limitation of this scheme is that it forces one to have $D$ and $v$ spline knots at the same locations, rather than possibly independent of each other. In the C-Mod plasmas examined in this work, having different knots for $D$ and $v$ appeared unnecessarily complicated, therefore sampling of physically-correlated $D$, $v$ and $v/D$ values constituted a valuable technical capability.

\begin{figure}[ht]
	\centering
	\includegraphics[width=\textwidth]{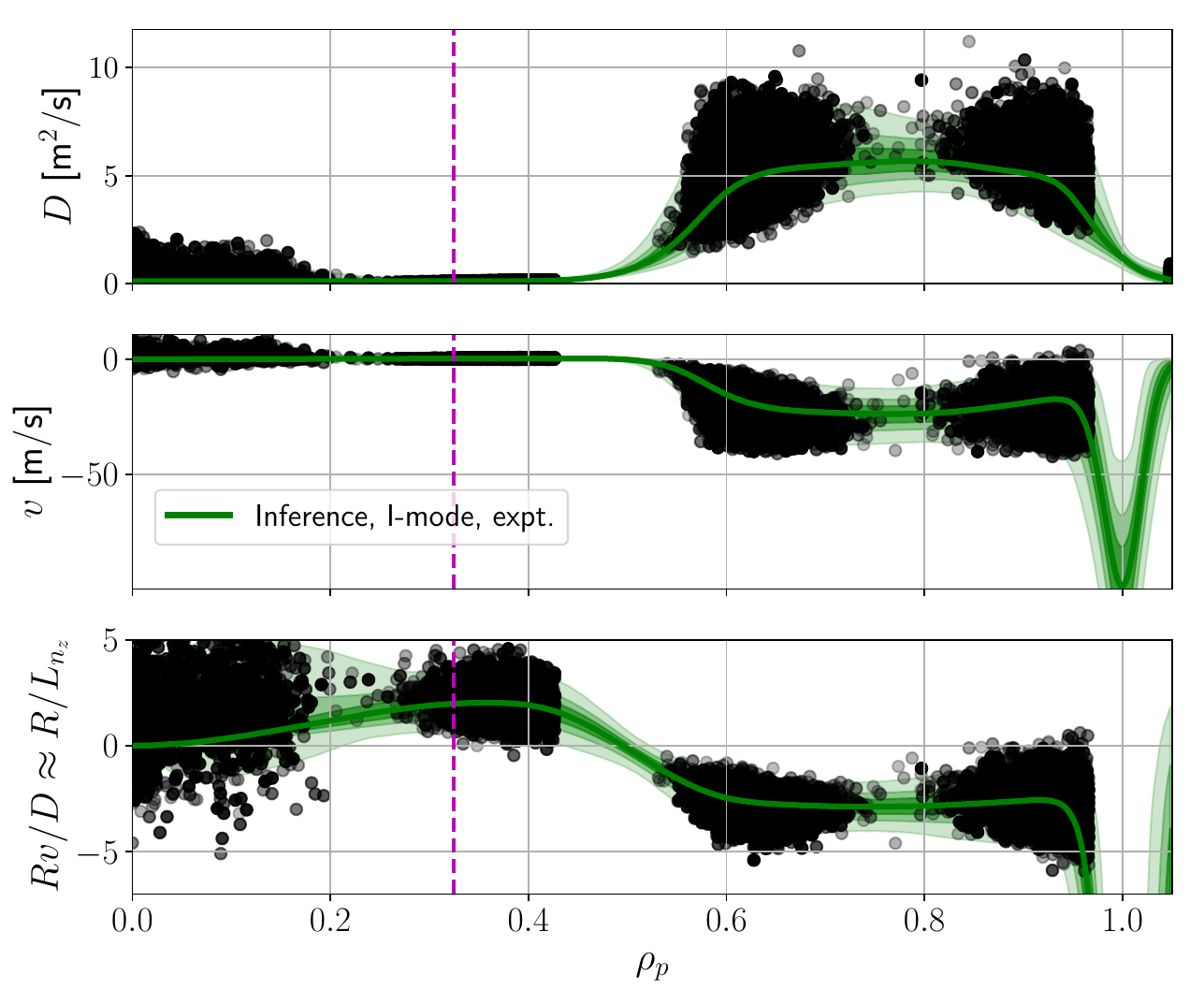}
	\caption[Inference of Ca particle transport coefficients in the C-Mod I-mode discharge 1101014030, showing the distribution of free knots that were sampled with highest posterior probability via MultiNest.]{Inference of Ca particle transport coefficients in the C-Mod I-mode discharge 1101014030, showing the distribution of free knots that were sampled with highest posterior probability via MultiNest. Each spline knot sample is shown by a grey circle; the overlap between many samples gives darker points.} 
	\label{fig:inference_imode_knots}
\end{figure}


\section{Non-Separable Priors for Free Spline Knots} \label{app:knots}
Previous research on impurity transport inferences on Alcator C-Mod~\cite{Chilenski2017ExperimentalSimulations} found issues with the \emph{identifiability} of free spline knots when these are allowed to be free within the given domain -- in the case of interest, the radial range of impurity transport simulations. This problem can be understood as the result of knots not having an individual identity when they are all being sampled within the same range and then being sorted from smallest to largest values, since two knots can in this case effectively swap places with no practical difference for the forward model. This leads to complications in learning the posterior landscape since parameters are not uniquely defined.

We address this problem by ``forcing'' identifiability of sampled knots by remapping the unit hypercube samples for spline knot locations from a sampling algorithm in such a way as to always retain their numerical order. To explain this method, first let us recall the standard use of separable priors for each sampled dimension:
\begin{equation}
    \pi(\theta)=\prod_{i} \pi_{i}\left(\theta_{i}\right)
\end{equation}

In order to work in a ``whitened'' space of prior samples in a unit hypercube, we make use of inverse cumulative distribution functions (CDFs), which for a general 1-dimensional distribution $f(\theta)$ are defined as 
\begin{equation}
    F(\theta) = \int_{-\infty}^\theta f(\theta') d\theta'
\end{equation}
The inverse of the CDF function $F(\theta)$ is well defined for some of the most common prior distributions (e.g. uniform or Gaussian). This allows uniform random variables $x\sim U(0,1)$ to be easily mapped to samples from arbitrary distributions, if one knows the inverse CDF. In a more general N-dimensional case, this so-called \emph{inverse transform sampling} applied to prior distributions $\pi(\theta)$ requires use of N conditional distributions
\begin{equation} \label{eq:marg_prior}
    \pi_{i}\left(\theta_{i} \mid \theta_{i-1}, \ldots, \theta_{0}\right)=\frac{\int \pi_{i}(\theta) \mathrm{d} \theta_{i+1}, \ldots, \mathrm{d} \theta_{N}}{\int \pi_{i}(\theta) \mathrm{d} \theta_{i}, \ldots, \mathrm{d} \theta_{N}}
\end{equation}
which when integrated over each dimension of interest give
\begin{equation}  \label{eq:cdf}
    F_{i}\left(\theta_{i} \mid \theta_{i-1}, \ldots, \theta_{0}\right)=\int_{-\infty}^{\theta_{i}} \pi_{i}\left(\theta_{i}^{\prime} \mid \theta_{i-1}, \ldots, \theta_{1}\right) \mathrm{d} \theta_{i}^{\prime}    
\end{equation}
The relation of $\theta_i = F_i^{-1}(x_i\mid \theta_{i-1}, \dots, \theta_1)$ allows mapping of N uniformly distributed random variables $\{x_i\}$ into samples in the space of interest $\{\theta_i\}$, distributed according to $\pi_i(\theta)$. This is the basis on which it is possible to use \emph{prior space whitening} by using samples from a unit hypercube in the context of many Bayesian inference algorithms, including the nested sampling ones adopted in this work.

The method adopted in this work to force identifiability of free spline knots builds on the work by Handley~\cite{Handley2015POLYCHORD:Sampling} and makes use of a uniform prior in the hyper-triangle defined by $\theta_{min}<\theta_1<\dots< \theta_n<\theta_{max}$ with
\begin{equation} \label{eq:hypertriangle_uniform}
    \pi(\theta)=\left\{\begin{array}{cl}\frac{1}{n !\left(\theta_{\max }-\theta_{\min }\right)^{n}} & \text { for } \theta_{\min }<\theta_{1}<\cdots<\theta_{n}<\theta_{\max } \\ 0 & \text { otherwise. }\end{array}\right.
\end{equation}
Evaluating the marginalized prior of Eq.~\ref{eq:marg_prior} for this uniform hyper-triangle n-dimensional prior, one obtains~\cite{Handley2015POLYCHORD:Sampling}
\begin{equation}
    \pi_{i}\left(\theta_{i} \mid \theta_{i-1}, \ldots, \theta_{0}\right)=\frac{(n-i+1)\left(\theta_{i}-\theta_{i-1}\right)^{n-i}}{\left(\theta_{\max }-\theta_{\min }\right)^{n-i+1}}
\end{equation}
and the cumulative distribution function (CDF) is then found to be
\begin{equation}  \label{eq:knots_cdf}
    F_{i}\left(\theta_{i} \mid \theta_{i-1}, \ldots, \theta_{0}\right)=\left(\frac{\theta_{i}-\theta_{i-1}}{\theta_{\max }-\theta_{i-1}}\right)^{n-i+1}.
\end{equation}
Each value of the CDF, $x_i \equiv F_{i}\left(\theta_{i} \mid \theta_{i-1}, \ldots, \theta_{0}\right)$, can be taken to correspond to a unit hypercube sample; by inverting this relationship (i.e. using the inverse CDF), we can map unit hypercube samples $\{x_i\}$ to parameter samples $\{\theta_i\}$. A simple re-arrangement of Eq.~\ref{eq:knots_cdf} gives
\begin{equation} \label{eq:handley_free_knots}
    \theta_i = \theta_{i-1}+(\theta_{max} - \theta_{i-1}) x_i^{1/(n-i+1)}
\end{equation}
Eq.~\ref{eq:handley_free_knots} offers the mapping between unit hypercube samples and sorted uniformly-distributed spline knots between $\theta_{min}$ and $\theta_{max}$~\cite{Handley2015POLYCHORD:Sampling}. Comparing this to the equivalent linear remapping that would be obtained for a 1-dimensional uniform prior
\begin{equation}
    \theta_i = \theta_{i-1}+(\theta_{max} - \theta_{i-1}) x_i
\end{equation}
one can clearly notice the similarity.

In transport inferences, one is interested in finding a distribution for free knots between the magnetic axis at $\rho_p=0$ and the end of the radial grid at $\rho_p=\rho_{p,max}$, so one can set $\theta_{min} = 0$ and $\theta_{max}=\rho_{p,max}$. To ensure that knots are not sampled too close to each other, possibly leading to large gradients in transport coefficients that could lead to numerical instability, we wish to also impose a minimum distance, $\Delta\rho$, between sampled knots. In order to do this, we artificially reduce the value of $\theta_{max}$ from the end of the grid, $\rho_{p,max}$, to a value closer to the axis and dependent on the number of free spline knots being sampled, i.e.
\begin{equation}
    \theta_{\text{max}} = \rho_{p,max} - \Delta\rho \times (N-1)
\end{equation}
where $N$ is the number of free knots. Applying the sampling procedure of Eq.~\ref{eq:handley_free_knots} with this value of $\theta_{\text{max}}$, one can then shift each sampled knot location by $\Delta \rho \times i$, where $i$ is the index of each ordered knot value. This maintains the attractive feature of this technique, that is to enforce identifiability of samples, while demanding that a minimum distance of $\Delta \rho$ is respected between knots at all times.


\section{CGYRO Modeling} \label{app:cgyro}

In this appendix, we provide additional details and results from our CGYRO nonlinear, ion-scale, heat-flux-matched simulations. The material presented here complements results presented in Ref.~\cite{Sciortino2020InferenceSelection}, where the I-mode case was already discussed. The L-mode and EDA H-mode simulations, whose predictions for Ca transport coefficients are shown in Figs.~\ref{fig:inference_lmode} and~\ref{fig:inference_hmode}, are presented in this paper for the first time.

Simulations for each shot are focused on the flux tube at $r/a=0.6$, where experimental profiles are well determined and the effect of sawteeth is expected to be small. We use a domain size with $L_x/\rho_s\approx 100$ and $L_y/\rho_s\approx 100$, radial grid spacing of $\Delta x/\rho_s= 0.061$ and $\Delta y/\rho_s= 0.065$, with 344 radial modes and 22 toroidal modes, giving $\max(k_x \rho_s) \approx 10.5$ and  $\max(k_y \rho_s) \approx 1.4$. 
In velocity space, we use a grid with 16 pitch angles, 8 energy, and 24 poloidal points. We adopt experimental profile inputs, Miller geometry, electromagnetic ($\tilde{\phi}$ and $\tilde{A}_{||}$) effects, as well as gyrokinetic electrons. We apply the sonic rotation scheme available in CGYRO~\cite{Belli2018ImpactTransport}, known to be important for modeling of heavy impurities~\cite{Casson2015TheoreticalUpgrade}. Averages and uncertainties on simulation outputs are computed from moving averages over windows of length equal to 3 times the estimated correlation time. 

\begin{figure}[!htbp]
    \begin{center}
    \begin{sideways}
    \noindent \begin{minipage}{0.9\textheight}
        \begin{subfigure}{.33\textwidth}
            \centering
            \includegraphics[width=\linewidth]{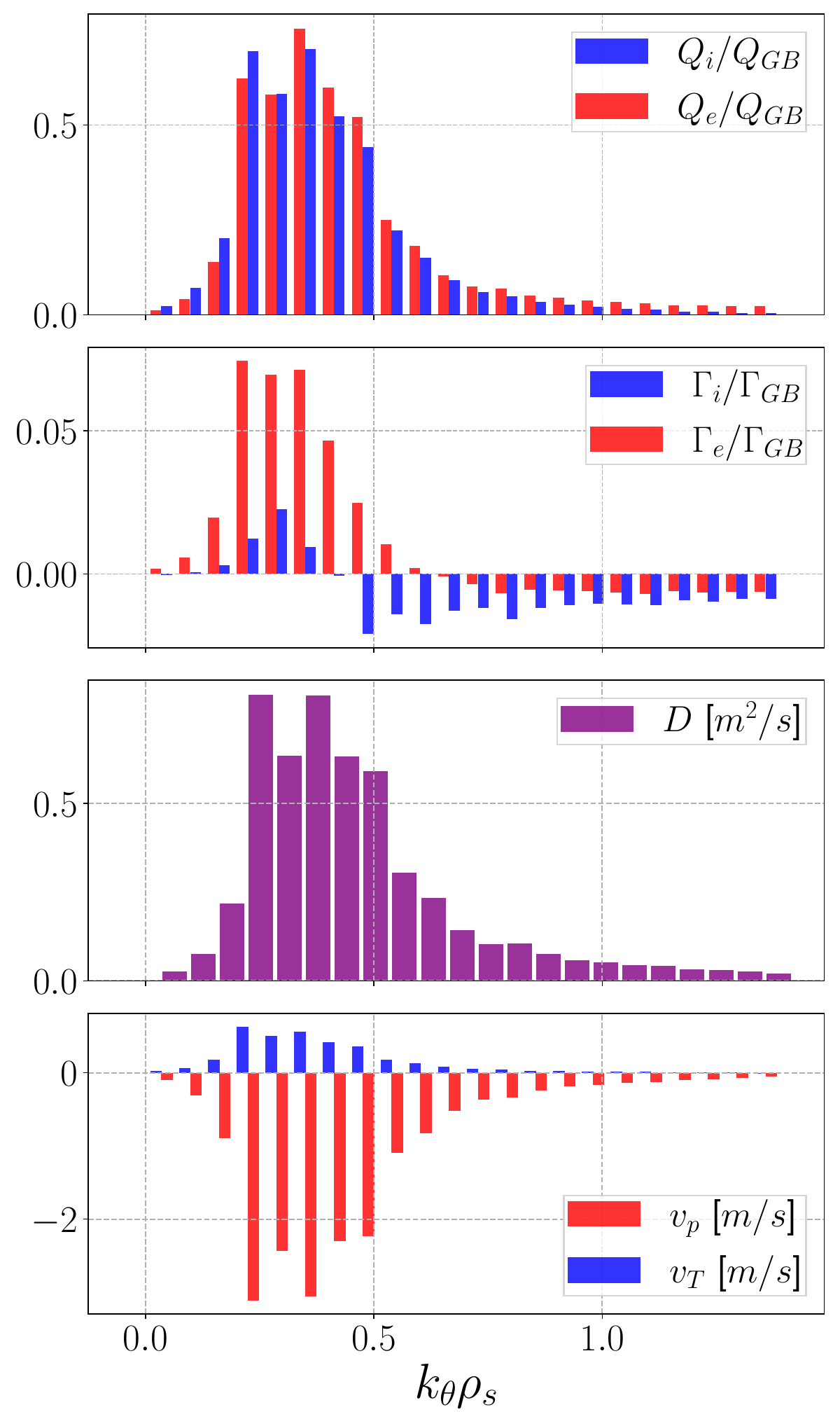}
        \end{subfigure}%
        \begin{subfigure}{.33\textwidth}
          \centering
          \includegraphics[width=\linewidth]{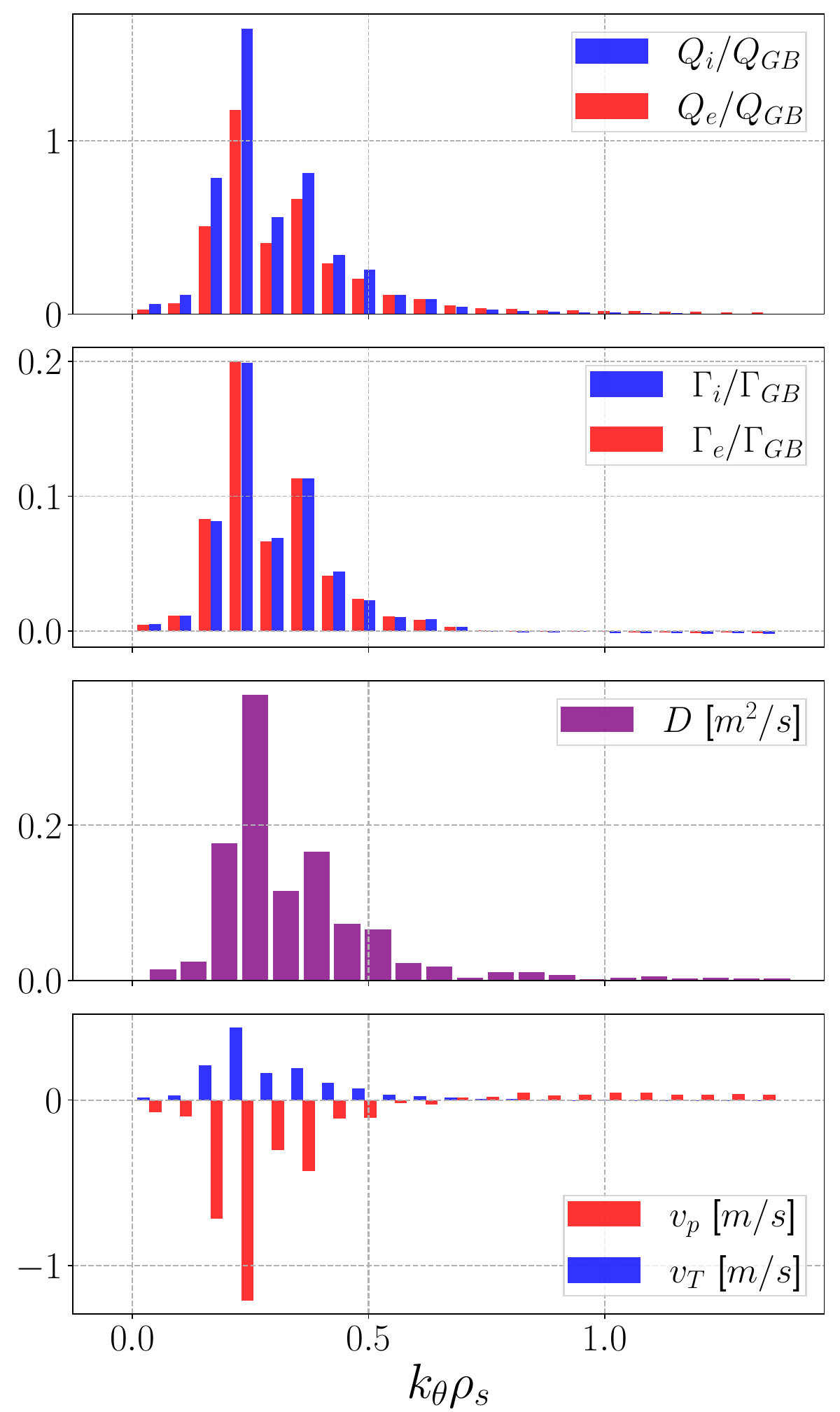}
        \end{subfigure}%
        \begin{subfigure}{.33\textwidth}
          \centering
          \includegraphics[width=\linewidth]{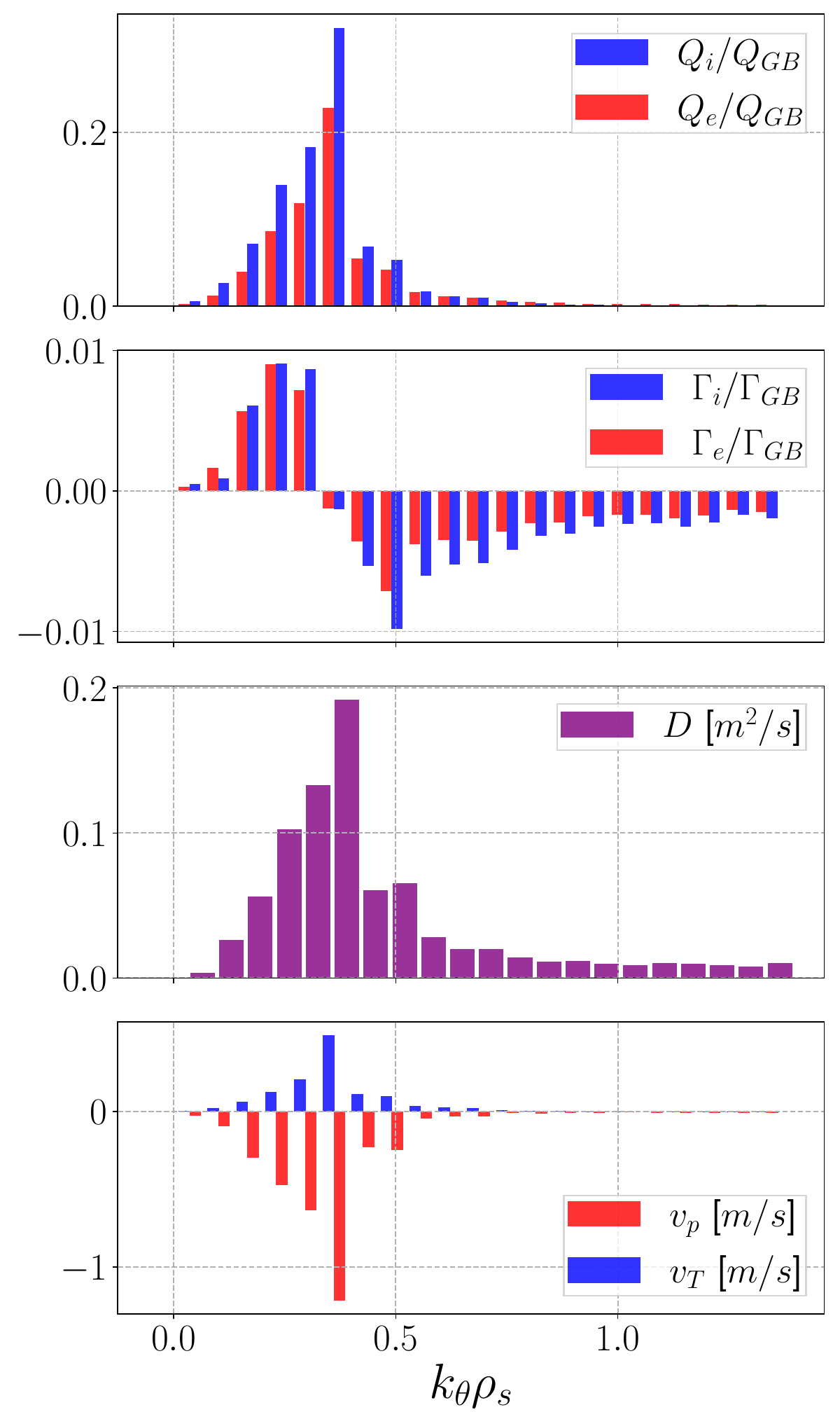}
        \end{subfigure}
    \end{minipage}%

    \end{sideways}
    \end{center}
    \caption[CGYRO spectra for L-mode (left), EDA H-mode (center), and I-mode (right) shots.]{CGYRO spectra for the L-mode (left), EDA H-mode (center), and I-mode (right) shots discussed in Section~\ref{sec:inferences}.}
    \label{fig:CGYRO_DV_spectra}
\end{figure}

Fig.~\ref{fig:CGYRO_DV_spectra} shows spectra of heat and particle fluxes, for both D ions and electrons, as well as $D$ and $v$ components for Ca trace impurities at the simulated radial location ($r/a=0.6$). From the left, results are shown for the L-mode, EDA H-mode, and I-mode cases. $D$ spectra show a remarkable similarity to the heat flux coefficients (of both D ions and electrons), whereas particle fluxes show more complex features. Interestingly, in the L-mode and I-mode cases the particle fluxes invert sign (direction) at $k_\theta \rho_s\approx 0.45$ and $0.35$, respectively. On the other hand, the EDA H-mode has $\Gamma>0$ (outward) for both D ions and electrons. The sign of total Ca convection is found to be negative in the bottom panel. 
Thermal convection ($v_T$, or equivalently thermodiffusion $D_T$) always gives $v>0$ (outwards) contributions, as expected for ITG-dominated regimes~\cite{Angioni2012Off-diagonalAspects}. The remaining pure pinch, $v_p$, is always larger and negative, resulting in predictions of peaked profiles in all the 3 shots at this location ($r/a=0.6$).

\begin{figure}
\begin{center}
\begin{sideways}
     \begin{minipage}{0.9\textheight}
        \begin{subfigure}{.33\textwidth}
            \centering
            \includegraphics[width=\linewidth]{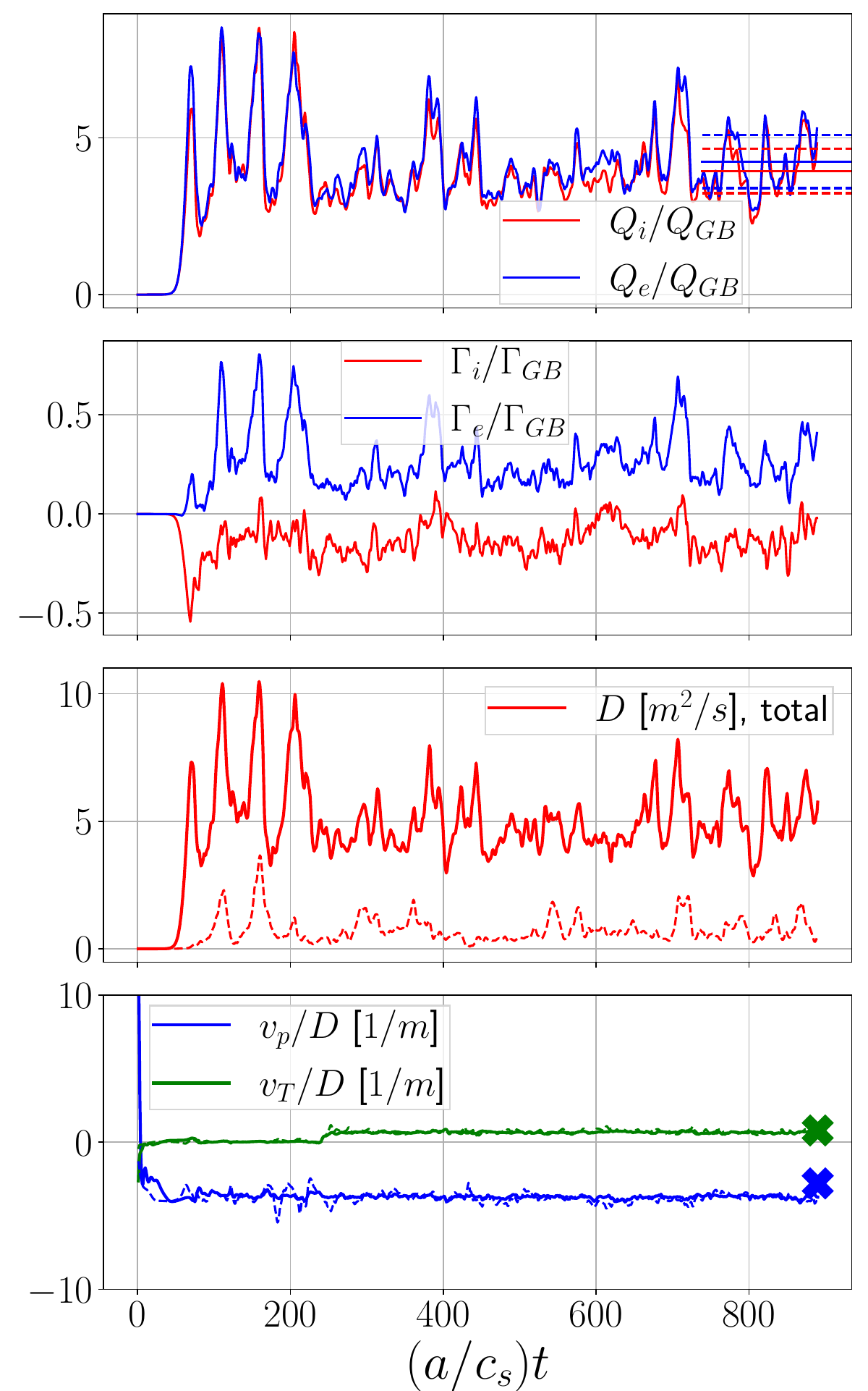}
        \end{subfigure}%
        \begin{subfigure}{.33\textwidth}
          \centering
          \includegraphics[width=\linewidth]{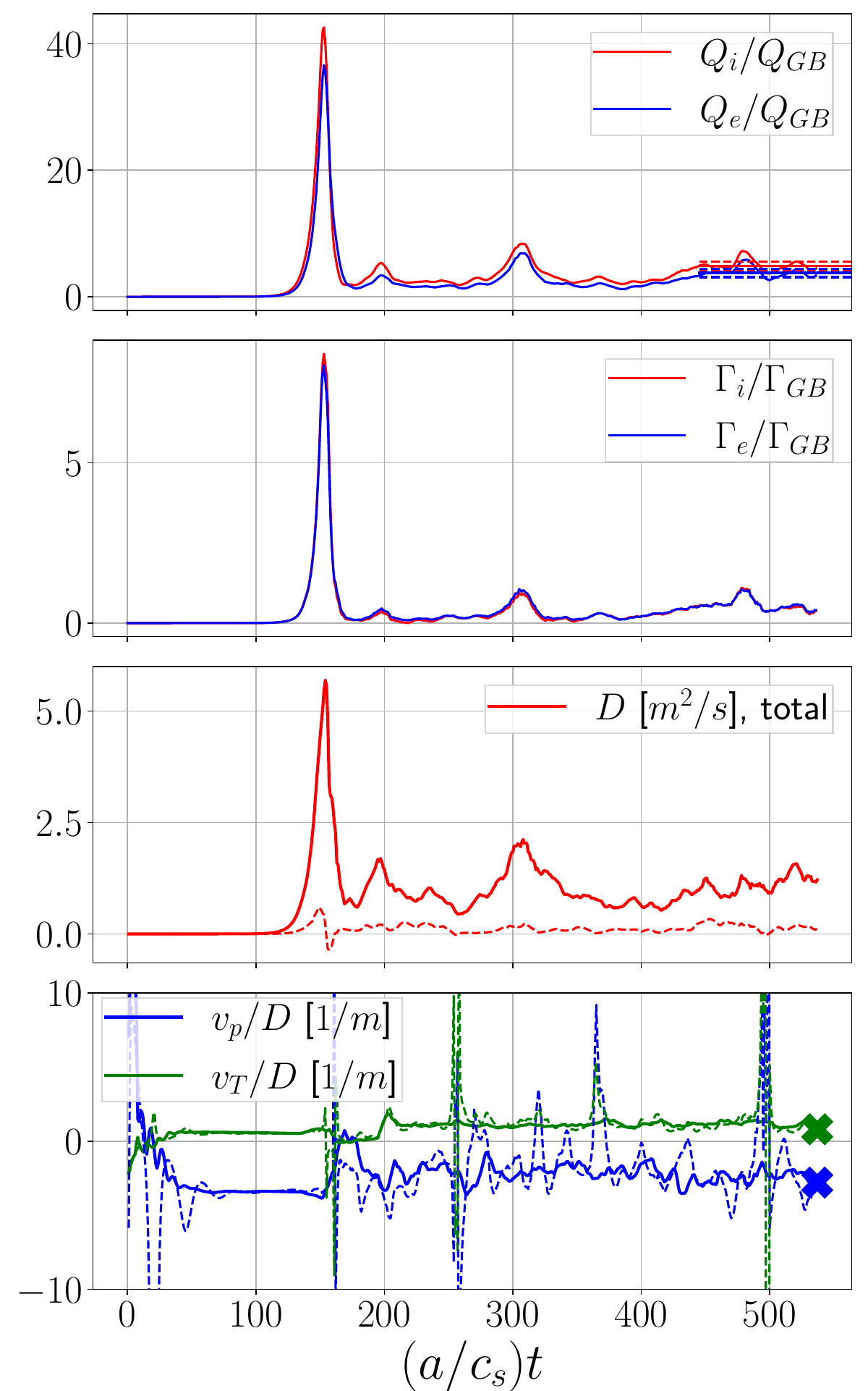}
        \end{subfigure}%
        \begin{subfigure}{.33\textwidth}
          \centering
          \includegraphics[width=\linewidth]{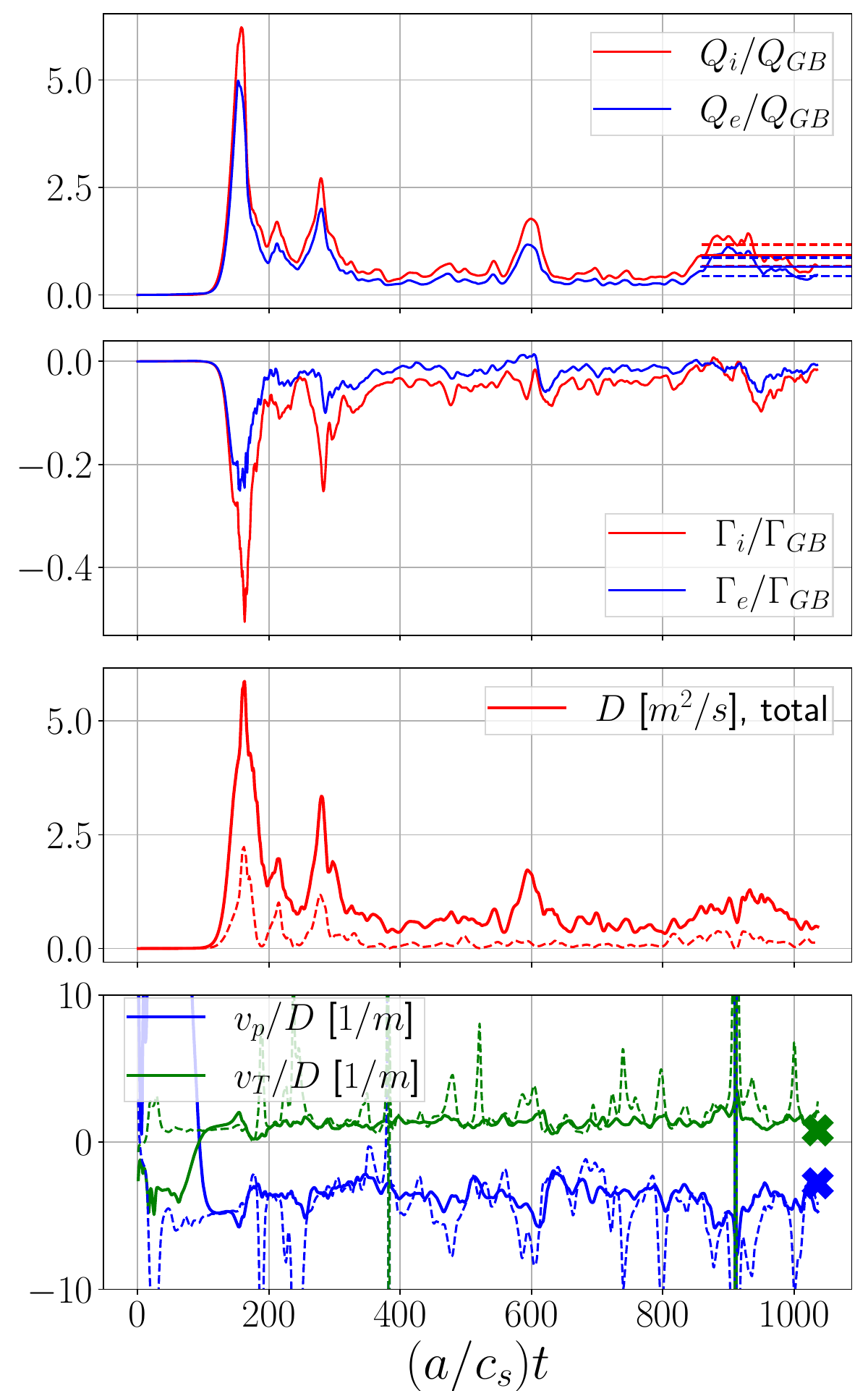}

        \end{subfigure}
     \end{minipage}
\end{sideways}
\end{center}
\caption[Time histories of heat and particle fluxes (for D ions and electrons) and transport coefficients ($D$ and $v/D$ components) from ion-scale heat-flux matched nonlinear CGYRO simulations for the L-mode (left), EDA H-mode (center), and I-mode (right) shots discussed in Section~\ref{sec:inferences}.]{Time histories of heat and particle fluxes (for D ions and electrons) and transport coefficients ($D$ and $v/D$ components) from ion-scale heat-flux matched nonlinear CGYRO simulations for the L-mode (left), EDA H-mode (center), and I-mode (right) shots discussed in Section~\ref{sec:inferences}.}
\label{fig:cgyro_histories}
\end{figure}

Fig.~\ref{fig:cgyro_histories} shows time histories during the CGYRO simulations for heat and particle fluxes, and $D$ and $v/D$ components. As in Fig.~\ref{fig:CGYRO_DV_spectra}, left plots refer to the L-mode, those in the center to the EDA H-mode, and those on the right to the I-mode. Significantly stronger and more intermittent turbulence is seen in the L-mode case. As shown in the comparison to the experimental inference in Fig.~\ref{fig:inference_imode}, CGYRO appears to under-predict the transport levels in the I-mode case. The fact that TGLF gives larger $D$ and $v$ cannot be the result of it being a ``better model'' than CGYRO, since it is a quasilinear and gyrofluid approximation to the gyrokinetic model implemented in the latter. The choice of displaying $v/D$ rather than $v$ in Fig.~\ref{fig:cgyro_histories} was made to allow comparison of the nonlinear CGYRO predictions with quasi-linear CGYRO runs, from which an estimate of $v/D$ can be obtained under the assumption that both $D$ and $v$ saturate in a quasi-linear manner. If so, the turbulence saturation amplitude, setting the scale of both $D$ and $v$, should cancel out. Continuous lines in Fig.~\ref{fig:cgyro_histories} are the quantities obtained by summing over all toroidal modes; dashed lines show results obtained from only the strongest mode, found via linear simulations to be at $k_y \rho_s\approx 0.4$ in each case. We note that the dashed red line in the $D$ panel does not match the time history (and, of course, not the amplitude) of the continuous red lines. On the other hand, dashed and continuous lines for $v/D$ match closely for each convection component separately; discrepancies between single-mode and overall traces are due to transient growth of other toroidal modes. Quasilinear CGYRO results at $k_y \rho_s=0.4$ are shown by blue and green crosses for thermodiffusion ($v_T/D$) and pure convection ($v_p/D$), respectively, at the end time of the nonlinear simulation. In each case, a good match of $v/D$ components between quasilinear and fully-nonlinear theory is found, suggesting that quasilinear impurity peaking predictions such as those of TGLF may be appropriate.

\end{appendices}

\clearpage

\bibliographystyle{unsrt}
\bibliography{references, references2}

\begin{thebibliography}{100}

\bibitem{Putterich2019DeterminationFactors}
T.~P{\"{u}}tterich, E.~Fable, R.~Dux, M.~O'Mullane, R.~Neu, and M.~Siccinio.
\newblock {Determination of the tolerable impurity concentrations in a fusion
  reactor using a consistent set of cooling factors}.
\newblock {\em Nuclear Fusion}, 59(5):ab0384, 2019.

\bibitem{Reinke2017HeatTokamaks}
M.~L. Reinke.
\newblock {Heat flux mitigation by impurity seeding in high-field tokamaks}.
\newblock {\em Nuclear Fusion}, 57(3):034004, 3 2017.

\bibitem{Mordijck2020OverviewTransport}
S.~Mordijck.
\newblock {Overview of density pedestal structure: Role of fueling versus
  transport}.
\newblock {\em Nuclear Fusion}, 60(8), 2020.

\bibitem{Dunne2017ImpactELMs}
M.~G. Dunne.
\newblock {Impact of impurity seeding and divertor conditions on transitions,
  pedestal structure and ELMs}.
\newblock {\em Nuclear Fusion}, 57(2):25002, 2017.

\bibitem{Sciortino2020InferenceSelection}
F.~Sciortino, N.~T. Howard, E~S. Marmar, T.~Odstrcil, N.~M Cao, R.~Dux, A.~E
  Hubbard, J.~W Hughes, J.~H Irby, Y.~M. Marzouk, L.~M. Milanese, M.~L. Reinke,
  J.~E Rice, and P.~Rodriguez-Fernandez.
\newblock {Inference of experimental radial impurity transport on Alcator
  C-Mod: Bayesian parameter estimation and model selection}.
\newblock {\em Nuclear Fusion}, 60(12), 8 2020.

\bibitem{Dux2003ImpurityUpgrade}
R.~Dux.
\newblock {Impurity transport in ASDEX upgrade}.
\newblock {\em Fusion Science and Technology}, 44(3):708--715, 2003.

\bibitem{Dux2003InfluenceUpgrade}
R.~Dux, R.~Neu, A.~G. Peeters, G.~Pereverzev, A.~M{\"{u}}ck, F.~Ryter, and
  J.~Stober.
\newblock {Influence of the heating profile on impurity transport in ASDEX
  Upgrade}.
\newblock {\em Plasma Physics and Controlled Fusion}, 45(9):1815--1825, 2003.

\bibitem{Putterich2011ELMUpgrade}
T.~P{\"{u}}tterich, R.~Dux, M.~A. Janzer, and R.~M. McDermott.
\newblock {ELM flushing and impurity transport in the H-mode edge barrier in
  ASDEX Upgrade}.
\newblock In {\em Journal of Nuclear Materials}, volume 415, pages 334--339,
  2011.

\bibitem{Bruhn2020Erratum:10.1088/1361-6587/aac870}
C.~Bruhn, R.~M. McDermott, C.~Angioni, J.~Ameres, V.~Bobkov, M.~Cavedon,
  R.~Dux, A.~Kappatou, A.~Lebschy, P.~Manas, and R.~Ochoukov.
\newblock {Erratum: A novel method of studying the core boron transport at
  ASDEX Upgrade (Plasma Physics and Controlled Fusion (2018) 60 (085011) DOI:
  10.1088/1361-6587/aac870)}, 2020.

\bibitem{Grierson2015ImpurityDIII-D}
B.~A. Grierson, K.~H. Burrell, R.~M. Nazikian, W.~M. Solomon, A.~M. Garofalo,
  E.~A. Belli, G.~M. Staebler, M.~E. Fenstermacher, G.~R. McKee, T.~E. Evans,
  D.~M. Orlov, S.~P. Smith, C.~Chrobak, and C.~Chrystal.
\newblock {Impurity confinement and transport in high confinement regimes
  without edge localized modes on DIII-D}.
\newblock {\em Physics of Plasmas}, 22(5):055901, 2015.

\bibitem{Odstrcil2020DependenceTokamak}
T.~Odstr{\v{c}}il, N.~T. Howard, F.~Sciortino, C.~Chrystal, C.~Holland,
  E.~Hollmann, G.~McKee, K.~E. Thome, and T.~M. Wilks.
\newblock {Dependence of the impurity transport on the dominant turbulent
  regime in ELM-y H-mode discharges on the DIII-D tokamak}.
\newblock {\em Physics of Plasmas}, 27(8):082503, 8 2020.

\bibitem{Cui2018StudyDepositions}
Z.~Y. Cui, K.~Zhang, S.~Morita, X.~Q. Ji, X.~T. Ding, Y.~Xu, P.~Sun, J.~M. Gao,
  C.~F. Dong, D.~L. Zheng, Y.~G. Li, M.~Jiang, D.~Li, W.~L. Zhong, Y.~Liu,
  Y.~B. Dong, S.~D. Song, L.~M. Yu, Z.~B. Shi, B.~Z. Fu, P.~Lu, M.~Huang, B.~S.
  Yuan, Q.~W. Yang, and X.~R. Duan.
\newblock {Study of impurity transport in HL-2A ECRH L-mode plasmas with
  radially different ECRH power depositions}.
\newblock {\em Nuclear Fusion}, 58(5), 2018.

\bibitem{Dux2004ImpurityJET}
R.~Dux, C.~Giroud, and K.~D. Zastrow.
\newblock {Impurity transport in internal transport barrier discharges on JET}.
\newblock {\em Nuclear Fusion}, 44(2):260--264, 2004.

\bibitem{Giroud2007MethodJET}
C.~Giroud, R.~Barnsley, P.~Buratti, I.~H. Coffey, M.~Von~Hellermann,
  C.~Jup{\'{e}}n, K.~D. Lawson, A.~Meigs, M.~O'Mullane, A.~D. Whiteford, and
  K.~D. Zastrow.
\newblock {Method for experimental determination of Z dependence of impurity
  transport on JET}.
\newblock {\em Nuclear Fusion}, 47(4):313--330, 4 2007.

\bibitem{Henderson2015ChargeMAST}
S.~S. Henderson, L.~Garzotti, F.~J. Casson, D.~Dickinson, M.~O'Mullane,
  A.~Patel, C.~M. Roach, H.~P. Summers, H.~Tanabe, and M.~Valovi{\v{c}}.
\newblock {Charge dependence of neoclassical and turbulent transport of light
  impurities on MAST}.
\newblock {\em Plasma Physics and Controlled Fusion}, 57(9):95001, 2015.

\bibitem{Delgado-Aparicio2009ImpurityPlasmas}
L.~Delgado-Aparicio, D.~Stutman, K.~Tritz, M.~Finkenthal, S.~Kaye, R.~Bell,
  R.~Kaita, B.~Leblanc, F.~Levinton, J.~Menard, S.~Paul, D.~Smith, and H.~Yuh.
\newblock {Impurity transport studies in NSTX neutral beam heated H-mode
  plasmas}.
\newblock {\em Nuclear Fusion}, 49(8), 2009.

\bibitem{Scavino2003EffectsTCV}
E.~Scavino, J.~S. Bakos, R.~Dux, and H.~Weisen.
\newblock {Effects of plasma shape on laser blow-off injected impurity
  transport in TCV}.
\newblock {\em Plasma Physics and Controlled Fusion}, 45(11):1961--1974, 2003.

\bibitem{Parisot2008ExperimentalPlasma}
T.~Parisot, R.~Guirlet, C.~Bourdelle, X.~Garbet, N.~Dubuit, F.~Imbeaux, and
  P.~R. Thomas.
\newblock {Experimental impurity transport and theoretical interpretation in a
  Tore Supra lower-hybrid heated plasma}.
\newblock {\em Plasma Physics and Controlled Fusion}, 50(5):1--18, 2008.

\bibitem{Villegas2014ExperimentalSupra}
D.~Villegas, R.~Guirlet, C.~Bourdelle, X.~Garbet, G.T. Hoang, R.~Sabot,
  F.~Imbeaux, and J.L. Segui.
\newblock {Experimental and theoretical study of nickel transport dependence on
  gradients in Tore Supra}.
\newblock {\em Nuclear Fusion}, 54:073011, 2014.

\bibitem{Geiger2019ObservationIron}
B.~Geiger, T.~Wegner, C.~D. Beidler, R.~Burhenn, B.~Buttensch{\"{o}}n, R.~Dux,
  A.~Langenberg, N.~A. Pablant, T.~P{\"{u}}tterich, Y.~Turkin, T.~Windisch,
  V.~Winters, M.~Beurskens, C.~Biedermann, K.~J. Brunner, G.~Cseh, H.~Damm,
  F.~Effenberg, G.~Fuchert, O.~Grulke, J.~H. Harris, C.~Killer, J.~Knauer,
  G.~Kocsis, A.~Kr{\"{a}}mer-Flecken, T.~Kremeyer, M.~Krychowiak, O.~Marchuk,
  D.~Nicolai, K.~Rahbarnia, G.~Satheeswaran, J.~Schilling, O.~Schmitz,
  T.~Schr{\"{o}}der, T.~Szepesi, H.~Thomsen, H.~Trimino~Mora, P.~Traverso, and
  D.~Zhang.
\newblock {Observation of anomalous impurity transport during low-density
  experiments in W7-X with laser blow-off injections of iron}.
\newblock {\em Nuclear Fusion}, 59(4):046009, 2019.

\bibitem{Rice1997XPlasmas}
J.~E. Rice, J.~L. Terry, E.~S. Marmar, and F~Bombarda.
\newblock {X ray observations of up-down impurity density asymmetries in
  alcator C-Mod plasmas}.
\newblock {\em Nuclear Fusion}, 37(2):241--249, 2 1997.

\bibitem{Rice2000ImpurityPlasmas}
J.~E. Rice, J.~A. Goetz, R.~S. Granetz, M.~J. Greenwald, A.~E. Hubbard, I.~H.
  Hutchinson, E.~S. Marmar, D.~Mossessian, T.~Sunn Pedersen, J.~A. Snipes,
  J.~L. Terry, and S.~M. Wolfe.
\newblock {Impurity toroidal rotation and transport in Alcator C-Mod ohmic high
  confinement mode plasmas}.
\newblock {\em Physics of Plasmas}, 7(5):1825--1830, 5 2000.

\bibitem{Pedersen2000RadialC-Mod}
T.~S. Pedersen, R.~S. Granetz, A.~E. Hubbard, I.~H. Hutchinson, E.~S. Marmar,
  J.~E. Rice, and J.~Terry.
\newblock {Radial impurity transport in the H mode transport barrier region in
  Alcator C-Mod}.
\newblock {\em Nuclear Fusion}, 40(10):1795--1804, 2000.

\bibitem{Rice2007ImpurityPlasmas}
J.~E. Rice, J.~L. Terry, E.~S. Marmar, R.~S. Granetz, M.~J. Greenwald, A.~E.
  Hubbard, J.~H. Irby, S.~M. Wolfe, and T.~Sunn~Pedersen.
\newblock {Impurity transport in alcator C-Mod plasmas}.
\newblock {\em Fusion Science and Technology}, 51(3):357--368, 2007.

\bibitem{Howard2012QuantitativePlasma}
N.~T. Howard, M.~Greenwald, D.~R. Mikkelsen, M.~L. Reinke, A.~E. White,
  D.~Ernst, Y.~Podpaly, and J.~Candy.
\newblock {Quantitative comparison of experimental impurity transport with
  nonlinear gyrokinetic simulation in an Alcator C-Mod L-mode plasma}.
\newblock {\em Nuclear Fusion}, 52(6), 2012.

\bibitem{Chilenski2018EfficientExperiments}
M.~A. Chilenski, M.~J. Greenwald, Y.~M. Marzouk, J.~E. Rice, and A.~E. White.
\newblock {Efficient design and verification of diagnostics for impurity
  transport experiments}.
\newblock {\em Review of Scientific Instruments}, 89(1):013504, 2018.

\bibitem{Howard2011CharacterizationSystem}
N.~T. Howard, M.~Greenwald, and J.~E. Rice.
\newblock {Characterization of impurity confinement on Alcator C-Mod using a
  multi-pulse laser blow-off system}.
\newblock {\em Review of Scientific Instruments}, 82(3):0--5, 2011.

\bibitem{Marmar1975SystemPlasmas}
E.~S. Marmar, J.~L. Cecchi, and S.~A. Cohen.
\newblock {System for rapid injection of metal atoms into plasmas}.
\newblock {\em Review of Scientific Instruments}, 46(9):1149--1154, 1975.

\bibitem{Wiesen2015ThePackage}
S.~Wiesen, D.~Reiter, V.~Kotov, M.~Baelmans, W.~Dekeyser, A.~S. Kukushkin,
  S.~W. Lisgo, R.~A. Pitts, V.~Rozhansky, G.~Saibene, I.~Veselova, and
  S.~Voskoboynikov.
\newblock {The new SOLPS-ITER code package}.
\newblock {\em Journal of Nuclear Materials}, 463:480--484, 7 2015.

\bibitem{Bonnin2016PresentationModelling}
Xavier Bonnin, Wouter Dekeyser, Richard Pitts, David Coster, Serguey
  Voskoboynikov, and Sven Wiesen.
\newblock {Presentation of the New SOLPS-ITER Code Package for Tokamak Plasma
  Edge Modelling}.
\newblock {\em Plasma and Fusion Research}, 11:1--6, 2016.

\bibitem{Reiter2005TheCodes}
D.~Reiter, M.~Baelmans, and P.~B{\"{o}}rner.
\newblock {The eirene and B2-eirene codes}.
\newblock {\em Fusion Science and Technology}, 47(2):172--186, 2005.

\bibitem{Sciortino2021ModelingAurora_2}
F.~Sciortino, T.~Odstr{\v{c}}il, A.~Cavallaro, S.~Smith, O.~Meneghini,
  R.~Reksoatmodjo, O.~Linder, J.~D. Lore, N.~T. Howard, E.~S. Marmar, and
  S.~Mordijck.
\newblock {Modeling of Particle Transport, Neutrals and Radiation in
  Magnetically-Confined Plasmas with Aurora}.
\newblock {\em Plasma Physics and Controlled Fusion}, 9 2021.

\bibitem{Hutchinson1994FirstAlcator-C-MOD}
I.~H. Hutchinson, R.~Boivin, F.~Bombarda, P.~Bonoli, S.~Fairfax, C.~Fiore,
  J.~Goetz, S.~Golovato, R.~Granetz, M.~Greenwald, S.~Horne, A.~Hubbard,
  J.~Irby, B.~LaBombard, B.~Lipschultz, E.~Marmar, G.~McCracken, M.~Porkolab,
  J.~Rice, J.~Snipes, Y.~Takase, J.~Terry, S.~Wolfe, C.~Christensen,
  D.~Garnier, M.~Graf, T.~Hsu, T.~Luke, M.~May, A.~Niemczewski, G.~Tinios,
  J.~Schachter, and J.~Urbahn.
\newblock {First results from Alcator-C-MOD}.
\newblock {\em Physics of Plasmas}, 1(5):1511--1518, 1994.

\bibitem{Greenwald201420Tokamak}
M.~Greenwald, A.~Bader, S.~Baek, M.~Bakhtiari, H.~Barnard, W.~Beck,
  W.~Bergerson, I.~Bespamyatnov, P.~Bonoli, D.~Brower, D.~Brunner, W.~Burke,
  J.~Candy, M.~Churchill, I.~Cziegler, A.~Diallo, A.~Dominguez, B.~Duval,
  E.~Edlund, P.~Ennever, D.~Ernst, I.~Faust, C.~Fiore, T.~Fredian, O.~Garcia,
  C.~Gao, J.~Goetz, T.~Golfinopoulos, R.~Granetz, O.~Grulke, Z.~Hartwig,
  S.~Horne, N.~Howard, A.~Hubbard, J.~Hughes, I.~Hutchinson, J.~Irby, V.~Izzo,
  C.~Kessel, B.~Labombard, C.~Lau, C.~Li, Y.~Lin, B.~Lipschultz, A.~Loarte,
  E.~Marmar, A.~Mazurenko, G.~McCracken, R.~McDermott, O.~Meneghini,
  D.~Mikkelsen, D.~Mossessian, R.~Mumgaard, J.~Myra, E.~Nelson-Melby,
  R.~Ochoukov, G.~Olynyk, R.~Parker, S.~Pitcher, Y.~Podpaly, M.~Porkolab,
  M.~Reinke, J.~Rice, W.~Rowan, A.~Schmidt, S.~Scott, S.~Shiraiwa, J.~Sierchio,
  N.~Smick, J.~A. Snipes, P.~Snyder, B.~Sorbom, J.~Stillerman, C.~Sung,
  Y.~Takase, V.~Tang, J.~Terry, D.~Terry, C.~Theiler, A.~Tronchin-James,
  N.~Tsujii, R.~Vieira, J.~Walk, G.~Wallace, A.~White, D.~Whyte, J.~Wilson,
  S.~Wolfe, G.~Wright, J.~Wright, S.~Wukitch, and S.~Zweben.
\newblock {20 years of research on the Alcator C-Mod tokamak}.
\newblock {\em Physics of Plasmas}, 21(11):110501, 2014.

\bibitem{Kallne1985HighPlasmas}
E.~K{\"{a}}llne, J.~K{\"{a}}llne, E.~S. Marmar, and J.~E. Rice.
\newblock {High resolution x-ray spectroscopy diagnostics of high temperature
  plasmas}.
\newblock {\em Physica Scripta}, 31(6):551--564, 1985.

\bibitem{Rice1995X-rayTokamak}
J.~E. Rice, M.~A. Graf, J.~L. Terry, E.~S. Marmar, K.~Giesing, and F.~Bombarda.
\newblock {X-ray observations of helium-like scandium from the alcator c-mod
  tokamak}.
\newblock {\em Journal of Physics B: Atomic, Molecular and Optical Physics},
  28(5):893--905, 1995.

\bibitem{Bitter2003NewIons}
M.~Bitter, M.~F. Gu, L.~A. Vainshtein, P.~Beiersdorfer, G.~Bertschinger,
  O.~Marchuk, R.~Bell, B.~LeBlanc, K.~W. Hill, D.~Johnson, and L.~Roquemore.
\newblock {New benchmarks from tokamak experiments for theoretical calculations
  of the dielectronic satellite spectra of heliumlike ions}.
\newblock {\em Physical Review Letters}, 91(26), 2003.

\bibitem{Marchuk2006ComparisonTEXTOR}
O.~Marchuk, M.~Z. Tokar, G.~Bertschinger, A.~Urnov, H.~J. Kunze, D.~Pilipenko,
  X.~Loozen, D.~Kalupin, D.~Reiter, A.~Pospieszczyk, W.~Biel, M.~Goto, and
  F.~Goryaev.
\newblock {Comparison of impurity transport model with measurements of He-like
  spectra of argon at the tokamak TEXTOR}.
\newblock {\em Plasma Physics and Controlled Fusion}, 48(11):1633--1646, 11
  2006.

\bibitem{Rice2014X-rayPlasmas}
J.~E. Rice, M.~L. Reinke, J.~M.A. Ashbourn, C.~Gao, M.~M. Victora, M.~A.
  Chilenski, L.~Delgado-Aparicio, N.~T. Howard, A.~E. Hubbard, J.~W. Hughes,
  and J.~H. Irby.
\newblock {X-ray observations of Ca19 +, Ca18 + and satellites from Alcator
  C-Mod tokamak plasmas}.
\newblock {\em Journal of Physics B: Atomic, Molecular and Optical Physics},
  47(7):075701, 2014.

\bibitem{Rice2015X-rayPlasmas}
J.~E. Rice, M.~L. Reinke, J.~M.A. Ashbourn, C.~Gao, M.~Bitter,
  L.~Delgado-Aparicio, K.~Hill, N.~T. Howard, J.~W. Hughes, and U.~I.
  Safronova.
\newblock {X-ray observations of medium Z H- and He-like ions with satellites
  from C-Mod tokamak plasmas}.
\newblock {\em Journal of Physics B: Atomic, Molecular and Optical Physics},
  48(14):1--8, 2015.

\bibitem{Gabriel1969InterpretationIntensities}
A.~H. Gabriel and C.~Jordan.
\newblock {Interpretation of Solar Helium-Like Ion Line Intensities}.
\newblock {\em Monthly Notices of the Royal Astronomical Society},
  145(2):241--248, 7 1969.

\bibitem{Mewe19722InterpolationNe-Sequences}
R.~Mewe.
\newblock {Interpolation Formulae for the Electron Impact Excitation of Ions in
  the H-, He-, Li- and Ne-Sequences}.
\newblock {\em Astronomy and Astrophysics}, 20:215--221, 19722.

\bibitem{MeweR.andSchrijver1978Helium-likeIntensities}
R.~Mewe and J.~Schrijver.
\newblock {Helium-like Ion Line Intensities}.
\newblock {\em Astronomy and Astrophysics}, 65:99--114, 1978.

\bibitem{Porquet2000X-rayNuclei}
D.~Porquet and J.~Dubau.
\newblock {X-ray photoionized plasma diagnostics with helium-like ions.
  Application to warm absorber-emitter in active galactic nuclei}.
\newblock {\em Astronomy and Astrophysics Supplement Series}, 143(3):495--514,
  5 2000.

\bibitem{Porquet2001LinePlasmas}
D.~Porquet, R.~Mewe, J.~Dubau, A.~J.J. Raassen, and J.~S. Kaastra.
\newblock {Line ratios for helium-like ions: Applications to
  collision-dominated plasmas}.
\newblock {\em Astronomy and Astrophysics}, 376(3):1113--1122, 2001.

\bibitem{Porquet2010He-likeNuclei}
D.~Porquet, J.~Dubau, and N.~Grosso.
\newblock {He-like ions as practical astrophysical plasma diagnostics: From
  stellar coronae to active galactic nuclei}.
\newblock {\em Space Science Reviews}, 157(1-4):103--134, 2010.

\bibitem{Vainshtein1978WavelengthsIons}
L.~A. Vainshtein and U.~I. Safronova.
\newblock {Wavelengths and transition probabilities of satellites to resonance
  lines of H- and He-like ions}, 1978.

\bibitem{Bely-Dubau1982DielectronicSpectra}
F.~Bely-Dubau, J.~Dubau, P.~Faucher, A.~H. Gabriel, M.~Loulergue,
  L.~Steenman-Clark, S.~Volonte, E.~Antonucci, and C.~G. Rapley.
\newblock {Dielectronic satellite spectra for highly-charged helium-like ions -
  VII: Ca spectra}.
\newblock {\em Monthly Notices of the Royal Astronomical Society},
  201:1155--1169, 1982.

\bibitem{Ince-Cushman2008Spatiallyinvited}
A~Ince-Cushman, J~E Rice, M~Bitter, M~L Reinke, K~W Hill, M~F Gu, E~Eikenberry,
  Ch~Broennimann, S~Scott, Y~Podpaly, S~G Lee, and E~S Marmar.
\newblock {Spatially resolved high resolution x-ray spectroscopy for
  magnetically confined fusion plasmas (invited)}.
\newblock In {\em Review of Scientific Instruments}, volume~79, pages 10--302,
  2008.

\bibitem{Reinke2012X-rayResearch}
M.~L. Reinke, Y.~A. Podpaly, M.~Bitter, I.~H. Hutchinson, J.~E. Rice,
  L.~Delgado-Aparicio, and C.~Gao.
\newblock {X-ray imaging crystal spectroscopy for use in plasma transport
  research}.
\newblock {\em Review of Scientific Instruments}, 83(113504), 2012.

\bibitem{Foster2020PyAtomDB:Uncertainties}
A.~R. Foster and K.~Heuer.
\newblock {PyAtomDB: Extending the AtomDB atomic database to model new plasma
  processes and uncertainties}.
\newblock {\em Atoms}, 8(3):49, 9 2020.

\bibitem{Liang2011R-matrixDamping}
G.~Y. Liang and N.~R. Badnell.
\newblock {R-matrix electron-impact excitation data for the Li-like
  iso-electronic sequence including Auger and radiation damping}.
\newblock {\em Astronomy and Astrophysics}, 528:A69, 4 2011.

\bibitem{Badnell2006RadiativePlasmas}
N.~R. Badnell.
\newblock {Radiative Recombination Data for Modeling Dynamic Finite‐Density
  Plasmas}.
\newblock {\em The Astrophysical Journal Supplement Series}, 167(2):334--342,
  2006.

\bibitem{Bautista2007DielectronicSequence}
M.~A. Bautista and N.~R. Badnell.
\newblock {Dielectronic recombination data for dynamic finite-density plasmas
  XII. The helium isoelectronic sequence}.
\newblock {\em Astronomy and Astrophysics}, 466(2):755--762, 5 2007.

\bibitem{Palmeri2008RadiativeSequences}
P.~Palmeri, P.~Quinet, C.~Mendoza, M.~A. Bautista, J.~Garc{\'{i}}a, and T.~R.
  Kallman.
\newblock {Radiative and Auger Decay of K‐Vacancy Levels in the Ne, Mg, Si,
  S, Ar, and Ca Isonuclear Sequences}.
\newblock {\em The Astrophysical Journal Supplement Series}, 177(1):408--416, 7
  2008.

\bibitem{NIST_ASD}
A.~Kramida, {Yu.~Ralchenko}, J.~Reader, and {and NIST ASD Team}.
\newblock {NIST Atomic Spectra Database (ver. 5.8), [Online]. Available:
  {\tt{https://physics.nist.gov/asd}} [2021, May 26]. National Institute of
  Standards and Technology, Gaithersburg, MD.}, 2020.

\bibitem{Li2015Radiative42}
S.~Li, J.~Yan, C.~Y. Li, R.~Si, X.~L. Guo, M.~Huang, C.~Y. Chen, and Y.~M. Zou.
\newblock {Radiative rates and electron-impact excitation for the n ≤ 6
  fine-structure levels in H-like ions with 13 ≤ Z ≤ 42}.
\newblock {\em Astronomy and Astrophysics}, 583:A82, 11 2015.

\bibitem{Gu2008TheCode}
M.~F. Gu.
\newblock {The flexible atomic code}.
\newblock {\em Canadian Journal of Physics}, 86(5):675--689, 2008.

\bibitem{Schlummer2014ChargeTEXTOR}
T.~Schlummer.
\newblock {\em {Charge exchange recombination in X-ray spectra of He-like argon
  measured at the tokamak TEXTOR}}.
\newblock PhD thesis, Heinrich-Heine-Universit{\"{a}}t D{\"{u}}sseldorf, 2014.

\bibitem{Rice1987RadialTokamak}
J.~E. Rice, E.~S. Marmar, E.~K{\"{a}}llne, and J.~K{\"{a}}llne.
\newblock {Radial profiles of ground-state transitions of heliumlike argon from
  the Alcator-C tokamak}.
\newblock {\em Physical Review A}, 35(7):3033--3045, 4 1987.

\bibitem{Rice1999ThePlasmas}
J.~E. Rice, K.~B. Fournier, U.~I. Safronova, J.~A. Goetz, S.~Gutmann, A.~E.
  Hubbard, J.~Irby, B.~LaBombard, E.~S. Marmar, and J.~L. Terry.
\newblock {The Rydberg series of helium-like Cl, Ar and S and their high-n
  satellites in tokamak plasmas}.
\newblock {\em New Journal of Physics}, 1:19--20, 1999.

\bibitem{Reinke2010VacuumTokamak}
M.~L. Reinke, P.~Beiersdorfer, N.~T. Howard, E.~W. Magee, Y.~Podpaly, J.~E.
  Rice, and J.~L. Terry.
\newblock {Vacuum ultraviolet impurity spectroscopy on the Alcator C-Mod
  tokamak}.
\newblock In {\em Review of Scientific Instruments}, volume~81, page 10D736,
  2010.

\bibitem{Howard2014ImpurityPlasmas}
N.~T. Howard, A.~E. White, M.~Greenwald, C.~Holland, J.~Candy, and J.~E. Rice.
\newblock {Impurity transport, turbulence transitions and intrinsic rotation in
  Alcator C-Mod plasmas}.
\newblock {\em Plasma Physics and Controlled Fusion}, 56(12):124004, 12 2014.

\bibitem{Chilenski2017ExperimentalSimulations}
M.~A. Chilenski.
\newblock {\em {Experimental Data Analysis Techniques for Validation of Tokamak
  Impurity Transport Simulations}}.
\newblock PhD thesis, MIT, 2017.

\bibitem{Summers2006IonizationElements}
H.~P. Summers, W.~J. Dickson, M.~G. O'Mullane, N.~R. Badnell, A.~D. Whiteford,
  D.~H. Brooks, J.~Lang, S.~D. Loch, and D.~C. Griffin.
\newblock {Ionization state, excited populations and emission of impurities in
  dynamic finite density plasmas: I. The generalized collisional-radiative
  model for light elements}.
\newblock {\em Plasma Physics and Controlled Fusion}, 48(2):263--293, 2006.

\bibitem{Greenwald2000StudiesC-Mod}
M.~Greenwald, R.~Boivin, P.~Bonoli, C.~Fiore, J.~Goetz, R.~Granetz, A.~Hubbard,
  I.~Hutchinson, J.~Irby, Y.~Lin, E.~Marmar, A.~Mazurenko, D.~Mossessian,
  T.~Sunn Pedersen, J.~Rice, J.~Snipes, G.~Schilling, G.~Taylor, J.~Terry,
  S.~Wolfe, and S.~Wukitch.
\newblock {Studies of EDA H-mode in Alcator C-Mod}.
\newblock {\em Plasma Physics and Controlled Fusion}, 42(SUPPL. 5A):A263--A269,
  2000.

\bibitem{Whyte2010I-mode:C-Mod}
D.~G. Whyte, A.~E. Hubbard, J.~W. Hughes, B.~Lipschultz, J.~E. Rice, E.~S.
  Marmar, M.~Greenwald, I.~Cziegler, A.~Dominguez, T.~Golfinopoulos, N.~Howard,
  L.~Lin, R.~M. McDermottb, M.~Porkolab, M.~L. Reinke, J.~Terry, N.~Tsujii,
  S.~Wolfe, S.~Wukitch, and Y.~Lin.
\newblock {I-mode: An H-mode energy confinement regime with L-mode particle
  transport in Alcator C-Mod}.
\newblock {\em Nuclear Fusion}, 50(10):105005, 2010.

\bibitem{Hughes2001High-resolutionTokamak}
J.~W. Hughes, D.~A. Mossessian, A.~E. Hubbard, E.~S. Marmar, D.~Johnson, and
  D.~Simon.
\newblock {High-resolution edge Thomson scattering measurements on the Alcator
  C-Mod tokamak}.
\newblock {\em Review of Scientific Instruments}, 72(1 II):1107--1110, 1 2001.

\bibitem{Hughes2013PedestalC-Mod}
J.~W. Hughes, P.~B. Snyder, J.~R. Walk, E.~M. Davis, A.~Diallo, B.~Labombard,
  S.~G. Baek, R.~M. Churchill, M.~Greenwald, R.~J. Groebner, A.~E. Hubbard,
  B.~Lipschultz, E.~S. Marmar, T.~Osborne, M.~L. Reinke, J.~E. Rice,
  C.~Theiler, J.~Terry, A.~E. White, D.~G. Whyte, S.~Wolfe, and X.~Q. Xu.
\newblock {Pedestal structure and stability in H-mode and I-mode: A comparative
  study on Alcator C-Mod}.
\newblock {\em Nuclear Fusion}, 53(4), 2013.

\bibitem{Basse2007DiagnosticC-Mod}
N.~P. Basse, A.~Dominguez, E.~M. Edlund, C.~L. Fiore, R.~S. Granetz, A.~E.
  Hubbard, J~W Hughes, I~H Hutchinson, J~H Irby, B.~LaBombard, L~Lin, Y.~Lin,
  B~Lipschultz, J~E Liptac, E~S Marmar, D~A Mossessian, R~R Parker, M~Porkolab,
  J~E Rice, J~A Snipes, V~Tang, J~L Terry, S~M Wolfe, S~J Wukitch, K~Zhurovich,
  R~V Bravenec, P~E Phillips, W~L Rowan, G~J Kramer, G.~Schilling, S~D Scott,
  and S~J Zweben.
\newblock {Diagnostic systems on Alcator C-Mod}.
\newblock {\em Fusion Science and Technology}, 51(3):476--507, 2007.

\bibitem{Groebner1998CriticalDIII-D}
R.~J. Groebner and T.~M. Carlstrom.
\newblock {Critical edge parameters for H-mode transition in DIII-D}.
\newblock {\em Plasma Physics and Controlled Fusion}, 40(673), 1998.

\bibitem{Lao1985ReconstructionTokamaks}
L.~L. Lao, H.~St John, R.~D. Stambaugh, A.~G. Kellman, and W.~Pfeiffer.
\newblock {Reconstruction of current profile parameters and plasma shapes in
  tokamaks}.
\newblock {\em Nuclear Fusion}, 25(11):1611--1622, 1985.

\bibitem{stangeby_book}
P.C. Stangeby.
\newblock {\em The Plasma Boundary of Magnetic Fusion Devices}.
\newblock CRC Press, 1 edition, 2000.

\bibitem{Brunner2018High-resolutionTokamak}
D.~Brunner, B.~Labombard, A.~Q. Kuang, and J.~L. Terry.
\newblock {High-resolution heat flux width measurements at reactor-level
  magnetic fields and observation of a unified width scaling across confinement
  regimes in the Alcator C-Mod tokamak}.
\newblock {\em Nuclear Fusion}, 58(9):094002, 7 2018.

\bibitem{Boivin2000EffectsPlasmas}
R.~L. Boivin, J.~A. Goetz, A.~E. Hubbard, J.~W. Hughes, I.~H. Hutchinson, J.~H.
  Irby, B.~LaBombard, E.~S. Marmar, D.~Mossessian, C.~S. Pitcher, J.~L. Terry,
  B.~A. Carreras, and L.~W. Owen.
\newblock {Effects of neutral particles on edge dynamics in Alcator C-Mod
  plasmas}.
\newblock {\em Physics of Plasmas}, 7(5):1919--1926, 5 2000.

\bibitem{Hughes2006AdvancesTokamak}
J.~W. Hughes, B.~LaBombard, D.~A. Mossessian, A.~E. Hubbard, J.~Terry, and
  T.~Biewer.
\newblock {Advances in measurement and modeling of the high-confinement-mode
  pedestal on the Alcator C-Mod tokamak}.
\newblock In {\em Physics of Plasmas}, volume~13, page 056103. American
  Institute of PhysicsAIP, 5 2006.

\bibitem{Reksoatmodjo2021TheExperiments}
R.~Reksoatmodjo, S.~Mordijck, J.~W. Hughes, J.~D. Lore, and X.~Bonnin.
\newblock {The role of edge fueling in determining the pedestal density in high
  neutral opacity Alcator C-Mod experiments}.
\newblock {\em Nuclear Materials and Energy}, 27:100971, 6 2021.

\bibitem{Janev1993CrossIons}
R.~K. Janev and J.~J. Smith.
\newblock {Cross Sections for Collisions Processes of Hydrogen Atoms with
  Electrons, Protons and Multiply Charged Ions}.
\newblock {\em Nuclear Fusion}, 4:195, 1993.

\bibitem{RR_2021}
R.~Reksoatmodjo, F.~Sciortino, S.~Mordijck, J.W. Hughes, J.D. Lore, and
  X.~Bonnin.
\newblock {An Alcator C-Mod neutral model validation study of SOLPS-ITER and
  KN1D}.
\newblock {\em In preparation}, 2021.

\bibitem{Scotti2021OuterNSTX-U}
F~Scotti, D~P Stotler, R~E Bell, B.~P. LeBlanc, S~A Sabbagh, V~A Soukhanovskii,
  M~V Umansky, and S~J Zweben.
\newblock {Outer midplane neutral density measurements and H-mode fueling
  studies in NSTX-U}.
\newblock {\em Nuclear Fusion}, 61(3), 2021.

\bibitem{Rice1986ObservationPlasma}
J.~E. Rice, E.~S. Marmar, J.~L. Terry, E.~Kallne, and J.~Kallne.
\newblock {Observation of charge-transfer population of high-n levels in Ar+16
  from neutral hydrogen in the ground and excited states in a tokamak plasma}.
\newblock {\em Physical Review Letters}, 56(1):50--53, 1 1986.

\bibitem{Kato1991EffectsTokamak}
T.~Kato, K.~Masai, T.~Fujimoto, F.~Koike, E.~K{\"{a}}llne, E.~S. Marmar, and
  J.~E. Rice.
\newblock {Effects of state-selective charge-exchange processes on the He-like
  spectra from the Alcator C tokamak}.
\newblock {\em Physical Review A}, 44(10):6776--6784, 1991.

\bibitem{Rice2021ThePlasmas}
J.~E. Rice, F.~Sciortino, M.~Gu, N.~Cao, J.~W. Hughes, J.~H. Irby, E.~S.
  Marmar, S.~Mordijck, M.~L. Reinke, and R.~Reksoatmodjo.
\newblock {The Very High n Rydberg Series of Ar16+ in Alcator C-Mod Tokamak
  Plasmas}.
\newblock {\em Journal of Physics B: Atomic, Molecular and Optical Physics}, 9
  2021.

\bibitem{Dux2006STRAHLManual}
R.~Dux.
\newblock {STRAHL User Manual}.
\newblock Technical report, 2006.

\bibitem{Dux2004ImpurityPlasmas}
R.~Dux.
\newblock {Impurity Transport in Tokamak Plasmas}.
\newblock Technical report, Max-Planck-Institut fuer Plasmaphysik, 2004.

\bibitem{Meneghini2015IntegratedOMFIT}
O.~Meneghini, S.~P. Smith, L.~L. Lao, O.~Izacard, Q.~Ren, J.~M. Park, J.~Candy,
  Z.~Wang, C.~J. Luna, V.~A. Izzo, B.~A. Grierson, P.~B. Snyder, C.~Holland,
  J.~Penna, G.~Lu, P.~Raum, A.~McCubbin, D.~M. Orlov, E.~A. Belli, N.~M.
  Ferraro, R.~Prater, T.~H. Osborne, A.~D. Turnbull, and G.~M. Staebler.
\newblock {Integrated modeling applications for tokamak experiments with
  OMFIT}.
\newblock {\em Nuclear Fusion}, 55(8):083008, 8 2015.

\bibitem{Dux2020InfluenceUpgrade}
R.~Dux, M.~Cavedon, A.~Kallenbach, R.~M. McDermott, G.~Vogel, and The Asdex
  Upgrade~Team.
\newblock {Influence of CX-reactions on the radiation in the pedestal region at
  ASDEX Upgrade}.
\newblock {\em Nuclear Fusion}, 60(12), 2020.

\bibitem{Sciortino2021ParticleLines}
F.~Sciortino, N.~M. Cao, N.~T. Howard, E.~S. Marmar, and J.~E. Rice.
\newblock {Particle transport constraints via Bayesian spectral fitting of
  multiple atomic lines}.
\newblock {\em Review of Scientific Instruments}, 92:53508, 2021.

\bibitem{Chilenski2019OnProfiles}
M.~A. Chilenski, M.~J. Greenwald, Y.~Marzouk, J.~E. Rice, and A.~E. White.
\newblock {On the importance of model selection when inferring impurity
  transport coefficient profiles}.
\newblock {\em Plasma Physics and Controlled Fusion}, 61(12):125012, 2019.

\bibitem{Skilling2006NestedComputation}
J.~Skilling.
\newblock {Nested sampling for general Bayesian computation}.
\newblock {\em Bayesian Analysis}, 1(4):833--860, 2006.

\bibitem{Feroz2008MultimodalAnalyses}
F.~Feroz and M.~P. Hobson.
\newblock {Multimodal nested sampling: An efficient and robust alternative to
  Markov Chain Monte Carlo methods for astronomical data analyses}.
\newblock {\em Monthly Notices of the Royal Astronomical Society},
  384(2):449--463, 2008.

\bibitem{Feroz2013ImportanceAlgorithm}
F.~Feroz, M.~P. Hobson, E.~Cameron, and A.~N. Pettitt.
\newblock {Importance Nested Sampling and the MultiNest Algorithm}.
\newblock {\em arXiv}, pages 1--28, 2013.

\bibitem{Buchner2014X-rayCatalogue}
J.~Buchner, A.~Georgakakis, K.~Nandra, L.~Hsu, C.~Rangel, M.~Brightman,
  A.~Merloni, M.~Salvato, J.~Donley, and D.~Kocevski.
\newblock {X-ray spectral modelling of the AGN obscuring region in the CDFS:
  Bayesian model selection and catalogue}.
\newblock {\em Astronomy and Astrophysics}, 564:1--25, 2014.

\bibitem{Hughes2003ThomsonC-Mod}
J.~W. Hughes, D.~Mossessia, K.~Zhurovich, M.~DeMaria, K.~Jensen, and
  A.~Hubbard.
\newblock {Thomson scattering upgrades on Alcator C-Mod}.
\newblock In {\em Review of Scientific Instruments}, volume~74, pages
  1667--1670, 3 2003.

\bibitem{Chilenski2015ImprovedRegression}
M.~A. Chilenski, M.~Greenwald, Y.~Marzouk, N.~T. Howard, A.~E. White, J.~E.
  Rice, and J.~R. Walk.
\newblock {Improved profile fitting and quantification of uncertainty in
  experimental measurements of impurity transport coefficients using Gaussian
  process regression}.
\newblock {\em Nuclear Fusion}, 55(2):23012, 2015.

\bibitem{Belli2008KineticDynamics}
E.~A. Belli and J.~Candy.
\newblock {Kinetic calculation of neoclassical transport including
  self-consistent electron and impurity dynamics}.
\newblock {\em Plasma Physics and Controlled Fusion}, 50(9):0--32, 2008.

\bibitem{Staebler2005Gyro-LandauParticles}
G~M Staebler, J~E Kinsey, and R~E Waltz.
\newblock {Gyro-Landau fluid equations for trapped and passing particles}.
\newblock {\em Physics of Plasmas}, 12(10):1--24, 2005.

\bibitem{Staebler2016TheTurbulence}
G.~M. Staebler, J.~Candy, N.~T. Howard, and C.~Holland.
\newblock {The role of zonal flows in the saturation of multi-scale gyrokinetic
  turbulence}.
\newblock {\em Physics of Plasmas}, 23(6), 2016.

\bibitem{Candy2016APlasmas}
J.~Candy, E.~A. Belli, and R.~V. Bravenec.
\newblock {A high-accuracy Eulerian gyrokinetic solver for collisional
  plasmas}.
\newblock {\em Journal of Computational Physics}, 324:73--93, 2016.

\bibitem{Candy2003AnomalousSimulation}
J.~Candy and R.~E. Waltz.
\newblock {Anomalous Transport Scaling in the DIII-D Tokamak Matched by
  Supercomputer Simulation}.
\newblock {\em Physical Review Letters}, 91(4), 2003.

\bibitem{Candy2003AnSolver}
J.~Candy and R.~E. Waltz.
\newblock {An Eulerian gyrokinetic-Maxwell solver}.
\newblock {\em Journal of Computational Physics}, 186(2):545--581, 4 2003.

\bibitem{Rice2015CorePlasmas}
J.~E. Rice, M.~L. Reinke, C.~Gao, N.~T. Howard, M.~A. Chilenski,
  L.~Delgado-Aparicio, R.~S. Granetz, M.~J. Greenwald, A.~E. Hubbard, J.~W.
  Hughes, J.~H. Irby, Y.~Lin, E.~S. Marmar, R.~T. Mumgaard, S.~D. Scott, J.~L.
  Terry, J.~R. Walk, A.~E. White, D.~G. Whyte, S.~M. Wolfe, and S.~J. Wukitch.
\newblock {Core impurity transport in Alcator C-Mod L-, I- and H-mode plasmas}.
\newblock {\em Nuclear Fusion}, 55(3):033014, 2015.

\bibitem{Hughes2007H-modeC-Mod}
J.~W. Hughes, A.~E. Hubbard, D.~A. Mossessian, B.~LaBombard, T.~M. Biewer,
  R.~S. Granetz, M.~Greenwald, I.~H. Hutchinson, J.~H. Irby, Y.~Lin, E.~S.
  Marmar, M.~Porkolab, J.~E. Rice, J.~A. Snipes, J.~L. Terry, S.~Wolfe, and
  K.~Zhurovich.
\newblock {H-mode pedestal and L-H transition studies on Alcator C-Mod}.
\newblock {\em Fusion Science and Technology}, 51(3):317--341, 2007.

\bibitem{Breslau2018TRANSP}
J.~Breslau, M.~Gorelenkova, F.~Poli, J.~Sachdev, X.~Yuan, and USDOE Office~of
  Science.
\newblock {TRANSP}, 2018.

\bibitem{Candy2009TokamakSimulation}
J.~Candy, C.~Holland, R.~E. Waltz, M.~R. Fahey, and E.~Belli.
\newblock {Tokamak profile prediction using direct gyrokinetic and neoclassical
  simulation}.
\newblock {\em Physics of Plasmas}, 16(6), 2009.

\bibitem{Angioni2007ScalingObservations}
C.~Angioni, H.~Weisen, O.~J.W.F. Kardaun, M.~Maslov, A.~Zabolotsky, C.~Fuchs,
  L.~Garzotti, C.~Giroud, B.~Kurzan, P.~Mantica, A.~G. Peeters, and J.~Stober.
\newblock {Scaling of density peaking in H-mode plasmas based on a combined
  database of AUG and JET observations}.
\newblock {\em Nuclear Fusion}, 47(9):1326--1335, 2007.

\bibitem{Casson2010GyrokineticPlasma}
F.~J. Casson, A.~G. Peeters, C.~Angioni, Y.~Camenen, W.~A. Hornsby, A.~P.
  Snodin, and G.~Szepesi.
\newblock {Gyrokinetic simulations including the centrifugal force in a
  rotating tokamak plasma}.
\newblock {\em Physics of Plasmas}, 17(10), 2010.

\bibitem{Angioni2012Off-diagonalAspects}
C.~Angioni, Y.~Camenen, F.~J. Casson, E.~Fable, R.~M. McDermott, A.~G. Peeters,
  and J.~E. Rice.
\newblock {Off-diagonal particle and toroidal momentum transport: A survey of
  experimental, theoretical and modelling aspects}, 2012.

\bibitem{Angioni2014TungstenModelling}
C.~Angioni, P.~Mantica, T.~P{\"{u}}tterich, M.~Valisa, M.~Baruzzo, E.~A. Belli,
  P.~Belo, F.~J. Casson, C.~Challis, P.~Drewelow, C.~Giroud, N.~Hawkes, T.~C.
  Hender, J.~Hobirk, T.~Koskela, L.~Lauro~Taroni, C.~F. Maggi, J.~Mlynar,
  T.~Odstrcil, M.~L. Reinke, and M.~Romanelli.
\newblock {Tungsten transport in JET H-mode plasmas in hybrid scenario,
  experimental observations and modelling}.
\newblock {\em Nuclear Fusion}, 54(8):083028, 2014.

\bibitem{Greenwald2007DensityScalings}
M.~Greenwald, C.~Angioni, J.~W. Hughes, J.~Terry, and H.~Weisen.
\newblock {Density profile peaking in low collisionality H-modes: Comparison of
  Alcator C-Mod data to ASDEX Upgrade/JET scalings}.
\newblock {\em Nuclear Fusion}, 47(9):L26–L29, 2007.

\bibitem{Wilkie2018FirstIons}
G.~J. Wilkie, A.~Iantchenko, I.~G. Abel, E.~Highcock, and I.~Pusztai.
\newblock {First principles of modelling the stabilization of microturbulence
  by fast ions}.
\newblock {\em Nuclear Fusion}, 58(8), 2018.

\bibitem{Eriksson2019ImpactModeling}
F.~Eriksson, M.~Oberparleiter, A.~Skyman, H.~Nordman, P.~Strand, A.~Salmi, and
  T.~Tala.
\newblock {Impact of fast ions on density peaking in JET: Fluid and gyrokinetic
  modeling}.
\newblock {\em Plasma Physics and Controlled Fusion}, 61(7):075008, 5 2019.

\bibitem{Handley2015POLYCHORD:Sampling}
W.~J. Handley, M.~P. Hobson, and A.~N. Lasenby.
\newblock {POLYCHORD: Next-generation nested sampling}.
\newblock {\em Monthly Notices of the Royal Astronomical Society},
  453(4):4384--4398, 2015.

\bibitem{Belli2018ImpactTransport}
E.~A. Belli and J.~Candy.
\newblock {Impact of centrifugal drifts on ion turbulent transport}.
\newblock {\em Phys. Plasmas}, 032301(December 2017), 2018.

\bibitem{Casson2015TheoreticalUpgrade}
F.~J. Casson, C.~Angioni, E.~A. Belli, R.~Bilato, P.~Mantica,
  T.~Odstr{\v{c}}il, T.~P{\"{u}}tterich, M.~Valisa, L.~Garzotti, C.~Giroud,
  J.~Hobirk, C.~F. Maggi, J.~Mlynar, and M.~L. Reinke.
\newblock {Theoretical description of heavy impurity transport and its
  application to the modelling of tungsten in JET and ASDEX upgrade}.
\newblock {\em Plasma Physics and Controlled Fusion}, 57(1):14031, 2015.

\end{thebibliography}

\end{document}